# Formation of the four terrestrial planets in the Jupiter–Saturn chaotic excitation scenario: fundamental properties and water delivery

Patryk Sofia Lykawka[1] & Takashi Ito[2,3]

[1] Kindai University, Shinkamikosaka 228-3, Higashiosaka, Osaka, 577-0813, Japan; patryksan@gmail.com

[2] Center for Computational Astrophysics, National Astronomical Observatory of Japan, Osawa 2-21-1, Mitaka, Tokyo 181-8588, Japan

[3] Planetary Exploration Research Center, Chiba Institute of Technology, 2-17-1 Tsudanuma, Narashino, 275-0016, Chiba, Japan

## Highlights

- Our scenario yielded four terrestrial planets with similar orbits and masses to ours
- All terrestrial planets acquired comparable water abundances during their formation
- Terrestrial planets accreted their bulk water from objects beyond 1 au within ~40 Myr
- Earth accreted most of its CC material from within 2 au during protracted accretion
- Mars analogs matched Mars' mass, eccentricity, inclination, and distance from Earth

**Abstract**

The Jupiter–Saturn chaotic excitation (JSCE) scenario proposes that the protoplanetary disk was dynamically excited and depleted beyond ~1–1.5 au in a few Myr, offering a new and plausible explanation for several observed properties of the inner solar system. Here, we expanded our previous work by conducting a comprehensive analysis of 37 optimal terrestrial planet systems obtained in the context of the JSCE scenario. Each optimal system harbored exactly four terrestrial planets analogs to Mercury, Venus, Earth, and Mars. We further investigated water delivery, feeding zones, and accretion history for the planet analogs, which allowed us to better constrain the water distribution in the disk. The main findings of this work are as follows: 1) the formation of four terrestrial planets with orbits and masses similar to those observed in our solar system in most of our sample, as evidenced by the dynamically colder and hotter orbits of Venus–Earth and Mercury–Mars analogs, and the high success rates of similar mutual orbital separations (~40–85%) and mass ratios of the planets (~70–90%) among the 37 systems; and 2) water was delivered to all terrestrial planets during their formation through the accretion of water-bearing disk objects from beyond ~1–1.5 au. The achievement of Earth's estimated bulk water content required the disk to contain sufficient water mass distributed within those objects initially. This requirement implies that Mercury, Venus, and Mars acquired water similar to the amount on Earth during their formation. Several of our planet analogs also matched additional constraints, such as the timing of Moon formation by a giant impact, Earth's late accretion mass and composition, and Mars's formation timescale.

**1 Introduction**

A common view in models of the inner solar system formation is that the terrestrial planets formed by the accretion of planetesimals and embryos (lunar-Mars-mass objects) within the protoplanetary disk (hereinafter, 'disk') (Chambers & Wetherill 1998; Kokubo & Ida 2000; Morishima et al. 2010; Morbidelli et al. 2012; Raymond & Morbidelli 2022). The formation of planets similar to Venus and Earth in terms of mass is relatively common, despite the different initial conditions and assumptions used in the models. However, the formation of a planet such as Mars, with a small mass (0.107 ME, where 1 ME = Earth's mass), is a typical outcome only in models that consider the disk to be mass-depleted beyond ~1 au (the disk's outer region); otherwise, excessively large Mars-like planets form (Raymond et al. 2009; Lykawka & Ito 2013; Clement et al. 2018). The origin of this mass depletion in the outer region is currently debated. For example, some models suggest that the outer region was nearly "empty" during planetesimal formation in the first Myr of solar system history when the disk gas persisted (Drazkowska et al. 2016; Raymond & Izidoro 2017a; Ogihara et al. 2018). Other models suggest that this region was initially massive, but that the mass was subsequently depleted by perturbations from Jupiter and Saturn before (Walsh et al. 2011) or during the giant-planet instability in the early solar system (Clement et al. 2018; Clement et al. 2021c). Inspired by Izidoro et al.'s (2016) findings on the asteroid belt, Lykawka & Ito (2023) (henceforth referred to as "Paper I") demonstrated that the chaotic excitation induced by the Jupiter–Saturn pair in their near 2:1 mean motion resonance (MMR) could effectively deplete the disk's outer region, before the giant planets' instability. In this Jupiter–Saturn chaotic excitation (JSCE) scenario, Saturn's nu16 and nu6 secular resonances[1] exhibit chaotic behavior, wandering across distinct locations within the disk, which can dynamically excite and deplete disk objects beyond ~1–1.5 au in only a few Myr. In summary, several models claim to explain the formation of Mars based on the premise that its small mass was the outcome of a mass-depleted outer region.

Most prior studies addressing the small-mass Mars problem disregarded Mercury's formation. However, if giant planet perturbations shaped Mars formation, they might have sculpted the formation of Mercury and Venus through Jupiter's nu5 secular resonance. Furthermore, the formation of Venus and Earth probably impacted the surrounding disk regions through secular interactions (e.g., Levison & Agnor 2003), possibly influencing the formation of Mercury and Mars as well. Therefore, as discussed in Lykawka & Ito (2019) and Paper I, a successful model should not be limited to the mass of Mars-like planets; it should also consider the formation of Mars in tandem with the other terrestrial planets and the asteroid belt, with the goal of reproducing their main properties. Regarding Mars, in addition to Mars' mass, other vital constraints are the planet's moderately excited orbit and mutual distance from Earth.

The formation of Mercury is an outstanding problem often neglected in terrestrial-planet formation models. Although some models have addressed Mercury formation using N-body simulations (Chambers 2001; O'Brien et al. 2006; Hansen 2009; Lykawka & Ito 2017), only more recent models have attempted to resolve the issue of Mercury formation along with Venus–Earth or all planets in a terrestrial system (Lykawka & Ito 2019; Clement et al. 2019a; Lykawka & Ito 2020; Clement & Chambers 2021; Clement et al. 2021a; Clement et al. 2021b; Paper I). In this work, consistent with the latter models, we limit our discussion of forming Mercury within the context of the simultaneous formation of the four terrestrial planets. However, in contrast to the connection of the disk's properties and the small mass of Mars discussed above, it is currently unclear which conditions in the disk's inner region within ~1 au would have been required to form a small-mass Mercury (0.055 ME or ~half the mass of Mars). Indeed, no consensus has been established regarding the best scenario that could explain Mercury's main properties (Clement et al. 2023b). Specifically, the small mass, the relatively excited orbit with eccentricity of $e$ ~ 0.2, inclination of $i$ ~ 7 deg, and orbital separation with Venus of $a_2$-$a_1$ ~ 0.34 au are challenging properties to explain ($a_2$ = 0.723 au and $a_1$ = 0.387 au are the semimajor axes of Venus and Mercury, respectively). Another important problem is explaining why Mercury's iron core comprises a significant fraction of the planet. This problem is discussed elsewhere (e.g., Asphaug & Reufer 2014; Chau et al. 2018; Clement et al. 2019a; Kruss & Wurm 2020; Johansen & Dorn 2022; Clement et al. 2023b). Thus, the reproduction of a small-mass Mercury on a relatively excited orbit well separated from Venus remains an outstanding unsolved problem in planetary sciences. We argue that a successful model should simultaneously produce Mercury and the other terrestrial planets in orbital-mass space.

---

[1] Further details on these resonances can be found in Williams & Faulkner (1981), Knezevic et al. (1991), Morbidelli & Henrard (1991), and Haghighipour & Winter (2016).

Outstanding questions persist regarding the formation of Venus and Earth. While past works can produce similar planets, they tend to place Venus-like planets too close to the Sun at ~0.5-0.6 au, and yield Venus–Earth pairs in dynamically overly excited and mutually distant orbits (e.g., Lykawka & Ito 2013; Brasser et al. 2016; Clement et al. 2019b; Nesvorny et al. 2021; Izidoro et al. 2022; Clement et al. 2023a). Lykawka & Ito (2020)'s main results using narrow disks (0.7–1.0 au) also encounter similar limitations (e.g., their Fig. 2). As discussed below, it is also unlikely that other recent models can explain the dynamically cold orbits of Venus and Earth after accounting for the effects of terrestrial- and giant-planet instabilities (Broz et al. 2021; Johansen et al. 2021). The Gyr-long stability and strong dynamical coupling between Venus and Earth represent additional properties that warrant further investigation (e.g., Tanikawa & Ito 2007). Future research should focus on discriminating their Venus and Earth analogs to evaluate whether the dynamically cold and compact orbits of the Venus–Earth pair are unique constraints beyond their masses.

As discussed previously, addressing the simultaneous formation of the terrestrial planets necessitates a robust classification scheme to properly identify analogs of each terrestrial planet in the system (the '4-P systems' discussed below). In this work, the planets obtained in simulations that successfully passed our algorithm's classification criteria are called planet analogs (Section 2.3). Regarding the results of other groups, we consider systems obtained after at least 100 Myr of evolution and refer to their planets as planet-like to distinguish them from our classified planets. In accordance with our previous work (Lykawka & Ito 2019; Lykawka & Ito 2020), we define terrestrial systems containing exactly one analog of Venus and Earth and one representative analog of Mercury and Mars within 2 au as 4-P systems (i.e., additional objects with masses 1–10 times that of the Moon might exist in the system). We argue that a better understanding of the main properties of each planet, such as the orbit, mass, bulk water content, accretion history (e.g., giant impacts), and formation timescale, should be preferentially based on 4-P systems (see Lykawka & Ito 2019 for more details). As discussed above, it is challenging to reproduce the orbital properties and masses of the Venus–Earth pair *simultaneously*, which are the planets at the core of a 4-P system (Clement et al. 2019b; Lykawka & Ito 2020). Notably, the difficulty of forming 4-P systems is also connected to the problem of forming reasonable Mercury and Mars analogs.

Although other models have yet to describe the likelihood of 4-P system formation, we provisionally discuss below some published results for completeness. Earlier models reported a few systems containing four planets within 2 au (Chambers 2001; O'Brien et al. 2006; Chambers 2013; Izidoro et al. 2015). However, these systems generally formed excessively large Mercury-like and Mars-like planets, so they do not qualify as adequate 4-P systems. Although one potential 4-P system may have formed in Chambers (2013) (the second system in the 'Frag' batch of their Fig. 7), the lack of details about the individual systems and planets formed precludes further discussion. Assuming the disk was described by a narrow annulus consistent with the premise of an "empty" outer region, Hansen (2009) obtained two candidate 4-P systems ('Sim1 and Sim23'). The Sim23 system contained an object with insufficient mass to be considered a Mercury-like planet. Sim1 appears to qualify as a 4-P; however, the system's orbital configuration is highly compact, and the Venus-like planet's location is within 0.6 au (i.e., closer to the Sun than observed). The representative system discussed in Clement et al. (2019c) is probably a 4-P system; however, the Venus-/Earth-/Mars-like planets acquired dynamically excited orbits, and two Mars-like planets are present. Among the best individual systems of Clement et al. (2019a), one appears close to being a 4-P system ('Annulus 2'), but it formed two Mars-like planets; additionally, the orbital separations of the Mercury–Venus and Venus–Earth pairs are insufficient and excessive, respectively. Clement et al. (2021b) obtained one representative 4-P system in a new scenario focusing on Mercury's formation. However, the mutual orbital separations of the terrestrial planet pairs are inconsistent with observations, and the Mercury-like planet's orbit appears to be overly excited. Notably, an optimal 4-P system with exactly four planets was obtained and discussed in Clement & Chambers (2021) (e.g., their Fig. 8); the acquisition of this system was probably related to the disk initial conditions that considered an inner region component. As discussed in Paper I, disks containing low-mass inner regions can boost the production of Mercury and 4-P systems. Studies regarding the Mercury–Venus pair formation also support this conclusion (Lykawka & Ito 2019; Clement et al. 2023b). The best individual system reported by Nesvorny et al. (2021) is not a 4-P system because there is a Venus-class planet with ~0.4 ME located at $a \sim 0.45$ au within the Mercury region of the system. Thus, this system would be a 3-P system containing only planets representing Venus, Earth, and Mars. The best individual system of Broz et al. (2021) (their Fig. 3) is probably a 4-P system; however, it is spatially very compact, with five planets packed at ~0.6-1.4 au and two Mars-like planets present. Broz et al. (2021) reported another potential 4-P system in their supplementary material; however, the four planets formed within ~0.6-1.1 au, which is too spatially compact. These systems are likely dynamically unstable, suggesting they may undergo a late terrestrial-planet orbital instability affecting the orbits of the planets (not modeled in Broz et al.). In short, whether the resulting post-instability systems would be compatible with real planets is unclear. In the supplementary material of Izidoro et al. (2022), it is difficult to determine whether the system labeled 'Analog-1' would represent a 4-P system because that work did not discuss individual systems, and it neglected Mercury formation. While Joiret et al. (2023) obtained a 4-P system labeled 'CL2 big3' after 100 Myr, the Mercury–Venus pair appears to have overly excited and mutually close orbits, suggesting it might be unstable on slightly longer timescales. Among the terrestrial formation studies mentioned above, only those conducted by Clement et al. (2019a, 2019c, 2021b, 2023b), Clement & Chambers (2021), Nesvorny et al. (2021), and Joiret et al. (2023) modeled the giant-planet instability (henceforth 'instability'). Thus, many 4-P-like systems found in other studies might be disrupted during the instability if it occurred late (e.g., Kaib & Chambers 2016; Clement et al. 2023a). Finally, Clement et al. (2023b) described an optimal 4-P system that satisfied some inner solar system constraints; however, as noted in that work, all formed planets acquired overly excited orbits after 200 Myr of simulation time.

Overall, previous models have difficulties with the formation of Mercury or 4-P systems that can meet the inner solar system constraints, particularly regarding the orbits and masses of the four terrestrial planets. There are additional difficulties. The analysis in Paper I demonstrated that the production rates of Mercury analogs and 4-P systems for the well-known annulus model of a narrow disk with a mass confined to 0.7–1.0 au (Hansen 2009; Walsh & Levison 2016) were 6% and 4%, respectively[2]. Furthermore, the Grand Tack and early instability models (Walsh et al. 2011; Jacobson & Morbidelli 2014; Brasser et al. 2016; Clement et al. 2018; Clement et al. 2019b) yielded < 5% (≤ 1%) and ~4–6% probabilities of forming Mercury-like planets (4-P-like systems), respectively. These probabilities are lower than those obtained for the favored disk models explored in Paper I: 34% and 13% for Mercury analogs and 4-P systems, respectively. In particular, the main reasons for these improved production rates were the inclusion of extended inner regions at 0.3–0.85 au and the dynamical depletion of the disk's outer region (see Section 2 for details). A caveat with this analysis is that these production rates, though derived using the same criteria, were obtained from different models, which employ diverse initial conditions and assumptions. A key challenge for all these models, including ours, is reproducing Mercury's small mass and the Mercury–Venus mutual distance (Section 3.1). In addition, Paper I assumed that the forming terrestrial planet systems underwent an early and dynamically non-violent instability by shifting Jupiter and Saturn to their near-current orbits (Section 2.1). While the dynamical cold orbits of Venus and Earth and the orbital stability of Mercury and Mars suggest a weaker instability, this simplification of the instability does not fully capture its influence on terrestrial planet formation. Future work should assess the validity of these assumptions by modeling the instability self-consistently[3]. While we simplified the instability, placing Jupiter and Saturn in more eccentric orbits than the current ones ensured that our terrestrial planets still experienced some instability-driven perturbations from both planets.

This work was performed to better understand the simultaneous formation of the four terrestrial planets and the origin of their main properties, as discussed below. Lykawka & Ito (2019) evaluated the properties of 4-P systems (17 in total) obtained via distinct terrestrial planet formation scenarios explored in that work. Paper I obtained 47 4-P systems based on disks with similar properties in a single evolutionary framework, the JSCE scenario. Notably, these 4-P systems yielded significant improvements compared with the results of Lykawka & Ito (2019). In this work, we selected 37 optimal analog systems that formed exactly four terrestrial planets and no remaining planetary bodies from the sample of systems obtained at the end of simulations performed in Paper I. Each optimal system contained only one planet analog of Mercury, Venus, Earth, and Mars. Here, we analyzed these systems to gain deeper insights into terrestrial planet formation while considering the following updated inner solar system constraints compared to Paper I.

---

[2] As discussed in the main text, Hansen (2009) obtained one system that might qualify as 4-P. Additionally, based on the results obtained from their 38 simulations, three planets may qualify as Mercury-like, reaching a minimum mass of 50% of Mercury's mass. Thus, rough estimates of Mercury and 4-P system production in that work are 3/38 ~ 8% and 1/38 ~ 3%, respectively, consistent with our analysis. However, in contrast to the present study and Paper I, recall that a proper classification scheme of planet analogs (systems) was not adopted in Hansen (2009). For example, it is unclear whether the systems that formed two of the Mercury-like planets also contained potential analogs of the other planets.

[3] We acknowledge that the terrestrial planet formation model presented in our previous work (Paper I) did not account for (1) the evolving secular resonances associated with Jupiter and Saturn during the instability and (2) the resonance dragging exerted during the giant planets' residual migration following the instability. However, it is important to note that Paper I did explore these effects self-consistently in the context of asteroid belt formation.

1.1 Fundamental inner solar system constraints

**A.** The orbits and masses of the four terrestrial planets. First, the final system must contain precisely four planet analogs. Additionally, it is necessary to explain the dynamically cold and close-in orbits of the Venus–Earth pair and the small masses of Mercury and Mars. The orbital separation of the Mercury–Venus pair is another challenging feature. Reproducing these properties is essential to better constrain the disk conditions that led to the formation of the planets and the asteroid belt (Constraint A is discussed and referenced in detail throughout this work).

**B.** The terrestrial planets' accretion history. This process includes statistics regarding giant impacts and the late accretion of remnant objects after the last giant impact experienced by a planet. Here, we focus on impactors with $\geq 0.02$–$0.10$ fractions of the proto-Earth's mass, which can represent Moon-forming giant impacts (e.g., Canup & Asphaug 2001; Cuk & Stewart 2012; Rufu et al. 2017). Studies suggest that proto-Earth's formation timescale is at least ~10-55 Myr (Rudge et al. 2010; Kleine & Walker 2017; Lammer et al. 2021). We also consider Mars' formation timescale, whose estimates based on analysis of meteoritic data vary from a few Myr (e.g., Dauphas & Pourmand 2011) to ~10–20 Myr (Mezger et al. 2013; Kruijer et al. 2017; Marchi et al. 2020; Zhang et al. 2021), or even ~60 Myr (Borg et al. 2016). The timescales longer than 10 Myr are consistent with a protracted accretion of Mars involving at least one giant impact. Investigating these properties can offer new insights about the Moon's origin, Mars formation, the role of giant impacts in terrestrial planet formation, and constraints on the building blocks of the planets (e.g., their feeding zones in space and time).

**C.** The terrestrial planets' bulk water contents and water delivery history. The water mass fractions on Mercury, Venus, Earth, and Mars should be reproduced simultaneously. These water contents were primarily acquired by impacts of water-bearing primordial disk objects within the inner solar system (>90%; e.g., Marty 2012). In addition, Earth's bulk water was probably delivered before the Moon-forming giant impact (Greenwood et al. 2018). Finally, we constrain the disk's initial water distribution by requiring our planet analogs to match the bulk water content of Mercury, Venus, Earth, and Mars. By addressing these properties, we can gain valuable insights into the water distribution in the disk and make predictions about the bulk water contents, water source regions, and water accretion history for the terrestrial planets (other details related to constraint C are discussed in Section 2.2).

The bulk composition of Earth and Mars and the accretion history of their building blocks can be constrained from mixing models of isotopic and elemental compositions. This modeling typically considers the EC, OC, and CC main groups associated with enstatite, ordinary, and carbonaceous chondrites materials, respectively[4] (Marty 2012; Fitoussi et al. 2016; Dauphas 2017; Burkhardt et al. 2021; Savage et al. 2022; Kleine et al. 2023; Dauphas et al. 2024). However, these models rely on distinct assumptions, have considerable uncertainties, and can yield conflicting results (in particular for Mars). Thus, we believe there is an ongoing debate about the bulk composition of Earth and Mars. Nevertheless, considering that Earth is better constrained than Mars and that there is some consensus about a small contribution of CC materials to our planet in those models, we discuss the implications of this constraint in Section 3.2.1. In this investigation, we considered only CC and NC materials, where NC consists of EC, OC, and other non-CC materials as typically discussed in the literature (e.g., Budde et al. 2016; Dauphas et al. 2024).

---

[4] For simplicity, we treated the main carbonaceous chondrite subgroups (CI, CO, CV, and CM) as a combined CC group. A more detailed analysis of these subgroups is outside the scope of this work.

The asteroid belt represents another fundamental constraint in the inner solar system. Paper I demonstrated that the main properties of the asteroid belt, specifically, the orbital architecture in $a$-$e$-$i$ space (including the peculiar concentration of asteroids at $i < 20$ deg), the small total mass, and the distributions of S-, C-, and D/P-type asteroids can be explained in the framework of the JSCE scenario; thus, it is not discussed in this study.

## 2 Methods

2.1 Basic setup

The 37 optimal 4-P systems considered in this work were obtained from the main N-body simulations of terrestrial-planet formation performed in the framework of the JSCE scenario (Paper I). These simulations were executed using a modified version of the MERCURY integrator[5] (Chambers 1999). The giant planets and embryos (forming terrestrial planets) gravitationally interacted with each other. The planetesimals experienced perturbations from the planets and embryos but did not mutually perturb each other. All simulations utilized perfect accretion, an acceptable assumption (see discussions in Lykawka & Ito 2019 and Paper I). The typical time steps used in these simulations were ~3.5–4.5 days.

Our model began at the time of disk gas dispersal. The model contained Jupiter and Saturn with their current masses on moderately eccentric orbits ($e = 0.08$ and 0.1, respectively) and an extended massive disk covering the inner solar system to 3.5 au. Jupiter and Saturn began with initial semimajor axes $a = 5.5$ au and ~8.7 au in their mutual near-2:1 MMR, an orbital configuration characterized by the orbital period ratio of Saturn and Jupiter (PS/PJ) ~ 2 and large libration amplitudes (~300–360 deg). Several lines of evidence indicate that Jupiter and Saturn could have acquired such an orbital configuration after their formation[6] (see Methods in Paper I). Additionally, because this configuration was probably metastable, both planets were susceptible to dynamical instabilities over short timescales, consistent with the notion that the instability event occurred early (Nesvorny 2018; Quarles & Kaib 2019; de Sousa et al. 2020; Liu et al. 2022). Thus, we assumed that Jupiter and Saturn remained in their near-2:1 MMR orbital configuration for ~5–10 Myr timescales before the instability. Importantly, secular resonances associated with the Jupiter–Saturn pair behaved chaotically under such a near-MMR state, leading to strong orbital excitation and the dynamical depletion of disk objects located beyond ~1–1.5 au. Because the instability is typically a rapid event (~10–100 kyr timescales), we changed the orbits of both planets to their near-current ones in the post-instability stage. Finally, after the initial stage of near-resonance interactions described above, we evolved all objects in our post-instability terrestrial systems for 400 Myr.

---

[5] This version incorporates real-time calculations of bulk density and radii for growing embryos, alongside general relativistic effects and minor refinements. See Hahn & Malhotra (2005) and Kaib & Chambers (2016) for further details.

[6] Note that other primordial configurations are possible and have advantages as well (e.g., Deienno et al. 2017; Clement et al. 2023b).

Our disk consisted of embryos concentrated within 1.15–1.5 au (typically separated by ~1-2 mutual Hill radii) and planetesimals placed within the entire disk. The results of several embryo formation models are consistent with these initial conditions (Morishima 2017; Walsh & Levison 2019; Clement et al. 2020; Woo et al. 2021). All objects began on nearly circular and coplanar orbits. Our disk models considered various extended disks because a narrow disk confined to 0.7–1.0 au cannot form the terrestrial planets (Lykawka & Ito 2019; Lykawka & Ito 2020; Woo et al. 2023). Overall, the disks were similar and consisted of three main regions: a core region centered at Earth's location with a high mass concentration within ~0.8–1.2 au, a less massive adjacent inner region with an inner edge near Mercury's location, and a broad relatively massive outer region located at ~1.2–3.5 au. The thin mass concentration in the core region could result from the gas-driven convergence of massive planetesimals or embryos (Ogihara et al. 2018; Broz et al. 2021). These disk initial conditions are within the possible outcomes of some planetesimal formation models (Izidoro et al. 2022). It is also conceivable that employing distinct gas dynamics parameters and initial planetesimal conditions in those models could generate disks with properties akin to those considered here. The disks' main differences in our simulations involved the inner region's properties, such as width and total mass. In addition, from a terrestrial-planet formation standpoint, a single annulus massive disk or an extended disk with mass following a simple power law is unlikely to reproduce the main properties of the terrestrial planets (O'Brien et al. 2006; Izidoro et al. 2015; Lykawka & Ito 2019; Lykawka & Ito 2020; Clement & Chambers 2021). Briefly, the inner, core, and outer regions likely played fundamental roles in forming Mercury, the Venus–Earth pair, and Mars and the asteroid belt, respectively. We also investigated water delivery to the terrestrial planets by assuming that our objects carried distinct water mass fractions according to their initial location in the disk (Section 2.2).

2.2 Initial water distribution in the disk constrained by the terrestrial planets

This study determined each terrestrial planet's water mass fraction (WMF) by considering the bulk water acquired by planet analogs formed in our optimal 4-P systems. These systems comprised a more suitable sample set for this analysis than Paper I, which used a sample of 87 systems that contained only 10 4-P systems. For this investigation, we tested various initial water distributions in the disks that formed these systems. In total, 556 distinct WMF models were considered, significantly expanding the work in Paper I that covered only a limited range of similar models. Here, the large number of models reflected the combinations of distinct WMFs assigned to objects according to their initial locations in specific regions of the disk at < 1.0 au, 1.0–1.5 au, 1.5–2.0 au, 2.0–2.5 au, and > 2.5 au (Table 1. See also Table A1 in the Appendix). By employing a broad range of initial WMFs, we were able to explore diverse possibilities for water distributions in the disk in great detail. Specifically, we constrained the disk's water distribution at the onset of terrestrial planet formation by considering a WMF model to be successful if the median WMFs of Venus, Earth, and Mars analogs simultaneously satisfied the estimated terrestrial planets' water contents[7]. We then obtained the WMF representative of each terrestrial planet by calculating the median WMF over the 37 analogs that represented each planet. We repeated these calculations for each of the 556 WMF models tested.

Earth's bulk water content is commonly estimated to lie within 0.05–0.5% (equivalent to ~2–20 times Earth's water in oceans), but these values might be higher due to uncertainties related to the difficulty of probing the water contents of the mantle and core (Marty 2012; Peslier et al. 2017; Hirschmann 2018; McCubbin & Barnes 2019; Piani et al. 2020; Venturini et al. 2020). Even with appreciable uncertainties, Earth's estimated WMF is sufficiently low to provide valuable constraints on the primordial WMF distribution in the disk. The bulk water contents on Venus and Mars are less constrained. In particular, Venus has likely lost a significant portion of its primordial water via atmospheric loss (Hamano et al. 2013; Greenwood et al. 2018; Salvador et al. 2023). Nevertheless, Venus' WMF estimates can be obtained based on atmospheric and geologic evolution models by assuming Venus formed dry or wet. In short, even possessing larger uncertainties, Venus' and Mars' estimated WMFs offer helpful complementary constraints. We considered a range of WMFs for Venus and Mars that include typical values estimated in the literature (Lunine et al. 2003; Kurokawa et al. 2014; Greenwood et al. 2018; McCubbin & Barnes 2019; Rickman et al. 2019; Salvador et al. 2023). Finally, we considered the suggestion that Mercury might contain considerable water in its interior (McCubbin & Barnes 2019). In conclusion, by requiring our planet analogs to simultaneously satisfy the WMFs of Mercury, Venus, Earth, and Mars, we could constrain the initial WMF distribution in the disk (Section 3.3).

Following the discussion above, we assumed two main scenarios of water contents for the terrestrial planets obtained after 400 Myr in our simulations. First, we considered the 'fiducial' hypothesis in which Earth and Mars analogs were required to acquire WMFs in the respective ranges of 0.05–0.5% (the same range estimated for Earth) and 0.005–0.5%. The assumed WMF range of values is within the estimated uncertainties for Mars. The minimum WMF required for Venus analogs was 0.005% or 0.05%, assuming that the newly formed Venus was 'dry' or 'wet', respectively. Conversely, the maximum WMF of dry or wet Venus should be within 0.05% or 0.5%, respectively. These WMF ranges bracket the uncertainties of Venus' estimated values. The possibility of early Venus possessing water contents comparable to Earth's oceans (~0.025% of Earth's mass) suggests that the wet Venus case would better represent the primordial planet. For Mercury, we assumed a minimum WMF of 0.005% (the same for dry Venus and Mars) and left the maximum WMF unconstrained. Second, we considered the 'water worlds' hypothesis by assuming that all terrestrial planets were wet after formation. This approach was motivated by the possibility that Earth is more water-rich than previously thought (Tagawa et al. 2021). Consistent with the latter study, the WMFs required for water-rich Mercury, Venus, Earth, and Mars analogs in the water worlds hypothesis were >0.05%, 0.5–1.5%, 0.5–1.5%, and 0.05–1.5%, respectively. Under this hypothesis, the minimum and maximum WMFs for all planets were 10 and 3 times larger than that considered in the fiducial hypothesis. Overall, we updated, improved, and expanded the WMF modeling of the terrestrial planets in many aspects in this work compared to Paper I.

---

[7] For simplicity, we ignored water loss through atmospheric and collisional processes (Burger et al. 2020). Therefore, the WMFs of our planet analogs represent upper limits. Although the WMF of Mercury is currently unknown, we considered the hypothesis that the planet is rich in volatiles.

2.3 Classification of terrestrial systems and planet analogs

We used Lykawka & Ito (2019)'s classification scheme to identify the analogs of each terrestrial planet formed in a given system. As discussed in that work, a dedicated classification of terrestrial systems is crucial for analyzing the results in terrestrial-planet formation models. However, similar classification schemes remain scarce in the literature; thus, most past models show their formed planets in a mixed manner or do not identify terrestrial analog systems. Additionally, the classification approach used in this study was more rigorous than the approach used in our previous work after updating the first two conditions below (Lykawka & Ito 2019; Lykawka & Ito 2020; Paper I). The conditions (1–3) applied to our terrestrial systems were as follows. 1) We required the final systems to be optimal 4-P systems; thus, we discarded systems where two or more analogs of Mercury or Mars were formed in a single system. 2) We discarded systems that formed a non-analog object with mass $m \geq 0.05$ ME (approximately a Mercury mass) in any region of the system. 3) Finally, the planetary masses considered in the algorithm's first search for analog candidates were 0.025–0.27 (Mercury), 0.4–1.5 (Venus), 0.5–1.5 (Earth), and 0.05–0.32 ME (Mars). As justified in our previous work, the main results and conclusions in this study were insensitive to the selection of these mass ranges. Finally, we evaluated the properties of our 37 optimal 4-P systems and the planet analogs formed in those systems with respect to several critical constraints of the terrestrial planets (Section 1).

**3 Results and Discussion**

In the following sections, we examine in detail the formation of 4-P systems and their representative analogs of Mercury, Venus, Earth, and Mars.

3.1 Orbits and masses

Our selected 37 optimal 4-P systems formed planets analogous to the four terrestrial planets in many ways. First, by definition, each terrestrial planet is represented by a single analog in the system. Second, as summarized in Figures 1 and 2 and Table 2, our main results reveal that the optimal systems produced planet analogs with orbits and masses similar to the real terrestrial system. Angular momentum deficit (AMD) and radial mass concentration (RMC) metrics are often used in the literature to measure the orbital excitation and mass distribution of terrestrial planet systems. However, because conclusions based on these metrics may be misleading (Clement et al.

2023a), we selected other metrics to evaluate our systems in the discussion below (AMDs and RMCs are given in Table 3 for completeness). Overall, the orbit–mass distribution of our planet analogs reproduced the dichotomy of two massive planets (Venus and Earth) surrounded by two much less massive planets (Mercury and Mars) on relatively distant orbits. Furthermore, these systems satisfied some fundamental constraints of the inner solar system, as discussed below.

The orbital properties of our planet analogs could explain the dynamically colder and hotter orbits of the Venus–Earth and Mercury–Mars pairs, respectively. We confirmed this result by verifying the orbital excitation of the planets based on their excitation factors, $Ef$. Following the methodology described in the literature (Nesvorny et al. 2021; Clement et al. 2023a), we calculated $Ef_{XY}$ as $(e_X + e_Y + \sin i_X + \sin i_Y)/4$, where X and Y refer to the pair's planets. We required our Venus–Earth pairs to be dynamically cold with $Ef_{23}$ within 1.5 times the observed value. We required $Ef_{14}$ to be within the observed value for the Mercury–Mars pairs. This stricter criterion accounts for the possibility that Mercury and Mars underwent some dynamical excitation during the instability or the subsequent residual migration of the giant planets (Laskar 2008; Brasser et al. 2013; Roig et al. 2016; Paper I). Nevertheless, these effects likely had a minimal impact on the final orbits of our planet analogs for the following reasons. First, limited excitation is expected in the system due to the weak, early instability and effective disk dynamical friction from the extended disks considered in our scenario (O'Brien et al. 2006). Second, negligible excitation is expected if Jupiter and Saturn begin their post-instability residual migration with an orbital period ratio greater than 2.2 (Brasser et al. 2013). Finally, our Mercury and Mars analogs experienced late (occurring after tens of Myr) and prolonged excitation due to gravitational interactions with neighboring embryos and forming planets, and secular effects from all planets. A more detailed examination of the dynamics of Venus–Earth and Mercury–Mars pairs is reserved for future work.

The high success fractions of our systems in satisfying orbital excitation constraints (84% and 76% for $Ef_{23}$ and $Ef_{14}$, respectively) and the similarity of median $a$-$e$-$i$ orbital elements of our analogs with the real planets indicate that our results are robust (Table 2). This similarity is also evident when the obtained systems are compared with the real system (Figures 1 and 2). However, a caveat is that even our best Mercury analogs obtained in disks with extended inner regions tended to form slightly farther from the Sun than the real planet (median $a_1 = 0.448$ au vs real $a_1 = 0.387$ au). Nevertheless, as discussed in Section 1 and as demonstrated in recent works (Clement & Chambers 2021; Clement et al. 2023b), further exploration of the properties of the disk's inner region may yield Mercury analogs closer to the Sun.

Concerning the orbital separations of each pair of terrestrial planets, similar to the criteria considered in recent models (e.g., Clement et al. 2023b), we required our planetary pairs to be within the success range $0.67$–$1.33 \times D_{XY}$, where $D_{XY} = |a_X - a_Y|$ represents the mutual orbital separation of the two planets (this criterion is stricter than the $0.67$–$1.33$ range of orbital period ratio of both planets). Our Venus–Earth and Earth–Mars pairs approximately matched observations with success fractions of about 50% and 85%, respectively (Table 2). In particular, our Mars analogs often formed at correct distances from their Earth counterparts. These results confirm the hypothesis that reproducing the Venus–Earth and Earth–Mars mutual distances may require initially narrow mass concentrations within the disk's core region and the existence of an outer region relative to the disk, respectively (Paper I). However, a caveat is that the median orbital separation of our Venus and Earth analogs remains higher than the observed value. This finding highlights the difficulty of replicating the spatially compact orbital configuration of Venus and Earth. Additionally, Mercury and Venus analogs tended to form closer to each other than in reality, resulting in a smaller success fraction (~40%). Only two systems (#7 and #15 in Table 2) contained Mercury–Venus pairs matching the observed mutual spacing. In short, replicating this orbital separation remains an outstanding issue for future studies. Further refining the disk modeling (e.g., properties of the inner region) could potentially explain this constraint on Mercury–Venus separation.

Regarding some fundamental properties of our optimal systems, Mercury analogs acquired excited orbits and small masses. However, whereas Venus, Earth, and Mars analogs acquired median masses close to the observed values, our Mercury analogs tended to acquire masses ~two to four times the mass of the real planet. The small-mass Mercury problem is challenging and affects even state-of-the-art models focusing on Mercury formation (Clement et al. 2023b and references therein). In particular, future work should consider distinct disk properties (e.g., inner regions with smaller initial masses) and more sophisticated modeling (e.g., inclusion of collisional fragmentation or cratering erosion) to improve Mercury formation in our scenario. Nevertheless, our results indicate that disk properties at < 1 au can play essential roles in improving Mercury formation in terms of its mass and orbital separation from Venus, consistent with previous similar investigations (Clement et al. 2019a; Clement & Chambers 2021; Clement et al. 2023b). In conclusion, our results support the notion that Mercury formed in situ within the inner region near its current location and accreted from both local and distant objects.

Our Venus and Earth analogs acquired orbits and masses in reasonable agreement with the real planets. In particular, our results indicate that these analogs could explain the dynamically cold and close-in orbits of the Venus–Earth pair, which have been difficult constraints to satisfy in previous representative studies (Clement et al. 2021c). However, a caveat is that Venus analogs exhibited a moderate tendency to acquire slightly larger final masses than their Earth counterparts. This tendency might be related to the initial mass distribution within the disk's core region. Thus, future studies should identify the key variables that determine the masses of both planets.

We confirmed that JSCE's mass depletion and dynamical excitation beyond ~1–1.5 au allowed the formation of many Mars analogs with properties similar to the properties of Mars. For instance, our Mars analogs statistically match the observed eccentricity and inclination (Table 2). Also, their dynamical evolution suggests that various competing processes (JSCE, mutual embryo perturbations, disk dynamical friction, secular effects from Jupiter and Saturn, and collisions with embryos/planetesimals) played a role in shaping their eccentricities and inclinations during the 400 Myr evolution. The existence of the outer region also caused our Mars analogs to form at correct orbital separations from their Earth counterparts. Thus, Mars's small mass and current orbit are consistent with the JSCE perturbations experienced by the disk's outer region during the early solar system. Overall, our optimal 4-P systems yielded successful Mars analogs with moderately excited orbits, small masses, and mutual separation from the Earth analogs.

Regarding the mutual mass ratios $M_{XY} = m_X/m_Y$ of our planet analogs, we found that they were mostly successful as defined by the ranges $0.5–2 \times M_{23}$ (Venus–Earth pair) and $0.5–3 \times M_{12(34)}$ (Mercury–Venus and Earth–Mars pairs), which are similar to criteria adopted in many previous works (Clement et al. 2018; Walsh & Levison 2019; Clement & Chambers 2021; Izidoro et al. 2022; Woo et al. 2022). In particular, the success fractions for the mass ratios were 70%, 81%, and 89% for the Mercury–Venus, Venus–Earth, and Earth–Mars pairs, respectively (Table 2). However, considering the above caveats, further investigations are needed to improve the Mercury–Venus and Venus–Earth mass ratios.

While our simulations at 400 Myr explore timescales exceeding those typically used in previous studies by two to four times, some resulting systems might contain a Mercury or Mars analog dynamically unstable over Gyr timescales. However, terrestrial planets dynamics over such timescales are highly chaotic, suggesting that a statistical approach is required using numerous Gyr-long simulations of the same terrestrial system (e.g., Ito & Tanikawa 2002; Laskar & Gastineau 2009).

3.2 Accretion history, formation timescales, and feeding zones

Firstly, a summary of additional properties of the 37 optimal 4-P systems and their planet analogs is given in Table 3. In the discussion below, a giant impact (hereinafter, GI) is defined as the impact of an object with at least 5% of the mass of the target unless explicitly stated otherwise (i.e., an impactor-to-target-mass ratio ITr = 0.05). This definition is conservative because an ITr of $\geq 0.01$ can be considered a GI based on its global consequences for the target (Gabriel & Cambioni 2023). First, our Mercury analogs experienced from none to a few GIs. Overall, the median GI was equal to 1, suggesting that Mercury experienced at least one GI during its formation. The median timing of the last GI experienced by these analogs was 24 Myr, but the individual values considerably varied. Although Earth likely experienced at least one GI, as indicated by the Moon's formation, the situation is less clear for Venus. The lack of a magnetic dynamo is associated with the lack of GIs on the planet (Jacobson et al. 2017). Conversely, the absence of Venusian satellites is associated with fortuitous GIs (Alemi & Stevenson 2006). Recent studies also indicate that early GIs were crucial in generating long-term volcanic resurfacing on Venus (Marchi et al. 2023). These events probably depend on particular details of the GIs and require specific timings during the formation of Venus. However, such a discussion is beyond the scope of this paper. Based on our results, Venus and Earth experienced a similar number of GIs (medians of 9 and 10, respectively). We also obtained the last GI timings of 24 Myr and 45 Myr for both planets (medians); the median numbers of GIs become 7 (2) and 8 (2) if an alternative ITr = 0.1 (0.5) is considered to probe more (highly) energetic GIs. In this case, the last GIs would have occurred early during the accretion history of both planets, with medians of 14 (3) Myr and 23 (1) Myr for Venus and Earth at the same alternative values of ITr. Finally, the results for our Mars analogs were similar to those of their Mercury counterparts. The median number of GIs and timing of the last GI experienced by the Mars analogs were 1 and 16 Myr, respectively.

Considering both lunar formation and the delivery of late-accreting material to Earth, the best results were achieved for ITr < 0.1, as outlined below. For ITr > 0.02, > 0.05, and > 0.1, our 37 optimal Earth analogs experienced successful late Moon-forming GIs in 70%, 60%, and 45% with medians of 54, 45, and 23 Myr, respectively. Moon-formation timings of 45-54 Myr are consistent with recent studies that suggest a Moon-formation window of approximately 50-110 Myr (Barboni et al. 2017; Thiemens et al. 2019; Greer et al. 2023). For the same values of ITr, the results satisfied Earth's late accretion mass in 50%, 40%, and 30% of systems with medians of 0.01, 0.025, and 0.042, respectively.

Our Earth and Mars analogs acquired 90% of their final masses within median formation timescales of 37.3 and 25.3 Myr, respectively. In particular, Martian analogs experienced one or more GIs during their accretion histories, with the last impact often pushing their final masses to >>90%. Therefore, these formation patterns align with independent studies that suggest GIs played an essential role in Mars formation (Section 1.1). However, this 25 Myr timescale remains longer than the conservative ~10–20 Myr protracted formation timescale for Mars. Mars formation timescale warrants further exploration in future work. Finally, although our numerical setup was not designed to obtain median formation timescales for Mercury and Venus analogs in this work, we discuss some related results for these planets below.

Figure 3 illustrates the individual analogs' combined accretion timescales and feeding zones for the four terrestrial planets in the 37 optimal 4-P systems. First, although Mercury and Mars are the closest and farthest planets from the Sun, our results revealed that Mercury and Mars analogs experienced the slowest and fastest accretion histories among the terrestrial planets, respectively. Venus analogs also exhibited accretion timescales similar to Earth's. Consistent with the discussion above, Mars analogs accreted source objects significantly faster than their Earth counterparts. Similarly, all planets rapidly accreted disk objects initially located beyond ~2 au. This fast accretion was caused by the JSCE mechanism that dynamically excited and depleted the disk outer region in only a few Myr. Because this region was probably water-rich, this result also implies that bulk water was delivered to the planets during their formation in less than ~15 Myr, and thus before the Moon-forming GI for Earth. This result is in line with findings from meteoritic studies (Carlson et al. 2018; Greenwood et al. 2023). Furthermore, disk objects initially in the region at ~1–2 au may have contributed additional water over a ~15–40 Myr timescale. Another notable feature is the faster accretion of objects initially located within ~1.2 au for Venus and Earth analogs. Therefore, the main building blocks of the Venus–Earth pair were rapidly incorporated by both planets during their formation. These results align with the findings of composition models on Earth's bulk composition and late accretion, as discussed in Section 3.2.1.

The feeding zones of our representative planet analogs reflect the disk's initial mass distribution. In particular, the smaller and larger contributions of the inner region within 0.8 au and the mass-enhanced zone at 0.8–1 au are evident. The contribution of the outer region beyond ~1 au is linked to the disk's dynamical stirring and depletion caused by the JSCE, which reduced the probability of accretion of objects from this region to the planets. Our Mercury analogs exhibited a substantial contribution of objects from the inner region to their final masses (30%). Conversely, although they possessed the closest orbits to the Sun, these analogs acquired a remaining high fraction of objects from the region beyond 0.8 au (70%). This finding is consistent with MESSENGER data (Peplowski et al. 2011; Rodriguez et al. 2023) and other studies (Greenwood et al. 2018; McCubbin & Barnes 2019) supporting a volatile-rich Mercury. The accretion fractions of objects located at ~1.2–2.5 au were similar for Mercury, Venus, and Earth analogs; however, they were higher for Mars analogs because such planets typically formed within that region of the disk. These results imply that all terrestrial planets acquired large amounts of water if water-rich objects were commonly present beyond 1.2 au at the onset of terrestrial planet formation (see Section 3.3). The 0.8–1 au zone significantly contributed, with ~55% of the final masses for Venus and Earth analogs. Although the Venus and Earth analogs had similar feeding zones, the inner and outer regions made greater contributions to the final masses of Venus and Earth, respectively. Finally, our Mars analogs exhibited feeding zones with smaller contributions from the core region and shifted towards distances beyond ~1 au, compared with their Venus–Earth counterparts.

### 3.2.1 Earth's bulk composition and late accretion

Reflecting Paper I's findings on the asteroid belt composition taxonomy that required the existence of water-rich objects in the disk inner regions, our modeled disks harbored ~10–20% and 50% CC-objects (representing objects containing CC materials) within 2 au and beyond 2 au, respectively. In particular, we examined various plausible fractions of NC- and CC-objects in distinct disk zones within 2 au (Table 4). Under these assumptions, the CC fractions acquired by our representative Earth (based on 37 Earth analogs formed in our optimal 4-P systems) concentrated within 1–14%, consistent with predictions of mixing models of a few to ~10–20% for Earth's composition (Section 1). However, like other published models, our scenario would require a more sophisticated analysis (e.g., considering varied initial embryo compositions) or new dedicated simulations to address the hypothesis that Earth acquired a larger CC fraction compared to Mars (Burkhardt et al. 2021; Kleine et al. 2023). For completeness, we also calculated the bulk water contents acquired by our representative Earth using the same NC/CC models employed in this investigation. Overall, we found that CC-objects are sufficient to deliver the water needed to explain Earth's water budget regardless of any potential contribution from NC-objects (Table 4).

According to our NC/CC models, the inner disk region within 2 au supplied most CC materials accreted by our representative Earth. Conversely, the outer region beyond 2 au contributed with <1%. Our analysis revealed that the CC delivery differed between the inner and outer regions of the disk. JSCE triggered rapid accretion of outer CC-objects within the first Myr, while inner CC-objects accreted over a few tens of Myr. This dual accretion of CC-objects hints at compositional variations in Earth's bulk composition. This result also supports the idea that late accretion of CC materials from the disk is possible without requiring a specific and stochastic delivery by the Moon-forming impactor (e.g., Budde et al. 2019).

Lastly, our results can constrain the nature of Earth's late accretion material. First, using Paper I's asteroid belt model, we identified the source regions within the disk for Earth's late accreting objects. Then, applying the composition models described above, we found that the fraction of NC materials delivered to our representative Earth during late accretion concentrated within 85–99% (Table 4). Again, these results are broadly consistent with predictions of mixing models (e.g., Marty 2012; Dauphas 2017; '4-stage model' in Dauphas et al. 2024) and other models about the feedstocks of late accretion in the inner solar system (Carlson et al. 2018; Zhu et al. 2021).

3.3 Water acquisition

Overall, the results demonstrate that the most widely used WMF models in the literature (Morbidelli et al. 2000; Raymond et al. 2009; Clement et al. 2018; O'Brien et al. 2018) cannot provide sufficient water to explain the bulk water contents of Earth (models 1 and 2 in Table A1 in the Appendix). This result implies that the disk was wetter than assumed in existing WMF models. Our finding is consistent with the evidence of higher water content in EC meteorites (the primary representative of Earth's building blocks) and S-type asteroids (Jin et al. 2019; Piani et al. 2020; Jin et al. 2021), and with models of implantation of water-rich objects into the inner solar system before terrestrial planet formation (Raymond & Izidoro 2017b; Nesvorny et al. 2024).

Assuming Earth acquired an estimated water content of ~2–20 times its current ocean mass (the fiducial hypothesis described in Section 2.2), we found that only 28 WMF models simultaneously satisfied the WMF constraints of Mercury, Venus, Earth, and Mars in the case where Venus was dry at the end of its formation. The main reason is that forming a dry Venus and a relatively wet Earth is difficult when both planets have similar feeding zones (Section 3.2). In contrast, if Venus formed wet, many more WMF models could explain the bulk water contents of the terrestrial planets. Specifically, 142 WMF models were successful, implying the following plausible initial WMFs in the disk[8]: 0.001–0.1%, 0.001–0.1%, 0.01–1%, 1–30%, and 10–50% within 1 au, 1–1.5 au, 1.5–2 au, 2–2.5 au, and beyond 2.5 au, respectively. Although the initial water distribution within 1.5 au was similar for dry and wet Venus constraints, disks with wetter components beyond 1.5 au were required to satisfy the wet Venus constraint. This result was observed in many successful WMF models with higher maximum WMFs. The result is also consistent with the notion that Venus and Earth acquired comparable amounts of water during their formation, which could significantly influence Venus' history (O'Rourke et al. 2023). Given these reasons, we refer to the wet Venus case when discussing the fiducial hypothesis below.

Some of our successful WMF models feature relatively high WMFs within 2 au of the disk. This result supports studies suggesting the disk inner regions contained a limited population of primordial water-rich objects (e.g., Raymond & Izidoro 2017b). For example, a 1% abundance of CC-objects (with 5–20% of water by mass; e.g., O'Brien et al. 2018) at < 1 au can yield a regional

---

[8] The typical values cover the initial conditions of > 90% of the successful models for a particular disk water distribution hypothesis.

WMF of 0.05–0.2% assuming negligible water in the remaining 99% NC-objects. Similar water content is achievable even without CC-objects (e.g., 50% fraction of water-enriched NC-objects with 0.1–0.5% of water by mass; e.g., Piani et al. 2020). This reasoning applies to any region of the disk. Finally, it is also possible that disk objects held more water than meteoritic data suggests (Alexander et al. 2018; O'Brien et al. 2018; Greenwood et al. 2023). This additional evidence strengthens the case for our wetter WMF models.

Alternatively, in the case where terrestrial planets acquired more significant amounts of water at the end of their formation (the water worlds hypothesis described in Section 2.2), we found the following plausible WMFs among 99 successful WMF models: 0.001–1%, 0.01–1%, 1–10%, 5–50%, and 10–50% within 1 au, 1–1.5 au, 1.5–2 au, 2–2.5 au, and beyond 2.5 au, respectively. Notably, none of these models were successful under the fiducial hypothesis for either a dry or wet Venus. Additionally, the water worlds hypothesis implies a disk (much) wetter beyond 1 au (1.5 au) compared with the fiducial hypothesis discussed above. In particular, the minimum WMFs in the 1–1.5 au, 1.5–2 au, and 2–2.5 au regions would be 10, 100, and 5 times higher, respectively. Furthermore, the successful WMF models indicate that the allowed maximum WMFs within 2 au would be 10 times higher. A comparison of typical WMFs required in distinct regions of the disk is summarized in Table 1.

In summary, the successful WMF models discussed above could explain the water contents of the terrestrial planets and constrain the likely initial water distributions in the disk. These models could also indicate unfavorable WMFs in some areas of the disk. Specifically, we found that it was implausible for the region within 1.5 au to contain a water fraction $\geq 1\%$ or $\geq 5\%$ under the fiducial Venus or water worlds hypotheses, respectively. Finally, whereas the fiducial Venus hypothesis disfavors WMFs $\geq 5\%$ within the 1.5–2 au region, the water world hypothesis disfavors drier ($\leq 0.001–0.1\%$) and highly wet WMFs ($\geq 30\%$) in the same region. Water-poorer WMFs ($\leq 1\%$) in the 2–2.5 au region are also incompatible with this hypothesis. The WMF models and results are extensively described in Table A1 in the Appendix.

Based on the successful WMF models of the fiducial and water worlds hypotheses, the median WMFs acquired by Mercury, Venus, Earth, and Mars analogs over the 37 optimal 4-P systems and the medians of these results over the respective 142 and 99 successful WMF models are given in Table 5. These results reveal linear correlations between the bulk water acquired by any pair of planets. Given the similar feeding zones of water-bearing objects for the four terrestrial planets, all planets acquired comparable amounts of water within a factor of a few for any successful WMF model adopted, irrespective of the success criteria considered (fiducial vs. water worlds). Our results indicate that the median amount of water delivered to an Earth analog in our optimal 4-P systems agrees with typical estimates of Earth's bulk water (Section 2.2). Furthermore, as discussed above, our findings indicate that water delivery is a single global process in the inner solar system. Therefore, a successful WMF model providing sufficient water to Earth also delivered comparable amounts to Mercury, Venus, and Mars during their formation. A significant amount of water delivered to Mercury confirms the expectation of a volatile-rich composition (Section 3.2).

These findings also support the notion that Venus and Mars had significant water inventories during their final stages of accretion.

In conclusion, our analysis of the water mass acquired by the planet analogs formed in our optimal 4-P systems indicates that a significant population of objects with individual WMFs $\geq 0.1\%$ (NC-like) and/or $\geq 5\%$ (CC-like) beyond ~1–1.5 au in the disk were required to satisfy the water contents of Mercury, Venus, Earth, and Mars. Therefore, this region was probably massive (not "empty") and contained a significant fraction of water-bearing objects.

### 3.4 Examples of representative individual 4-P systems

Aiming at providing a complementary discussion about the results, we illustrate the orbital and accretion evolution of the analogs of Mercury, Venus, Earth, and Mars formed in optimal 4-P systems #6, #10, and #16 in Figures 4, 5, and 6. The figures also show the evolution of the Earth analog's WMF during its formation. Confirming the discussion in Section 3.1, the planets obtained in those systems acquired orbits and masses similar to the properties of the actual planets (Figure 1 and Table 2). These systems simultaneously satisfied several inner solar system constraints, as probed by many representative parameters (Tables 2 and 3). These systems also illustrate some features commonly seen in our optimal 4-P systems. First, the Mercury analogs formed relatively isolated from the other planets, confirming the essential role of disks containing an extended inner region component (Disk Ix in Paper I). Second, the Venus and Earth analogs exhibited similar dynamical and accretional evolutions. Also, Mars analogs tended to acquire distant orbits from their Earth counterparts. Additionally, Mars analogs acquired orbits compatible with observed values in terms of $a$, $e$, and $i$ simultaneously. Finally, considering one of our WMF models that assumed the disk was dry within 2 au (initial WMF = 0.001%) and wet beyond it (initial WMF = 10%), all analogs began dry and later acquired at least two Earth's ocean of water in mass in less than 10 Myr. At the end of this evolution, the analogs accreted sufficient water mass to satisfy the Earth's bulk water constraint. See the captions of Figures 4-6 for other details about these systems.

### 4 Summary

The JSCE scenario is based on widespread dynamical excitation of the protoplanetary disk beyond ~1–1.5 au and the existence of an extended disk inner region within 0.8 au. Here, we evaluated the properties of 37 optimal terrestrial planet systems containing exactly one analog of Mercury, Venus, Earth, and Mars in each system, as obtained in the framework of the JSCE scenario. We found that these systems often satisfied some fundamental constraints of the inner solar system (Section 1.1), as evidenced by our main results (1–5) summarized below.

1) Four terrestrial planets within the same system (4-P system) formed with orbits and masses similar to the observed values, as indicated by the dynamically cold and mild orbits of the Venus–Earth and Mercury–Mars analog pairs and the relatively high success rates in explaining the mutual orbital separations (~40–85%) and mass ratios of the planets (~70–90%).
2) Water delivery to all terrestrial planets occurred through early accretion of water-bearing objects

in less than 40 Myr.

3) Earth analogs acquired water mass fractions consistent with estimates of Earth's bulk water only in disks that initially contained sufficient water mass in objects located beyond ~1–1.5 au. This constraint implies that Mercury, Venus, and Mars acquired water similar to the amount on Earth during their formation.

4) Consistent with our water/composition models positing the existence of volatile-rich objects throughout the disk, Earth accreted most of its carbonaceous chondrite (CC) material from within 2 au during its main and late accretion stages over a few tens to hundreds of Myr. Most Earth analogs also satisfied other constraints: Moon-forming last giant impacts within 25–250 Myr (~60% of cases) at a median of 45 Myr and non-CC dominated late accretion of 0.01 ME (~40–50% of cases).

5) Mars analogs typically accreted $\geq$ 80% and $\geq$ 90% of their final masses after ~5 Myr and a median of ~25 Myr (associated with the last giant impact), respectively. These results imply that Mars experienced protracted accretion.

In conclusion, although terrestrial planet formation is a highly stochastic process that prevents the acquisition of results matching all constraints every time, our model of terrestrial planet formation could produce representative systems that simultaneously satisfy several constraints of the inner solar system. These additional results suggest that the JSCE scenario can reproduce the orbits, masses, and other properties of the solar system's four terrestrial planets in a single evolutionary framework.

**Acknowledgments**

We are grateful to the reviewer Matthew Clement and one anonymous reviewer for several helpful comments, which allowed us to improve the overall presentation of this work. The simulations presented here were mainly performed using the general-purpose PC cluster at the Center for Computational Astrophysics (CfCA) in the National Astronomical Observatory of Japan (NAOJ). We are grateful for the generous time allocated to run the simulations.

**Data availability**

The presented data substantiate this study's findings. Supplementary data are available upon reasonable request from the corresponding author.

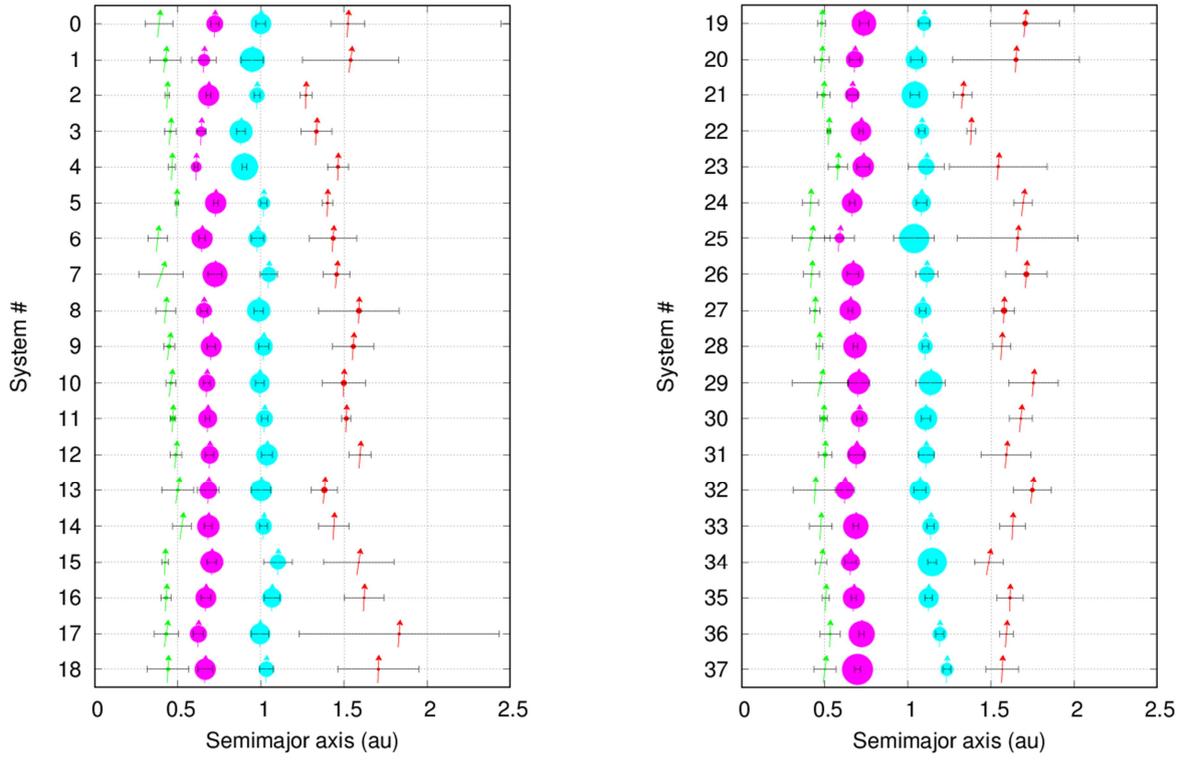

**Figure 1.** Thirty-seven individual optimal analog systems containing exactly one representative planet analog of each real terrestrial planet. System #0 shows the inner solar system planets. Green-, magenta-, cyan- and red-filled symbols indicate the planet analogs of Mercury, Venus, Earth, and Mars, respectively. The inclination $i$ of the planets is represented by the angle between the vector and the perpendicular (e.g., the vector points to the top for $i = 0$ deg). Error bars indicate variation in heliocentric distance based on the object's perihelion and aphelion. The radii of planet symbols scale in proportion to the mass.

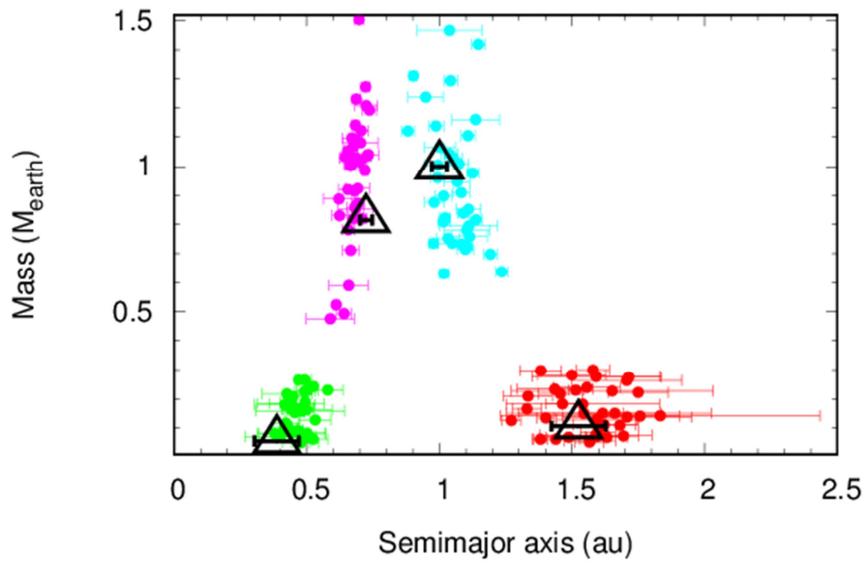

**Figure 2.** Planets formed in 37 optimal 4-P systems. Green-, magenta-, cyan- and red-filled symbols indicate the planet analogs of Mercury, Venus, Earth, and Mars, respectively. Error bars indicate variation in heliocentric distance based on the object's perihelion and aphelion. Large open triangles represent the terrestrial planets of the solar system.

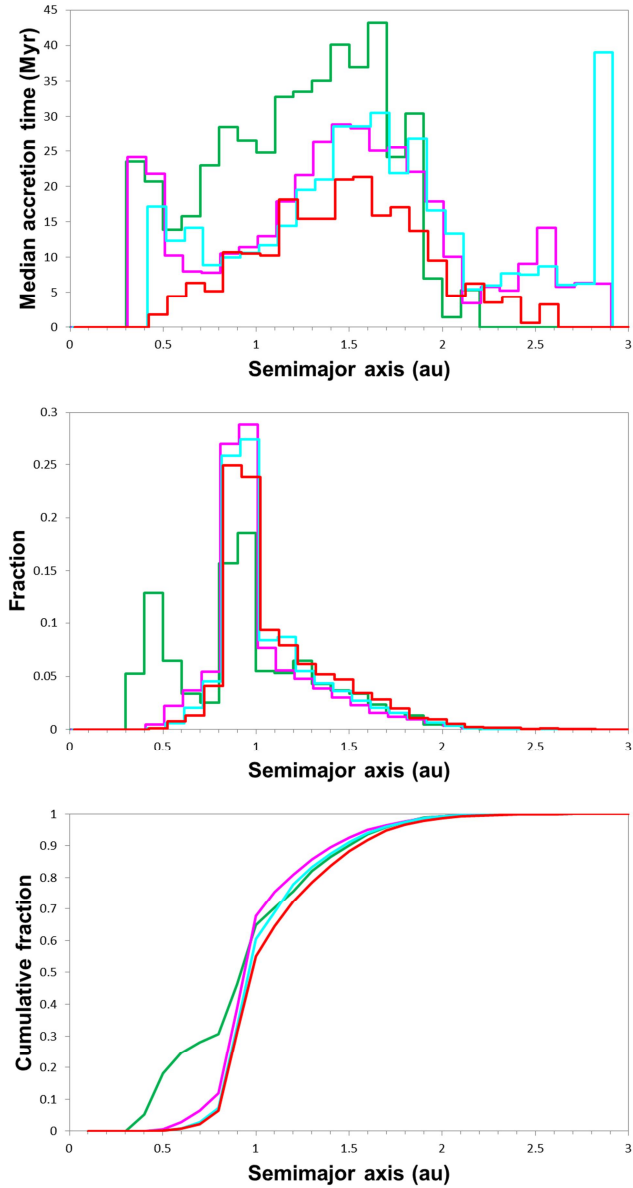

**Figure 3.** Feeding zones and their accretion timescales for planets formed in 37 optimal 4-P systems. The median accretion time for objects to accrete was determined for each 0.1 au zone, considering their initial location within that zone. The results for planet analogs of Mercury, Venus, Earth, and Mars are indicated by green, magenta, cyan, and red curves, respectively. The histograms were offset by 0.0075 au for clarity. The result in cyan at bin 2.9–3.0 au is an outlier (top panel).

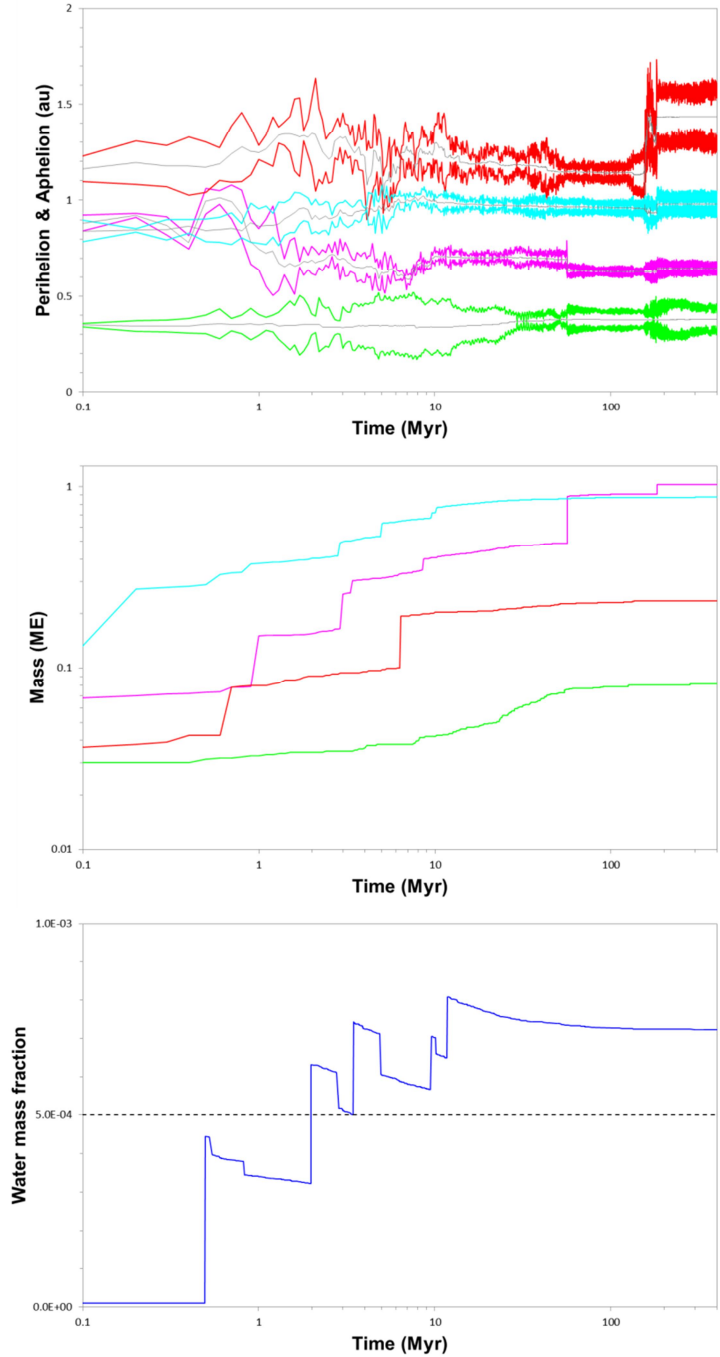

**Figure 4.** Orbital evolution (top) and mass accretion history (middle) of the planet analogs of Mercury (green curves), Venus (magenta curves), Earth (cyan curves), and Mars (red curves) formed in 4-P system #6 (Figure 1). The water mass fraction evolution of the Earth analog and the minimum Earth's bulk water amount (dashed line) is illustrated in the bottom panel. This system formed an excellent analog of Mercury in terms of orbit and mass. The other planets also represent suitable analogs, as the excitation factors $Ef_{23}$ and $Ef_{14}$ were reproduced, and all the planets' mutual distances and mass ratios were within the success criteria adopted in this work. The Earth analog experienced 12 giant impacts (for an ITr = 0.05). However, the last Moon-forming giant impact occurred too early (~10 Myr), resulting in excessive mass accretion afterward. See Section 3.4 and Tables 2 and 3 for other details.

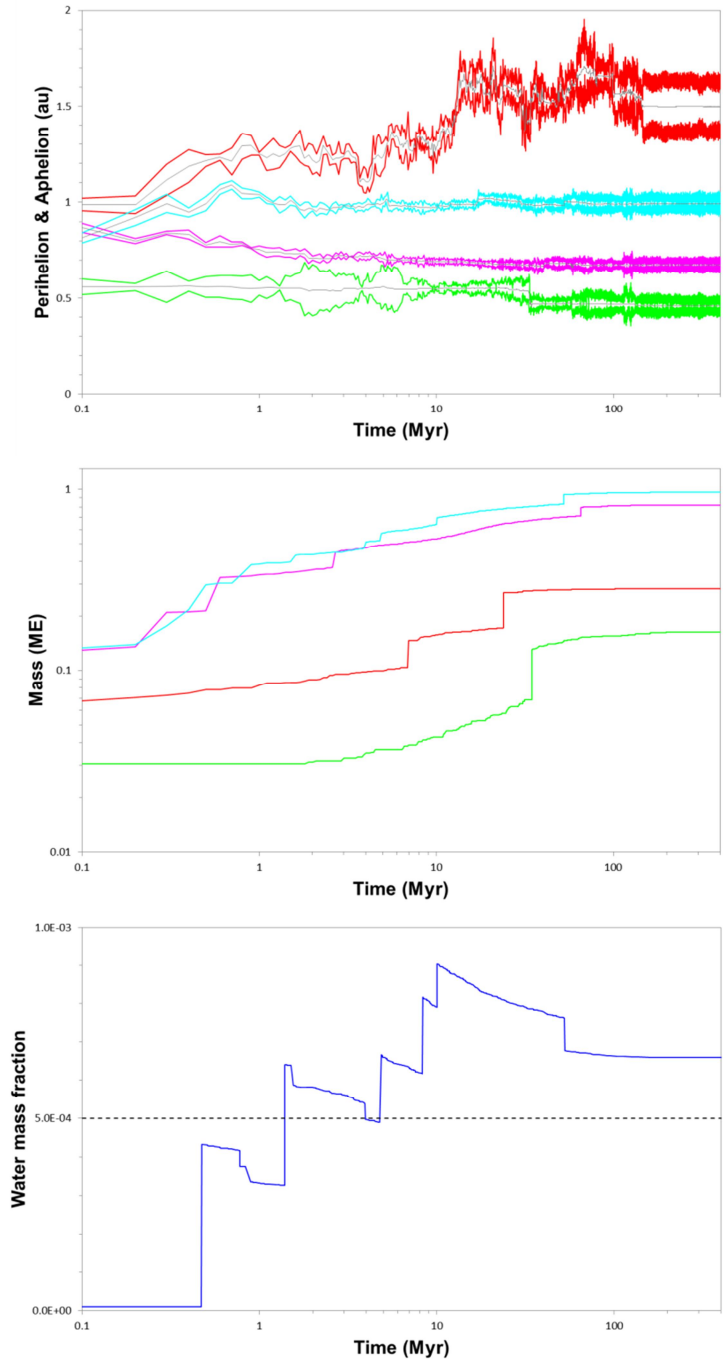

**Figure 5.** Same as in Figure 4, but illustrated for 4-P system #10. This system produced excellent analogs of Venus, Earth, and Mars in terms of orbit and mass, as well as an acceptable Mercury analog. In particular, the Venus and Earth analogs acquired dynamically cold orbits and masses that closely resembled those of the actual planets. Both planets also formed in mutually close orbits similar to the observed. The Mars analog formed at approximately the correct distance from the Earth analog. The excitation factors $Ef_{23}$ and $Ef_{14}$ were reproduced. However, the Mercury analog is overly massive (~3 times the observed). Although all the planets' mutual distances and mass ratios were within the success criteria adopted in this work, the results for the Mercury analog were marginal at best. The Earth analog experienced 13 GIs (for an ITr = 0.05), whereas the last was compatible with the Moon-forming timing. The planet's late accretion mass fraction was slightly higher than the estimated for Earth. See Section 3.4 and Tables 2 and 3 for other details.

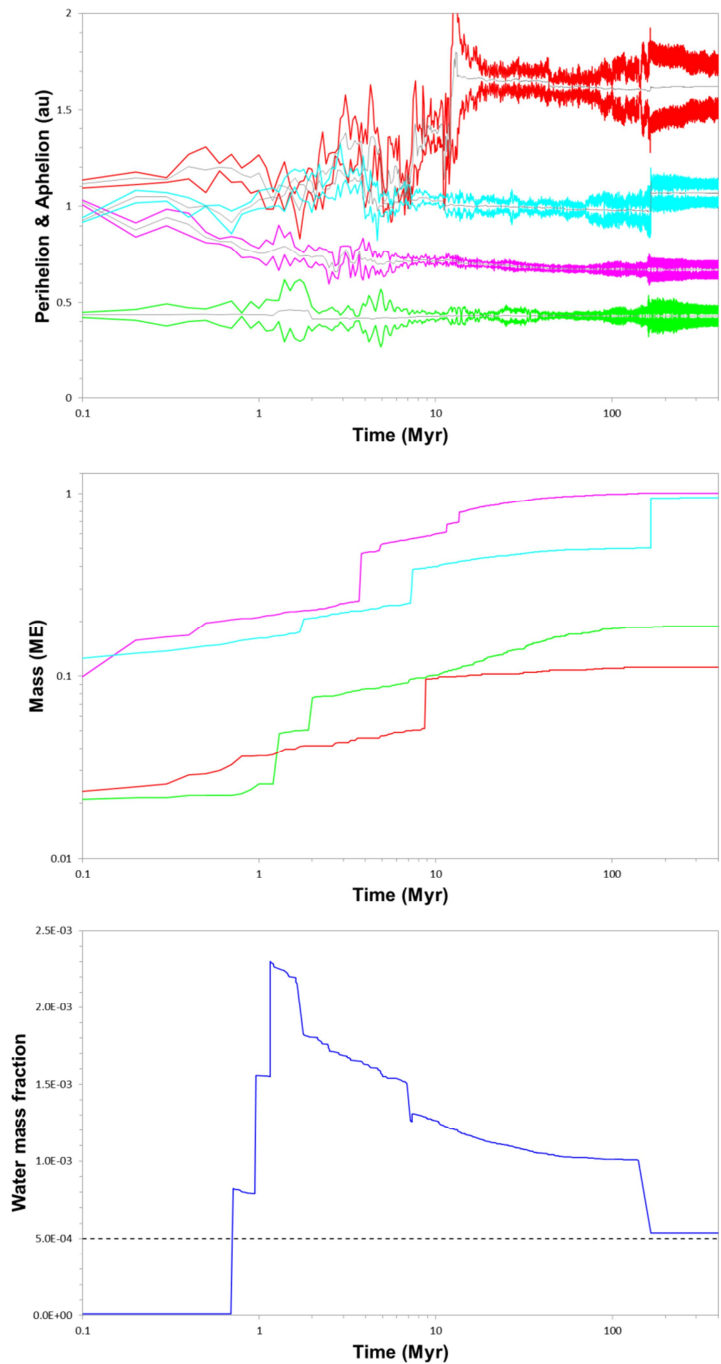

**Figure 6.** Same as in Figure 4, but illustrated for 4-P system #16. This system produced terrestrial planet analogs with properties resembling those of system #10. The following highlights the main differences. The Mars analog acquired an orbit and mass in excellent agreement with the observed. The orbital separation of the Earth–Mars analog pair was also reproduced. However, the orbital separation of the Venus–Earth analog pair was marginal at best. Earth analog's late accretion mass fraction was reproduced after the last giant impact. See Section 3.4 and Tables 2 and 3 for other details.

**Table 1.** Water mass fractions (WMFs) of objects according to their initial location in the protoplanetary disk and favored WMFs as constrained by models that simultaneously satisfied the bulk water contents of Mercury, Venus, Earth, and Mars

| Disk region (au) | Initial WMFs (%) | Fiducial hypothesis - dry Venus (%) | Fiducial hypothesis - wet Venus (%) | Water worlds hypothesis (%) |
|---|---|---|---|---|
| <1.0 | 0.001, 0.01, 0.1, 1, 5 | 0.001-0.01 | 0.001-0.1 | 0.001-1 |
| 1.0–1.5 | 0.001, 0.01, 0.1, 1, 5 | 0.001-0.1 | 0.001-0.1 | 0.01-1 |
| 1.5–2.0 | 0.001, 0.01, 0.1, 1, 5, 10, 30, 50 | 0.01-0.1 | 0.01-1 | 1-10 |
| 2.0–2.5 | 0.1, 1, 5, 10, 30, 50 | 0.1-1 | 1-30 | 5-50 |
| >2.5 | 5, 10, 30, 50 | 10-50 | 10-50 | 10-50 |

**Notes.** We established distinct water distributions in the disk by combining the initial WMFs above in the different regions. In this process, we assumed that the WMF of an outward region always contained a WMF equal to or larger than the WMF assumed for an adjacent inward region. For example, if the disk region at < 1.0 au initially contained a 0.1% WMF, the WMFs tested in the adjacent 1.0–1.5 au region were 0.1, 1, and 5%. If the disk region at 1.0–1.5 au initially contained a 0.1% WMF, the WMFs tested in the adjacent 1.5–2.0 au region were 0.1, 1, 5, 10, 30, and 50%. Based on this rule, all possible combinations of WMFs among the regions considered above resulted in 556 distinct WMF models. The rightmost three columns summarize the favored initial WMFs constrained by 28, 142, and 99 successful WMF models under the fiducial (dry Venus case), fiducial (wet Venus case), and water worlds hypotheses for the water distribution in the disk, respectively. See Sections 2.2 and 3.3 for more details. See also Table A1 (Appendix) for a complete list of the initial conditions and results among the 556 WMF models tested in this work.

**Table 2.** Fundamental properties of analogs of Mercury, Venus, Earth, and Mars formed in 37 optimal 4-P systems

| # | $m_1$ | $a_1$ | $e_1$ | $i_1$ | $m_2$ | $a_2$ | $e_2$ | $i_2$ | $m_3$ | $a_3$ | $e_3$ | $i_3$ | $m_4$ | $a_4$ | $e_4$ | $i_4$ | $Ef_{23}$ | $Ef_{14}$ | $D_{12}$ | $D_{23}$ | $D_{34}$ | $M_{12}$ | $M_{23}$ | $M_{43}$ |
|---|---|---|---|---|---|---|---|---|---|---|---|---|---|---|---|---|---|---|---|---|---|---|---|---|
| 1 | 0.218 | 0.425 | 0.220 | 7.4 | 0.589 | 0.658 | 0.112 | 3.6 | 1.236 | 0.948 | 0.070 | 2.9 | 0.182 | 1.540 | 0.188 | 7.9 | 7.4 | 16.9 | 0.23 | 0.29 | 0.59 | 0.37 | 0.48 | 0.15 |
| 2 | 0.092 | 0.437 | 0.031 | 2.7 | 1.029 | 0.686 | 0.020 | 1.8 | 0.735 | 0.977 | 0.019 | 1.8 | 0.126 | 1.271 | 0.028 | 2.4 | 2.6 | 3.7 | 0.25 | 0.29 | 0.29 | 0.09 | 1.40 | 0.17 |
| 3 | 0.153 | 0.455 | 0.077 | 5.5 | 0.493 | 0.641 | 0.044 | 3.3 | 1.121 | 0.881 | 0.029 | 2.0 | 0.210 | 1.334 | 0.070 | 4.5 | 4.2 | 8.0 | 0.19 | 0.24 | 0.45 | 0.31 | 0.44 | 0.19 |
| 4 | 0.091 | 0.465 | 0.043 | 2.9 | 0.523 | 0.611 | 0.019 | 1.9 | 1.308 | 0.902 | 0.016 | 1.8 | 0.183 | 1.464 | 0.043 | 1.8 | 2.5 | 4.2 | 0.15 | 0.29 | 0.56 | 0.17 | 0.40 | 0.14 |
| 5 | 0.190 | 0.494 | 0.020 | 2.3 | 1.036 | 0.728 | 0.019 | 1.8 | 0.630 | 1.017 | 0.018 | 1.8 | 0.135 | 1.401 | 0.023 | 1.9 | 2.5 | 2.9 | 0.23 | 0.29 | 0.38 | 0.18 | 1.64 | 0.21 |
| 6 | 0.081 | 0.379 | 0.157 | 6.5 | 1.032 | 0.646 | 0.028 | 2.5 | 0.876 | 0.980 | 0.038 | 2.6 | 0.234 | 1.435 | 0.099 | 4.6 | 3.9 | 11.2 | 0.27 | 0.33 | 0.46 | 0.08 | 1.18 | 0.27 |
| 7 | 0.071 | 0.400 | 0.334 | 20.0 | 1.207 | 0.724 | 0.058 | 3.0 | 0.735 | 1.048 | 0.050 | 3.3 | 0.218 | 1.456 | 0.056 | 5.9 | 5.4 | 20.9 | 0.32 | 0.32 | 0.41 | 0.06 | 1.64 | 0.30 |
| 8 | 0.093 | 0.427 | 0.138 | 5.9 | 0.782 | 0.657 | 0.034 | 2.2 | 1.138 | 0.987 | 0.029 | 2.0 | 0.278 | 1.590 | 0.152 | 1.8 | 3.4 | 10.6 | 0.23 | 0.33 | 0.60 | 0.12 | 0.69 | 0.24 |
| 9 | 0.209 | 0.448 | 0.074 | 8.4 | 1.015 | 0.701 | 0.035 | 2.4 | 0.898 | 1.016 | 0.030 | 2.7 | 0.240 | 1.556 | 0.080 | 4.7 | 3.9 | 9.6 | 0.25 | 0.32 | 0.54 | 0.21 | 1.13 | 0.27 |
| 10 | 0.163 | 0.459 | 0.066 | 8.5 | 0.819 | 0.675 | 0.031 | 2.4 | 0.961 | 0.993 | 0.026 | 2.6 | 0.283 | 1.499 | 0.086 | 3.0 | 3.6 | 8.8 | 0.22 | 0.32 | 0.51 | 0.20 | 0.85 | 0.29 |
| 11 | 0.265 | 0.470 | 0.032 | 3.1 | 0.915 | 0.680 | 0.019 | 2.0 | 0.823 | 1.022 | 0.019 | 1.9 | 0.231 | 1.514 | 0.019 | 2.2 | 2.7 | 3.6 | 0.21 | 0.34 | 0.49 | 0.29 | 1.11 | 0.28 |
| 12 | 0.171 | 0.489 | 0.072 | 5.6 | 0.873 | 0.692 | 0.037 | 2.2 | 1.043 | 1.037 | 0.031 | 2.3 | 0.085 | 1.597 | 0.041 | 4.5 | 3.6 | 7.2 | 0.20 | 0.35 | 0.56 | 0.20 | 0.84 | 0.08 |
| 13 | 0.163 | 0.500 | 0.193 | 9.5 | 0.854 | 0.684 | 0.096 | 3.6 | 1.065 | 1.002 | 0.058 | 2.8 | 0.298 | 1.382 | 0.057 | 5.1 | 6.6 | 12.6 | 0.18 | 0.32 | 0.38 | 0.19 | 0.80 | 0.28 |
| 14 | 0.062 | 0.524 | 0.107 | 8.7 | 1.093 | 0.684 | 0.034 | 3.1 | 0.809 | 1.016 | 0.023 | 3.9 | 0.062 | 1.439 | 0.063 | 3.7 | 4.5 | 9.7 | 0.16 | 0.33 | 0.42 | 0.06 | 1.35 | 0.08 |
| 15 | 0.107 | 0.424 | 0.050 | 2.1 | 1.123 | 0.705 | 0.039 | 1.9 | 0.780 | 1.103 | 0.078 | 1.9 | 0.076 | 1.589 | 0.133 | 8.7 | 4.6 | 9.3 | 0.28 | 0.40 | 0.49 | 0.10 | 1.44 | 0.10 |
| 16 | 0.188 | 0.430 | 0.073 | 4.7 | 1.007 | 0.669 | 0.044 | 1.8 | 0.948 | 1.067 | 0.044 | 2.1 | 0.111 | 1.621 | 0.073 | 3.3 | 3.9 | 7.1 | 0.24 | 0.40 | 0.55 | 0.19 | 1.06 | 0.12 |
| 17 | 0.183 | 0.431 | 0.171 | 9.2 | 0.830 | 0.624 | 0.047 | 2.6 | 1.004 | 0.995 | 0.054 | 2.1 | 0.142 | 1.832 | 0.329 | 5.0 | 4.6 | 18.6 | 0.19 | 0.37 | 0.84 | 0.22 | 0.83 | 0.14 |
| 18 | 0.183 | 0.441 | 0.286 | 3.3 | 1.029 | 0.665 | 0.066 | 2.5 | 0.752 | 1.033 | 0.041 | 2.4 | 0.138 | 1.707 | 0.143 | 2.4 | 4.8 | 13.2 | 0.22 | 0.37 | 0.67 | 0.18 | 1.37 | 0.18 |
| 19 | 0.158 | 0.482 | 0.050 | 4.6 | 1.192 | 0.736 | 0.037 | 2.2 | 0.714 | 1.098 | 0.029 | 2.2 | 0.264 | 1.706 | 0.122 | 6.0 | 3.6 | 8.9 | 0.25 | 0.36 | 0.61 | 0.13 | 1.67 | 0.37 |
| 20 | 0.168 | 0.482 | 0.091 | 5.1 | 0.859 | 0.681 | 0.046 | 2.2 | 1.030 | 1.051 | 0.033 | 1.9 | 0.228 | 1.651 | 0.231 | 3.3 | 3.8 | 11.7 | 0.20 | 0.37 | 0.60 | 0.20 | 0.83 | 0.22 |
| 21 | 0.226 | 0.492 | 0.078 | 5.3 | 0.711 | 0.666 | 0.047 | 2.6 | 1.292 | 1.042 | 0.025 | 1.9 | 0.166 | 1.330 | 0.042 | 6.1 | 3.7 | 7.9 | 0.17 | 0.38 | 0.29 | 0.32 | 0.55 | 0.13 |
| 22 | 0.243 | 0.525 | 0.019 | 3.0 | 0.986 | 0.719 | 0.017 | 1.9 | 0.732 | 1.084 | 0.016 | 1.9 | 0.062 | 1.381 | 0.019 | 2.6 | 2.5 | 3.4 | 0.19 | 0.37 | 0.30 | 0.25 | 1.35 | 0.09 |
| 23 | 0.231 | 0.579 | 0.100 | 3.9 | 1.041 | 0.732 | 0.053 | 3.1 | 0.795 | 1.111 | 0.098 | 3.8 | 0.150 | 1.544 | 0.190 | 5.0 | 6.8 | 11.1 | 0.15 | 0.38 | 0.43 | 0.22 | 1.31 | 0.19 |
| 24 | 0.098 | 0.415 | 0.116 | 3.6 | 1.002 | 0.665 | 0.026 | 1.8 | 0.910 | 1.082 | 0.029 | 1.9 | 0.073 | 1.694 | 0.033 | 8.0 | 3.0 | 8.8 | 0.25 | 0.42 | 0.61 | 0.10 | 1.10 | 0.08 |

| # | $m_1$ | $a_1$ | $e_1$ | $i_1$ | $m_2$ | $a_2$ | $e_2$ | $i_2$ | $m_3$ | $a_3$ | $e_3$ | $i_3$ | $m_4$ | $a_4$ | $e_4$ | $i_4$ | $Ef_{12}$ | $Ef_{23}$ | $D_{12}$ | $D_{23}$ | $D_{34}$ | $M_{12}$ | $M_{23}$ | $M_{43}$ |
|---|---|---|---|---|---|---|---|---|---|---|---|---|---|---|---|---|---|---|---|---|---|---|---|---|
| 25 | 0.181 | 0.418 | 0.274 | 11.2 | 0.475 | 0.589 | 0.154 | 6.4 | 1.467 | 1.038 | 0.117 | 4.0 | 0.151 | 1.661 | 0.219 | 5.1 | 11.3 | 19.4 | 0.17 | 0.45 | 0.62 | 0.38 | 0.32 | 0.10 |
| 26 | 0.117 | 0.421 | 0.114 | 4.8 | 1.097 | 0.670 | 0.053 | 1.8 | 0.758 | 1.114 | 0.060 | 2.1 | 0.276 | 1.713 | 0.073 | 4.4 | 4.5 | 8.7 | 0.25 | 0.44 | 0.60 | 0.11 | 1.45 | 0.36 |
| 27 | 0.165 | 0.442 | 0.068 | 2.6 | 1.053 | 0.654 | 0.021 | 1.8 | 0.839 | 1.090 | 0.017 | 2.1 | 0.300 | 1.579 | 0.040 | 2.5 | 2.7 | 4.9 | 0.21 | 0.44 | 0.49 | 0.16 | 1.26 | 0.36 |
| 28 | 0.091 | 0.467 | 0.041 | 3.1 | 1.141 | 0.683 | 0.020 | 1.9 | 0.723 | 1.105 | 0.017 | 2.0 | 0.052 | 1.565 | 0.035 | 4.2 | 2.6 | 5.1 | 0.22 | 0.42 | 0.46 | 0.08 | 1.58 | 0.07 |
| 29 | 0.159 | 0.475 | 0.355 | 12.8 | 1.081 | 0.703 | 0.093 | 3.2 | 1.160 | 1.137 | 0.078 | 3.1 | 0.141 | 1.756 | 0.085 | 5.5 | 7.0 | 18.9 | 0.23 | 0.43 | 0.62 | 0.15 | 0.93 | 0.12 |
| 30 | 0.265 | 0.493 | 0.047 | 3.3 | 0.820 | 0.708 | 0.024 | 2.2 | 1.105 | 1.108 | 0.025 | 1.9 | 0.110 | 1.680 | 0.041 | 6.3 | 3.0 | 6.4 | 0.22 | 0.40 | 0.57 | 0.32 | 0.74 | 0.10 |
| 31 | 0.229 | 0.502 | 0.079 | 2.9 | 0.924 | 0.692 | 0.066 | 2.0 | 0.852 | 1.110 | 0.042 | 2.2 | 0.130 | 1.593 | 0.094 | 6.0 | 4.5 | 8.2 | 0.19 | 0.42 | 0.48 | 0.25 | 1.09 | 0.15 |
| 32 | 0.082 | 0.442 | 0.295 | 3.7 | 0.888 | 0.621 | 0.093 | 2.0 | 1.009 | 1.073 | 0.035 | 2.0 | 0.223 | 1.749 | 0.065 | 7.6 | 4.9 | 13.9 | 0.18 | 0.45 | 0.68 | 0.09 | 0.88 | 0.22 |
| 33 | 0.050 | 0.476 | 0.144 | 3.9 | 1.229 | 0.687 | 0.023 | 1.9 | 0.816 | 1.138 | 0.019 | 1.9 | 0.068 | 1.631 | 0.047 | 3.5 | 2.7 | 8.0 | 0.21 | 0.45 | 0.49 | 0.04 | 1.51 | 0.08 |
| 34 | 0.051 | 0.477 | 0.073 | 10.5 | 0.920 | 0.655 | 0.053 | 2.2 | 1.419 | 1.147 | 0.023 | 1.8 | 0.069 | 1.487 | 0.058 | 10.5 | 3.6 | 12.4 | 0.18 | 0.49 | 0.34 | 0.06 | 0.65 | 0.05 |
| 35 | 0.070 | 0.506 | 0.043 | 3.6 | 1.067 | 0.675 | 0.022 | 2.4 | 0.977 | 1.125 | 0.021 | 1.9 | 0.151 | 1.615 | 0.049 | 2.0 | 3.0 | 4.7 | 0.17 | 0.45 | 0.49 | 0.07 | 1.09 | 0.16 |
| 36 | 0.128 | 0.532 | 0.112 | 2.3 | 1.271 | 0.722 | 0.022 | 1.8 | 0.697 | 1.192 | 0.021 | 2.1 | 0.071 | 1.593 | 0.026 | 4.9 | 2.8 | 6.6 | 0.19 | 0.47 | 0.40 | 0.10 | 1.82 | 0.10 |
| 37 | 0.083 | 0.501 | 0.133 | 7.8 | 1.504 | 0.697 | 0.025 | 1.8 | 0.636 | 1.235 | 0.019 | 2.0 | 0.057 | 1.568 | 0.062 | 3.2 | 2.8 | 9.7 | 0.20 | 0.54 | 0.33 | 0.06 | 2.37 | 0.09 |
| *median* | *0.159* | *0.467* | *0.079* | *4.7* | *1.007* | *0.681* | *0.037* | *2.2* | *0.898* | *1.051* | *0.029* | *2.1* | *0.150* | *1.579* | *0.063* | *4.5* | *3.7* | *8.8* | *0.21* | *0.37* | *0.49* | *0.17* | *1.10* | *0.15* |
| Solar system | 0.055 | 0.387 | 0.215 | 6.8 | 0.815 | 0.723 | 0.032 | 2.2 | 1.000 | 1.000 | 0.028 | 2.0 | 0.107 | 1.524 | 0.067 | 4.0 | 3.3 | 11.8 | 0.336 | 0.277 | 0.524 | 0.067 | 0.815 | 0.107 |
| SC | | | | | | | | | | | | | | | | | ≤5.0 | ≤11.8 | 0.22-0.45 | 0.18-0.37 | 0.35-0.70 | 0.03-0.20 | 0.41-1.63 | 0.05-0.32 |
| Sf (%) | | | | | | | | | | | | | | | | | 84 | 76 | 38 | 51 | 84 | 70 | 81 | 89 |

**Notes.** $m$, planet mass (ME); $a$, semimajor axis (au); $e$, eccentricity; $i$, inclination (deg); $Ef$, excitation factor combining the eccentricities and inclinations of a pair of planets (%); $D_{12}$, the mutual distance (difference of semimajor axes) of Mercury and Venus (au); $D_{23}$, the mutual distance of Venus and Earth (au); $D_{34}$, the mutual distance of Earth and Mars (au); $M_{12}$, the mass ratio of Mercury and Venus; $M_{23}$, the mass ratio of Venus and Earth; $M_{43}$, the mass ratio of Mars and Earth. Subscripts 1, 2, 3, and 4 refer to Mercury, Venus, Earth, and Mars, respectively. SC and Sf are the success criteria and success fraction, respectively. See Section 3.1 for more details. We averaged the orbital elements of the real planets over the last 100 Myr after integrating their orbits.

Table 3. Some properties of 37 optimal 4-P systems that simultaneously contained analogs of Mercury, Venus, Earth, and Mars

| # | LAf | $fT_3$ | $fT_4$ | $tLGI_1$ | $tLGI_2$ | $tLGI_3$ | $tLGI_4$ | $nGI_1$ | $nGI_2$ | $nGI_3$ | $nGI_4$ | AMD | RMC |
|---|---|---|---|---|---|---|---|---|---|---|---|---|---|
| 1 | 4.2 | 18.2 | 178.2 | 88.0 | 31.1 | 17.1 | 178.2 | 7 | 15 | 16 | 7 | 0.0094 | 53.4 |
| 2 | 11.1 | 19.0 | 138.9 | - | 2.1 | 17.8 | 138.8 | 0 | 10 | 7 | 1 | 0.0008 | 91.9 |
| 3 | 21.0 | 18.7 | 373.2 | 130.0 | 23.3 | 8.0 | 373.2 | 1 | 6 | 13 | 2 | 0.0024 | 75.9 |
| 4 | 0.7 | 75.3 | 33.9 | - | 3.6 | 146.4 | 31.6 | 0 | 4 | 9 | 1 | 0.0008 | 75.6 |
| 5 | 29.5 | 14.7 | 60.4 | 57.5 | 67.3 | 2.8 | 60.4 | 1 | 8 | 7 | 2 | 0.0007 | 80.4 |
| 6 | 12.3 | 12.2 | 22.9 | - | 182.4 | 10.1 | 6.3 | 0 | 7 | 12 | 2 | 0.0029 | 56.7 |
| 7 | 0.2 | 147.0 | 9.7 | - | 9.1 | 205.8 | 6.4 | 0 | 13 | 8 | 3 | 0.0059 | 71.0 |
| 8 | 0.0 | 365.5 | 11.8 | - | 41.0 | 365.5 | 6.0 | 0 | 7 | 10 | 3 | 0.0033 | 53.3 |
| 9 | 41.2 | 23.3 | 22.6 | 32.1 | 190.3 | 2.2 | 1.9 | 2 | 7 | 8 | 3 | 0.0030 | 52.2 |
| 10 | 2.5 | 52.6 | 24.0 | 34.7 | 65.4 | 52.6 | 23.9 | 1 | 8 | 13 | 3 | 0.0026 | 53.8 |
| 11 | 5.7 | 35.6 | 31.0 | 2.5 | 4.2 | 35.5 | 30.9 | 3 | 8 | 11 | 3 | 0.0009 | 49.2 |
| 12 | 0.7 | 150.7 | 282.3 | 5.7 | 14.4 | 150.7 | 282.2 | 1 | 17 | 14 | 1 | 0.0018 | 70.1 |
| 13 | 12.6 | 21.7 | 20.2 | 20.1 | 14.1 | 12.4 | 15.8 | 2 | 8 | 15 | 9 | 0.0060 | 68.4 |
| 14 | 4.0 | 16.1 | 6.7 | 14.0 | 24.3 | 17.5 | 0.5 | 1 | 9 | 14 | 1 | 0.0028 | 102.3 |
| 15 | 0.9 | 37.9 | 7.7 | 8.9 | 63.5 | 49.5 | 0.4 | 1 | 14 | 14 | 1 | 0.0032 | 63.9 |
| 16 | 0.4 | 166.4 | 15.0 | 2.0 | 13.6 | 166.3 | 8.8 | 2 | 7 | 6 | 1 | 0.0020 | 50.3 |
| 17 | 0.8 | 80.8 | 27.7 | 3.0 | 32.9 | 80.8 | 24.7 | 1 | 9 | 7 | 1 | 0.0092 | 43.0 |
| 18 | 5.2 | 22.2 | 21.7 | 0.6 | 30.9 | 22.2 | 3.3 | 2 | 9 | 10 | 1 | 0.0059 | 48.8 |
| 19 | 0.8 | 93.4 | 25.3 | 69.0 | 45.3 | 93.4 | 2.2 | 1 | 8 | 8 | 3 | 0.0033 | 50.2 |
| 20 | 0.9 | 100.0 | 154.2 | 95.7 | 7.1 | 99.9 | 154.2 | 1 | 7 | 9 | 1 | 0.0054 | 50.4 |
| 21 | 1.3 | 21.8 | 13.2 | 47.7 | 10.6 | 35.3 | 13.2 | 5 | 11 | 18 | 5 | 0.0021 | 64.7 |
| 22 | 4.4 | 44.8 | 33.9 | 14.3 | 11.6 | 44.8 | - | 2 | 7 | 9 | 0 | 0.0008 | 77.0 |
| 23 | 1.6 | 48.5 | 201.2 | 26.4 | 26.1 | 48.4 | 201.2 | 8 | 12 | 11 | 4 | 0.0066 | 70.7 |
| 24 | 8.1 | 25.4 | 57.7 | - | 149.2 | 25.3 | 49.5 | 0 | 10 | 11 | 1 | 0.0016 | 55.9 |

| | LAf | fT2 | fT3 | fT4 | tLGI2 | tLGI3 | tLGI4 | nGI1 | nGI2 | nGI3 | nGI4 | AMD | RMC |
|---|---|---|---|---|---|---|---|---|---|---|---|---|---|
| 25 | 0.0 | 396.1 | 46.2 | 66.6 | 13.3 | 396.0 | 46.1 | 1 | 4 | 9 | 1 | 0.0150 | 43.4 |
| 26 | 10.7 | 33.2 | 33.9 | 135.5 | 89.9 | 22.5 | 33.9 | 1 | 13 | 11 | 4 | 0.0030 | 38.5 |
| 27 | 3.7 | 53.8 | 27.2 | 5.1 | 47.5 | 53.7 | 27.1 | 1 | 6 | 8 | 3 | 0.0011 | 40.4 |
| 28 | 18.2 | 18.8 | 54.6 | - | 5.8 | 7.6 | - | 0 | 9 | 7 | 0 | 0.0009 | 69.5 |
| 29 | 1.0 | 50.7 | 13.5 | 42.1 | 35.3 | 50.6 | 13.4 | 3 | 15 | 17 | 5 | 0.0091 | 52.4 |
| 30 | 0.2 | 138.3 | 53.2 | 10.3 | 27.9 | 138.3 | 53.1 | 2 | 6 | 7 | 1 | 0.0015 | 52.5 |
| 31 | 0.3 | 37.3 | 12.2 | 9.6 | 6.1 | 166.3 | 2.7 | 2 | 6 | 9 | 1 | 0.0030 | 52.8 |
| 32 | 0.9 | 79.6 | 9.1 | - | 26.5 | 79.6 | 6.3 | 0 | 7 | 7 | 2 | 0.0050 | 40.0 |
| 33 | 12.2 | 25.7 | 4.4 | - | 5.9 | 8.2 | 1.5 | 0 | 12 | 11 | 1 | 0.0011 | 67.0 |
| 34 | 1.3 | 43.0 | 18.4 | - | 15.0 | 42.9 | 4.3 | 0 | 18 | 10 | 1 | 0.0021 | 61.1 |
| 35 | 3.2 | 30.1 | 8.1 | 27.5 | 24.8 | 30.0 | 5.0 | 1 | 15 | 16 | 4 | 0.0011 | 56.9 |
| 36 | 1.0 | 16.5 | 90.9 | 24.1 | 12.0 | 56.0 | 90.8 | 1 | 12 | 13 | 1 | 0.0012 | 67.8 |
| 37 | 3.1 | 20.4 | 6.9 | 17.3 | 22.4 | 20.3 | 6.7 | 1 | 16 | 18 | 1 | 0.0014 | 63.5 |
| *median* | *2.5* | *37.3* | *25.3* | *24.1* | *24.3* | *44.8* | *15.8* | *1* | *9* | *10* | *1* | *0.0026* | *56.7* |
| Solar system | ≤1 | ≥10-55 | ≤10-20 | ? | ? | 25-250 | ? | ? | ≥0-2? | ≥1 | ≥1? | 0.0018 | 89.7 |
| SC | ≤1 | ≥35 | ≤15 | | | 25-250 (25-150) | | | | | | | |
| Sf (%) | 41 | 54 | 32 | | | 59 (49) | | | | | | | |

**Notes.** *LAf*, Earth's late accretion fraction (%); *fT*, time the planet acquired ≥ 90% of its final mass (Myr); *tLGI*, time of the planet's last giant impact (Myr); *nGI*, number of giant impacts experienced by the planet during its formation. Subscripts 1, 2, 3, and 4 refer to Mercury, Venus, Earth, and Mars, respectively. An impact-to-target ratio of ITr = 0.05 was assumed when determining *tLGI* and *nGI*. SC and Sf are the success criteria and success fraction, respectively. AMD and RMC refer to the system's angular momentum deficit and radial mass concentration, respectively. See Sections 1 and 3.2 for more details.

**Table 4.** Distribution of non-carbonaceous (NC) and carbonaceous (CC) materials in objects according to their initial location in the protoplanetary disk and the results for our model Earth based on the analogs of Earth formed in 37 optimal 4-P systems

| Model | NC100 region (au) | NC+CC region 1 (au) | NC+CC region 2 (au) | CC fraction 1 (%) | CC fraction 2 (%) | Earth fraction in CC (%) | Earth's late accretion fraction in NC (%) | Earth's WMF (Earth oceans) |
|---|---|---|---|---|---|---|---|---|
| 1  | 0–0.8 | 0.8–2   |       | 10 |    | 9.6  | 89.7 | 39.8 |
| 2  | 0–0.8 | 0.8–2   |       | 20 |    | 18.8 | 80.1 | 76.4 |
| 3  | 0–0.8 | 0.8–1   | 1–2   | 5  | 10 | 6.9  | 91.2 | 29.1 |
| 4  | 0–0.8 | 0.8–1   | 1–2   | 5  | 20 | 10.8 | 84.6 | 44.4 |
| 5  | 0–0.8 | 0.8–1   | 1–2   | 10 | 20 | 13.4 | 83.1 | 55.1 |
| 6  | 0–0.8 | 0.8–1.2 | 1.2–2 | 5  | 10 | 6.1  | 92.1 | 25.7 |
| 7  | 0–0.8 | 0.8–1.2 | 1.2–2 | 5  | 20 | 8.2  | 87.3 | 34.2 |
| 8  | 0–0.8 | 0.8–1.2 | 1.2–2 | 10 | 20 | 11.7 | 84.9 | 48.3 |
| 9  | 0–0.8 | 0.8–1.5 | 1.5–2 | 5  | 10 | 5.4  | 93.6 | 23.1 |
| 10 | 0–0.8 | 0.8–1.5 | 1.5–2 | 5  | 20 | 6.2  | 91.7 | 26.2 |
| 11 | 0–0.8 | 0.8–1.5 | 1.5–2 | 10 | 20 | 10.4 | 87.8 | 43.0 |
| 12 | 0–0.8 | 0.8–1.8 | 1.8–2 | 5  | 10 | 5.1  | 94.4 | 21.8 |
| 13 | 0–0.8 | 0.8–1.8 | 1.8–2 | 5  | 20 | 5.2  | 94.1 | 22.5 |
| 14 | 0–0.8 | 0.8–1.8 | 1.8–2 | 10 | 20 | 9.8  | 89.4 | 40.5 |
| 15 | 0–1   | 1–2     |       | 10 |    | 4.3  | 92.7 | 18.4 |
| 16 | 0–1   | 1–2     |       | 20 |    | 8.1  | 86.1 | 33.7 |
| 17 | 0–1   | 1–1.2   | 1.2–2 | 5  | 10 | 3.4  | 93.6 | 15.0 |
| 18 | 0–1   | 1–1.2   | 1.2–2 | 5  | 20 | 5.5  | 88.8 | 23.6 |
| 19 | 0–1   | 1–1.2   | 1.2–2 | 10 | 20 | 6.4  | 87.9 | 26.9 |
| 20 | 0–1   | 1–1.5   | 1.5–2 | 5  | 10 | 2.7  | 95.0 | 12.4 |
| 21 | 0–1   | 1–1.5   | 1.5–2 | 5  | 20 | 3.5  | 93.1 | 15.6 |
| 22 | 0–1   | 1–1.5   | 1.5–2 | 10 | 20 | 5.1  | 90.8 | 21.6 |
| 23 | 0–1 | 1–1.8 | 1.8–2 | 5 | 10 | 2.4 | 95.8 | 11.1 |
| 24 | 0–1 | 1–1.8 | 1.8–2 | 5 | 20 | 2.6 | 95.6 | 11.8 |
| 25 | 0–1 | 1–1.8 | 1.8–2 | 10 | 20 | 4.4 | 92.4 | 19.1 |
| 26 | 0–1.2 | 1.2–2 | | 10 | | 2.5 | 94.5 | 11.7 |
| 27 | 0–1.2 | 1.2–2 | | 20 | | 4.7 | 89.7 | 20.2 |
| 28 | 0–1.2 | 1.2–1.5 | 1.5–2 | 5 | 10 | 1.9 | 96.0 | 9.0 |
| 29 | 0–1.2 | 1.2–1.5 | 1.5–2 | 5 | 20 | 2.7 | 94.0 | 12.2 |
| 30 | 0–1.2 | 1.2–1.5 | 1.5–2 | 10 | 20 | 3.3 | 92.6 | 14.8 |
| 31 | 0–1.2 | 1.2–1.8 | 1.8–2 | 5 | 10 | 1.6 | 96.8 | 7.7 |
| 32 | 0–1.2 | 1.2–1.8 | 1.8–2 | 5 | 20 | 1.7 | 96.5 | 8.4 |
| 33 | 0–1.2 | 1.2–1.8 | 1.8–2 | 10 | 20 | 2.7 | 94.2 | 12.3 |
| 34 | 0–1.5 | 1.5–2 | | 10 | | 1.2 | 97.4 | 6.3 |
| 35 | 0–1.5 | 1.5–2 | | 20 | | 2.0 | 95.5 | 9.5 |
| 36 | 0–1.5 | 1.5–1.8 | 1.8–2 | 5 | 10 | 0.9 | 98.2 | 5.1 |
| 37 | 0–1.5 | 1.5–1.8 | 1.8–2 | 5 | 20 | 1.1 | 97.9 | 5.8 |
| 38 | 0–1.5 | 1.5–1.8 | 1.8–2 | 10 | 20 | 1.4 | 97.1 | 7.0 |
| 39 | 0–1.8 | 1.8–2 | | 10 | | 0.6 | 99.0 | 3.8 |
| 40 | 0–1.8 | 1.8–2 | | 20 | | 0.7 | 98.7 | 4.5 |
| *median* | | | | | | *4.3* | *93.3* | *18.8* |

**Notes.** NC100 represents a region consisting of 100% of objects containing NC materials. CC fractions 1 and 2 represent the fractions of objects containing CC materials within NC+CC regions 1 and 2, respectively. Our model Earth combined the feeding zones of Earth analogs formed in 37 optimal 4-P systems. Late accretion fraction calculations combined the models above and the results of Paper I's asteroid belt model. Earth's WMF was obtained by assuming that NC-objects were dry (WMF = 0.001%) and water-enriched (WMF = 0.1%) within 1 au and at >1 au, respectively, and that CC-objects were water-rich (WMF = 10%). The contribution of NC-objects to Earth's WMF varied between 1.3 and 1.6 Earth oceans for all models (median 1.5 Earth oceans), where 1 Earth ocean = 0.025% of Earth's mass. Alternatively, if dry and wetter NC-objects existed at <1.5 au and >1.5 au, the resulting NC contribution would be roughly 0.3-0.4 Earth oceans. See Sections 2.2 and 3.2 for more details.

**Table 5.** Water mass fractions (WMFs) of analogs of Mercury, Venus, Earth, and Mars formed in 37 optimal 4-P systems that simultaneously satisfied the bulk water contents of Venus, Earth, and Mars

| Terrestrial planet (analog) | Fiducial hypothesis | | Water worlds hypothesis | |
|---|---|---|---|---|
| | range | median | range | median |
| Mercury | $3.4 \times 10^{-4}$–$3.2 \times 10^{-3}$ (1.4-12.8) | $1.2 \times 10^{-3}$ (4.8) | $5.3 \times 10^{-3}$–$1.4 \times 10^{-2}$ (21.2-56.0) | $9.0 \times 10^{-3}$ (36.0) |
| Venus | $5.1 \times 10^{-4}$–$3.3 \times 10^{-3}$ (2.0-13.2) | $1.2 \times 10^{-3}$ (4.8) | $5.2 \times 10^{-3}$–$1.3 \times 10^{-2}$ (20.8-52.0) | $7.3 \times 10^{-3}$ (29.2) |
| Earth | $5.5 \times 10^{-4}$–$4.4 \times 10^{-3}$ (2.2-17.6) | $1.5 \times 10^{-3}$ (6.0) | $5.1 \times 10^{-3}$–$1.3 \times 10^{-2}$ (20.4-52.0) | $8.1 \times 10^{-3}$ (32.4) |
| Mars | $5.4 \times 10^{-4}$–$5.0 \times 10^{-3}$ (2.2-20.0) | $1.8 \times 10^{-3}$ (7.2) | $8.0 \times 10^{-3}$–$1.5 \times 10^{-2}$ (32.0-60.0) | $1.2 \times 10^{-2}$ (48.0) |

**Notes.** Under a single WMF model, the WMF of a terrestrial planet is the median obtained over the WMFs acquired by 37 analogs of the respective planet. The column 'range' represents the interval of median WMFs based on the results of 142 and 99 successful WMF models for the fiducial and water worlds hypotheses, respectively. The 'median' columns indicate the median of the individual WMFs described in the range columns for each planet. The bulk water contents in Earth's ocean units were obtained via division of the WMFs by $2.5 \times 10^{-4}$ (shown within parentheses). More details are given in the caption of Table 1.

# Appendix

**Table A1.** Models of water mass fractions (WMFs) of objects according to their initial location in the protoplanetary disk and final WMFs acquired by planet analogs in 37 optimal 4-terrestrial planet analog systems

| Model | <1.0 au (%) | 1.0-1.5 au (%) | 1.5-2.0 au (%) | 2.0-2.5 au (%) | >2.5 au (%) | Mercury (%) | Venus (%) | Earth (%) | Mars (%) | Result (Fiducial/dry Venus) | Result (Fiducial/wet Venus) | Result (Water worlds) |
|---|---|---|---|---|---|---|---|---|---|---|---|---|
| 1 | 0.001 | 0.001 | 0.001 | 0.1 | 5 | $1.7 \times 10^{-3}$ | $3.7 \times 10^{-3}$ | $8.1 \times 10^{-3}$ | $2.5 \times 10^{-3}$ | - | - | - |
| 2 | 0.001 | 0.001 | 0.001 | 0.1 | 10 | $1.7 \times 10^{-3}$ | $6.0 \times 10^{-3}$ | $1.4 \times 10^{-2}$ | $2.5 \times 10^{-3}$ | - | - | - |
| 3 | 0.001 | 0.001 | 0.001 | 0.1 | 30 | $1.7 \times 10^{-3}$ | $1.5 \times 10^{-2}$ | $3.8 \times 10^{-2}$ | $2.5 \times 10^{-3}$ | - | - | - |
| 4 | 0.001 | 0.001 | 0.001 | 0.1 | 50 | $1.7 \times 10^{-3}$ | $2.4 \times 10^{-2}$ | $6.3 \times 10^{-2}$ | $2.5 \times 10^{-3}$ | - | - | - |
| 5 | 0.001 | 0.001 | 0.001 | 1 | 10 | $7.7 \times 10^{-3}$ | $1.2 \times 10^{-2}$ | $2.0 \times 10^{-2}$ | $1.7 \times 10^{-2}$ | - | - | - |
| 6 | 0.001 | 0.001 | 0.001 | 1 | 30 | $7.7 \times 10^{-3}$ | $1.9 \times 10^{-2}$ | $4.7 \times 10^{-2}$ | $1.7 \times 10^{-2}$ | - | - | - |
| 7 | 0.001 | 0.001 | 0.001 | 1 | 50 | $7.7 \times 10^{-3}$ | $2.8 \times 10^{-2}$ | $7.2 \times 10^{-2}$ | $1.7 \times 10^{-2}$ | OK | - | - |
| 8 | 0.001 | 0.001 | 0.001 | 5 | 10 | $3.4 \times 10^{-2}$ | $4.0 \times 10^{-2}$ | $4.1 \times 10^{-2}$ | $6.8 \times 10^{-2}$ | - | - | - |
| 9 | 0.001 | 0.001 | 0.001 | 5 | 30 | $3.4 \times 10^{-2}$ | $4.9 \times 10^{-2}$ | $6.8 \times 10^{-2}$ | $7.9 \times 10^{-2}$ | OK | - | - |
| 10 | 0.001 | 0.001 | 0.001 | 5 | 50 | $3.4 \times 10^{-2}$ | $5.4 \times 10^{-2}$ | $9.5 \times 10^{-2}$ | $7.9 \times 10^{-2}$ | - | OK | - |
| 11 | 0.001 | 0.001 | 0.001 | 10 | 10 | $6.8 \times 10^{-2}$ | $7.4 \times 10^{-2}$ | $7.0 \times 10^{-2}$ | $1.3 \times 10^{-1}$ | - | OK | - |
| 12 | 0.001 | 0.001 | 0.001 | 10 | 30 | $6.8 \times 10^{-2}$ | $8.2 \times 10^{-2}$ | $9.3 \times 10^{-2}$ | $1.6 \times 10^{-1}$ | - | OK | - |
| 13 | 0.001 | 0.001 | 0.001 | 10 | 50 | $6.8 \times 10^{-2}$ | $8.8 \times 10^{-2}$ | $1.2 \times 10^{-1}$ | $1.6 \times 10^{-1}$ | - | OK | - |
| 14 | 0.001 | 0.001 | 0.001 | 30 | 30 | $2.0 \times 10^{-1}$ | $2.2 \times 10^{-1}$ | $2.1 \times 10^{-1}$ | $3.8 \times 10^{-1}$ | - | OK | - |
| 15 | 0.001 | 0.001 | 0.001 | 30 | 50 | $2.0 \times 10^{-1}$ | $2.4 \times 10^{-1}$ | $2.2 \times 10^{-1}$ | $4.0 \times 10^{-1}$ | - | OK | - |
| 16 | 0.001 | 0.001 | 0.001 | 50 | 50 | $3.3 \times 10^{-1}$ | $3.7 \times 10^{-1}$ | $3.5 \times 10^{-1}$ | $6.3 \times 10^{-1}$ | - | - | - |
| 17 | 0.001 | 0.001 | 0.01 | 0.1 | 10 | $2.6 \times 10^{-3}$ | $6.3 \times 10^{-3}$ | $1.5 \times 10^{-2}$ | $3.5 \times 10^{-3}$ | - | - | - |
| 18 | 0.001 | 0.001 | 0.01 | 0.1 | 30 | $2.6 \times 10^{-3}$ | $1.5 \times 10^{-2}$ | $3.9 \times 10^{-2}$ | $3.5 \times 10^{-3}$ | - | - | - |
| 19 | 0.001 | 0.001 | 0.01 | 0.1 | 50 | $2.6 \times 10^{-3}$ | $2.5 \times 10^{-2}$ | $6.4 \times 10^{-2}$ | $3.5 \times 10^{-3}$ | - | - | - |
| 20 | 0.001 | 0.001 | 0.01 | 1 | 10 | $8.7 \times 10^{-3}$ | $1.2 \times 10^{-2}$ | $2.0 \times 10^{-2}$ | $1.7 \times 10^{-2}$ | - | - | - |
| 21 | 0.001 | 0.001 | 0.01 | 1 | 30 | $8.7 \times 10^{-3}$ | $1.9 \times 10^{-2}$ | $4.8 \times 10^{-2}$ | $1.7 \times 10^{-2}$ | - | - | - |
| 22 | 0.001 | 0.001 | 0.01 | 1 | 50 | $8.7 \times 10^{-3}$ | $2.8 \times 10^{-2}$ | $7.3 \times 10^{-2}$ | $1.7 \times 10^{-2}$ | OK | - | - |
| 23 | 0.001 | 0.001 | 0.01 | 5 | 10 | $3.5 \times 10^{-2}$ | $4.1 \times 10^{-2}$ | $4.1 \times 10^{-2}$ | $6.9 \times 10^{-2}$ | - | - | - |
| 24 | 0.001 | 0.001 | 0.01 | 5 | 30 | $3.5 \times 10^{-2}$ | $4.9 \times 10^{-2}$ | $6.9 \times 10^{-2}$ | $8.0 \times 10^{-2}$ | OK | - | - |
| 25 | 0.001 | 0.001 | 0.01 | 5 | 50 | $3.5 \times 10^{-2}$ | $5.4 \times 10^{-2}$ | $9.5 \times 10^{-2}$ | $8.0 \times 10^{-2}$ | - | OK | - |
| 26 | 0.001 | 0.001 | 0.01 | 10 | 10 | $6.8 \times 10^{-2}$ | $7.4 \times 10^{-2}$ | $7.1 \times 10^{-2}$ | $1.3 \times 10^{-1}$ | - | OK | - |
| 27 | 0.001 | 0.001 | 0.01 | 10 | 30 | $6.8 \times 10^{-2}$ | $8.3 \times 10^{-2}$ | $9.3 \times 10^{-2}$ | $1.6 \times 10^{-1}$ | - | OK | - |
| 28 | 0.001 | 0.001 | 0.01 | 10 | 50 | $6.8 \times 10^{-2}$ | $8.9 \times 10^{-2}$ | $1.2 \times 10^{-1}$ | $1.6 \times 10^{-1}$ | - | OK | - |
| 29 | 0.001 | 0.001 | 0.01 | 30 | 30 | $2.0 \times 10^{-1}$ | $2.2 \times 10^{-1}$ | $2.1 \times 10^{-1}$ | $3.8 \times 10^{-1}$ | - | OK | - |
| 30 | 0.001 | 0.001 | 0.01 | 30 | 50 | $2.0 \times 10^{-1}$ | $2.4 \times 10^{-1}$ | $2.2 \times 10^{-1}$ | $4.0 \times 10^{-1}$ | - | OK | - |
| 31 | 0.001 | 0.001 | 0.01 | 50 | 50 | $3.3 \times 10^{-1}$ | $3.7 \times 10^{-1}$ | $3.5 \times 10^{-1}$ | $6.3 \times 10^{-1}$ | - | - | - |
| 32 | 0.001 | 0.001 | 0.1 | 0.1 | 10 | $9.9 \times 10^{-3}$ | $1.0 \times 10^{-2}$ | $2.4 \times 10^{-2}$ | $1.4 \times 10^{-2}$ | - | - | - |
| 33 | 0.001 | 0.001 | 0.1 | 0.1 | 30 | $9.9 \times 10^{-3}$ | $1.9 \times 10^{-2}$ | $5.0 \times 10^{-2}$ | $1.4 \times 10^{-2}$ | - | - | - |
| 34 | 0.001 | 0.001 | 0.1 | 0.1 | 50 | $9.9 \times 10^{-3}$ | $2.8 \times 10^{-2}$ | $7.5 \times 10^{-2}$ | $1.4 \times 10^{-2}$ | OK | - | - |
| 35 | 0.001 | 0.001 | 0.1 | 1 | 10 | $1.8 \times 10^{-2}$ | $1.8 \times 10^{-2}$ | $2.7 \times 10^{-2}$ | $2.7 \times 10^{-2}$ | - | - | - |
| 36 | 0.001 | 0.001 | 0.1 | 1 | 30 | $1.8 \times 10^{-2}$ | $2.3 \times 10^{-2}$ | $5.7 \times 10^{-2}$ | $2.7 \times 10^{-2}$ | OK | - | - |
| 37 | 0.001 | 0.001 | 0.1 | 1 | 50 | $1.8 \times 10^{-2}$ | $3.2 \times 10^{-2}$ | $8.2 \times 10^{-2}$ | $2.7 \times 10^{-2}$ | OK | - | - |
| 38 | 0.001 | 0.001 | 0.1 | 5 | 10 | $4.4 \times 10^{-2}$ | $4.7 \times 10^{-2}$ | $4.7 \times 10^{-2}$ | $7.9 \times 10^{-2}$ | - | - | - |
| 39 | 0.001 | 0.001 | 0.1 | 5 | 30 | $4.4 \times 10^{-2}$ | $5.4 \times 10^{-2}$ | $7.6 \times 10^{-2}$ | $8.7 \times 10^{-2}$ | - | OK | - |
| 40 | 0.001 | 0.001 | 0.1 | 5 | 50 | $4.4 \times 10^{-2}$ | $6.1 \times 10^{-2}$ | $1.0 \times 10^{-1}$ | $8.7 \times 10^{-2}$ | - | OK | - |
| 41 | 0.001 | 0.001 | 0.1 | 10 | 10 | $7.9 \times 10^{-2}$ | $8.0 \times 10^{-2}$ | $7.6 \times 10^{-2}$ | $1.4 \times 10^{-1}$ | - | OK | - |
| 42 | 0.001 | 0.001 | 0.1 | 10 | 30 | $7.9 \times 10^{-2}$ | $8.6 \times 10^{-2}$ | $9.9 \times 10^{-2}$ | $1.6 \times 10^{-1}$ | - | OK | - |
| 43 | 0.001 | 0.001 | 0.1 | 10 | 50 | $7.9 \times 10^{-2}$ | $9.5 \times 10^{-2}$ | $1.3 \times 10^{-1}$ | $1.6 \times 10^{-1}$ | - | OK | - |
| 44 | 0.001 | 0.001 | 0.1 | 30 | 30 | $2.1 \times 10^{-1}$ | $2.2 \times 10^{-1}$ | $2.1 \times 10^{-1}$ | $3.9 \times 10^{-1}$ | - | OK | - |
| 45 | 0.001 | 0.001 | 0.1 | 30 | 50 | $2.1 \times 10^{-1}$ | $2.4 \times 10^{-1}$ | $2.2 \times 10^{-1}$ | $4.1 \times 10^{-1}$ | - | OK | - |
| 46 | 0.001 | 0.001 | 0.1 | 50 | 50 | $3.4 \times 10^{-1}$ | $3.7 \times 10^{-1}$ | $3.5 \times 10^{-1}$ | $6.4 \times 10^{-1}$ | - | - | - |
| 47 | 0.001 | 0.001 | 1 | 1 | 10 | $9.1 \times 10^{-2}$ | $7.5 \times 10^{-2}$ | $9.2 \times 10^{-2}$ | $1.3 \times 10^{-1}$ | - | OK | - |
| 48 | 0.001 | 0.001 | 1 | 1 | 30 | $9.1 \times 10^{-2}$ | $8.7 \times 10^{-2}$ | $1.2 \times 10^{-1}$ | $1.4 \times 10^{-1}$ | - | OK | - |
| 49 | 0.001 | 0.001 | 1 | 1 | 50 | $9.1 \times 10^{-2}$ | $9.1 \times 10^{-2}$ | $1.5 \times 10^{-1}$ | $1.4 \times 10^{-1}$ | - | OK | - |
| 50 | 0.001 | 0.001 | 1 | 5 | 10 | $1.3 \times 10^{-1}$ | $1.1 \times 10^{-1}$ | $1.1 \times 10^{-1}$ | $1.7 \times 10^{-1}$ | - | OK | - |
| 51 | 0.001 | 0.001 | 1 | 5 | 30 | $1.3 \times 10^{-1}$ | $1.1 \times 10^{-1}$ | $1.4 \times 10^{-1}$ | $1.8 \times 10^{-1}$ | - | OK | - |
| 52 | 0.001 | 0.001 | 1 | 5 | 50 | $1.3 \times 10^{-1}$ | $1.2 \times 10^{-1}$ | $1.7 \times 10^{-1}$ | $1.8 \times 10^{-1}$ | - | OK | - |
| 53 | 0.001 | 0.001 | 1 | 10 | 10 | $1.7 \times 10^{-1}$ | $1.4 \times 10^{-1}$ | $1.4 \times 10^{-1}$ | $2.5 \times 10^{-1}$ | - | OK | - |
| 54 | 0.001 | 0.001 | 1 | 10 | 30 | $1.7 \times 10^{-1}$ | $1.5 \times 10^{-1}$ | $1.7 \times 10^{-1}$ | $2.6 \times 10^{-1}$ | - | OK | - |
| 55 | 0.001 | 0.001 | 1 | 10 | 50 | $1.7 \times 10^{-1}$ | $1.6 \times 10^{-1}$ | $1.9 \times 10^{-1}$ | $2.6 \times 10^{-1}$ | - | OK | - |
| 56 | 0.001 | 0.001 | 1 | 30 | 30 | $2.9 \times 10^{-1}$ | $2.8 \times 10^{-1}$ | $2.8 \times 10^{-1}$ | $5.0 \times 10^{-1}$ | - | OK | - |
| 57 | 0.001 | 0.001 | 1 | 30 | 50 | $2.9 \times 10^{-1}$ | $3.0 \times 10^{-1}$ | $2.9 \times 10^{-1}$ | $5.1 \times 10^{-1}$ | - | - | - |
| 58 | 0.001 | 0.001 | 1 | 50 | 50 | $4.3 \times 10^{-1}$ | $4.3 \times 10^{-1}$ | $4.2 \times 10^{-1}$ | $7.6 \times 10^{-1}$ | - | - | - |
| 59 | 0.001 | 0.001 | 5 | 5 | 10 | $4.5 \times 10^{-1}$ | $3.6 \times 10^{-1}$ | $4.0 \times 10^{-1}$ | $5.9 \times 10^{-1}$ | - | - | - |
| 60 | 0.001 | 0.001 | 5 | 5 | 30 | $4.5 \times 10^{-1}$ | $3.7 \times 10^{-1}$ | $4.2 \times 10^{-1}$ | $6.3 \times 10^{-1}$ | - | - | - |
| 61 | 0.001 | 0.001 | 5 | 5 | 50 | $4.5 \times 10^{-1}$ | $3.7 \times 10^{-1}$ | $4.6 \times 10^{-1}$ | $6.3 \times 10^{-1}$ | - | - | - |
| 62 | 0.001 | 0.001 | 5 | 10 | 10 | $4.8 \times 10^{-1}$ | $3.9 \times 10^{-1}$ | $4.1 \times 10^{-1}$ | $6.4 \times 10^{-1}$ | - | - | - |
| 63 | 0.001 | 0.001 | 5 | 10 | 30 | $4.8 \times 10^{-1}$ | $4.0 \times 10^{-1}$ | $4.4 \times 10^{-1}$ | $6.9 \times 10^{-1}$ | - | - | - |
| 64 | 0.001 | 0.001 | 5 | 10 | 50 | $4.8 \times 10^{-1}$ | $4.0 \times 10^{-1}$ | $4.8 \times 10^{-1}$ | $6.9 \times 10^{-1}$ | - | - | - |
| 65 | 0.001 | 0.001 | 5 | 30 | 30 | $6.7 \times 10^{-1}$ | $5.7 \times 10^{-1}$ | $5.6 \times 10^{-1}$ | $9.2 \times 10^{-1}$ | - | - | OK |
| 66 | 0.001 | 0.001 | 5 | 30 | 50 | $6.7 \times 10^{-1}$ | $5.7 \times 10^{-1}$ | $5.9 \times 10^{-1}$ | $9.5 \times 10^{-1}$ | - | - | OK |
| 67 | 0.001 | 0.001 | 5 | 50 | 50 | $8.3 \times 10^{-1}$ | $7.2 \times 10^{-1}$ | $7.0 \times 10^{-1}$ | 1.2 | - | - | OK |
| 68 | 0.001 | 0.001 | 10 | 10 | 10 | $9.0 \times 10^{-1}$ | $7.1 \times 10^{-1}$ | $7.8 \times 10^{-1}$ | 1.2 | - | - | OK |
| 69 | 0.001 | 0.001 | 10 | 10 | 30 | $9.0 \times 10^{-1}$ | $7.1 \times 10^{-1}$ | $8.0 \times 10^{-1}$ | 1.2 | - | - | OK |
| 70 | 0.001 | 0.001 | 10 | 10 | 50 | $9.0 \times 10^{-1}$ | $7.3 \times 10^{-1}$ | $8.3 \times 10^{-1}$ | 1.2 | - | - | OK |
| 71 | 0.001 | 0.001 | 10 | 30 | 30 | 1.1 | $8.6 \times 10^{-1}$ | $8.8 \times 10^{-1}$ | 1.4 | - | - | OK |
| 72 | 0.001 | 0.001 | 10 | 30 | 50 | 1.1 | $8.6 \times 10^{-1}$ | $9.2 \times 10^{-1}$ | 1.4 | - | - | OK |
| 73 | 0.001 | 0.001 | 10 | 50 | 50 | 1.3 | 1.0 | 1.0 | 1.7 | - | - | - |
| 74 | 0.001 | 0.001 | 30 | 30 | 30 | 2.7 | 2.1 | 2.3 | 3.4 | - | - | - |
| 75 | 0.001 | 0.001 | 30 | 30 | 50 | 2.7 | 2.1 | 2.4 | 3.5 | - | - | - |
| 76 | 0.001 | 0.001 | 30 | 50 | 50 | 2.8 | 2.3 | 2.4 | 3.7 | - | - | - |
| 77 | 0.001 | 0.001 | 50 | 50 | 50 | 4.5 | 3.6 | 3.9 | 5.7 | - | - | - |
| 78 | 0.001 | 0.01 | 0.01 | 0.1 | 10 | $4.8 \times 10^{-3}$ | $8.3 \times 10^{-3}$ | $1.8 \times 10^{-2}$ | $6.7 \times 10^{-3}$ | - | - | - |
| 79 | 0.001 | 0.01 | 0.01 | 0.1 | 30 | $4.8 \times 10^{-3}$ | $1.7 \times 10^{-2}$ | $4.2 \times 10^{-2}$ | $6.7 \times 10^{-3}$ | - | - | - |
| 80 | 0.001 | 0.01 | 0.01 | 0.1 | 50 | $4.8 \times 10^{-3}$ | $2.7 \times 10^{-2}$ | $6.7 \times 10^{-2}$ | $6.7 \times 10^{-3}$ | - | - | - |
| 81 | 0.001 | 0.01 | 0.01 | 1 | 10 | $1.1 \times 10^{-2}$ | $1.4 \times 10^{-2}$ | $2.3 \times 10^{-2}$ | $2.0 \times 10^{-2}$ | - | - | - |
| 82 | 0.001 | 0.01 | 0.01 | 1 | 30 | $1.1 \times 10^{-2}$ | $2.1 \times 10^{-2}$ | $5.1 \times 10^{-2}$ | $2.0 \times 10^{-2}$ | OK | - | - |
| 83 | 0.001 | 0.01 | 0.01 | 1 | 50 | $1.1 \times 10^{-2}$ | $3.0 \times 10^{-2}$ | $7.7 \times 10^{-2}$ | $2.0 \times 10^{-2}$ | OK | - | - |
| 84 | 0.001 | 0.01 | 0.01 | 5 | 10 | $3.9 \times 10^{-2}$ | $4.4 \times 10^{-2}$ | $4.4 \times 10^{-2}$ | $7.1 \times 10^{-2}$ | - | - | - |
| 85 | 0.001 | 0.01 | 0.01 | 5 | 30 | $3.9 \times 10^{-2}$ | $5.1 \times 10^{-2}$ | $7.1 \times 10^{-2}$ | $8.3 \times 10^{-2}$ | - | OK | - |
| 86 | 0.001 | 0.01 | 0.01 | 5 | 50 | $3.9 \times 10^{-2}$ | $5.7 \times 10^{-2}$ | $9.8 \times 10^{-2}$ | $8.3 \times 10^{-2}$ | - | OK | - |
| 87 | 0.001 | 0.01 | 0.01 | 10 | 10 | $7.3 \times 10^{-2}$ | $7.7 \times 10^{-2}$ | $7.4 \times 10^{-2}$ | $1.3 \times 10^{-1}$ | - | OK | - |
| 88 | 0.001 | 0.01 | 0.01 | 10 | 30 | $7.3 \times 10^{-2}$ | $8.5 \times 10^{-2}$ | $9.5 \times 10^{-2}$ | $1.6 \times 10^{-1}$ | - | OK | - |
| 89 | 0.001 | 0.01 | 0.01 | 10 | 50 | $7.3 \times 10^{-2}$ | $9.1 \times 10^{-2}$ | $1.3 \times 10^{-1}$ | $1.6 \times 10^{-1}$ | - | OK | - |
| 90 | 0.001 | 0.01 | 0.01 | 30 | 30 | $2.1 \times 10^{-1}$ | $2.2 \times 10^{-1}$ | $2.1 \times 10^{-1}$ | $3.8 \times 10^{-1}$ | - | OK | - |
| 91 | 0.001 | 0.01 | 0.01 | 30 | 50 | $2.1 \times 10^{-1}$ | $2.4 \times 10^{-1}$ | $2.2 \times 10^{-1}$ | $4.1 \times 10^{-1}$ | - | OK | - |
| 92 | 0.001 | 0.01 | 0.01 | 50 | 50 | $3.4 \times 10^{-1}$ | $3.7 \times 10^{-1}$ | $3.5 \times 10^{-1}$ | $6.3 \times 10^{-1}$ | - | - | - |
| 93 | 0.001 | 0.01 | 0.1 | 0.1 | 10 | $1.2 \times 10^{-2}$ | $1.2 \times 10^{-2}$ | $2.7 \times 10^{-2}$ | $1.8 \times 10^{-2}$ | - | - | - |
| 94 | 0.001 | 0.01 | 0.1 | 0.1 | 30 | $1.2 \times 10^{-2}$ | $2.1 \times 10^{-2}$ | $5.2 \times 10^{-2}$ | $1.8 \times 10^{-2}$ | OK | - | - |
| 95 | 0.001 | 0.01 | 0.1 | 0.1 | 50 | $1.2 \times 10^{-2}$ | $3.0 \times 10^{-2}$ | $7.8 \times 10^{-2}$ | $1.8 \times 10^{-2}$ | OK | - | - |
| 96 | 0.001 | 0.01 | 0.1 | 1 | 10 | $2.0 \times 10^{-2}$ | $2.0 \times 10^{-2}$ | $3.0 \times 10^{-2}$ | $3.0 \times 10^{-2}$ | - | - | - |
| 97 | 0.001 | 0.01 | 0.1 | 1 | 30 | $2.0 \times 10^{-2}$ | $2.5 \times 10^{-2}$ | $6.0 \times 10^{-2}$ | $3.0 \times 10^{-2}$ | OK | - | - |
| 98  | 0.001 | 0.01 | 0.1 | 1  | 50 | $2.0 \times 10^{-2}$ | $3.4 \times 10^{-2}$ | $8.6 \times 10^{-2}$ | $3.0 \times 10^{-2}$ | OK | -  | - |
| 99  | 0.001 | 0.01 | 0.1 | 5  | 10 | $4.6 \times 10^{-2}$ | $4.9 \times 10^{-2}$ | $4.9 \times 10^{-2}$ | $8.1 \times 10^{-2}$ | -  | -  | - |
| 100 | 0.001 | 0.01 | 0.1 | 5  | 30 | $4.6 \times 10^{-2}$ | $5.7 \times 10^{-2}$ | $7.9 \times 10^{-2}$ | $9.0 \times 10^{-2}$ | -  | OK | - |
| 101 | 0.001 | 0.01 | 0.1 | 5  | 50 | $4.6 \times 10^{-2}$ | $6.3 \times 10^{-2}$ | $1.0 \times 10^{-1}$ | $9.0 \times 10^{-2}$ | -  | OK | - |
| 102 | 0.001 | 0.01 | 0.1 | 10 | 10 | $8.3 \times 10^{-2}$ | $8.3 \times 10^{-2}$ | $7.9 \times 10^{-2}$ | $1.4 \times 10^{-1}$ | -  | OK | - |
| 103 | 0.001 | 0.01 | 0.1 | 10 | 30 | $8.3 \times 10^{-2}$ | $8.8 \times 10^{-2}$ | $1.0 \times 10^{-1}$ | $1.7 \times 10^{-1}$ | -  | OK | - |
| 104 | 0.001 | 0.01 | 0.1 | 10 | 50 | $8.3 \times 10^{-2}$ | $9.7 \times 10^{-2}$ | $1.3 \times 10^{-1}$ | $1.7 \times 10^{-1}$ | -  | OK | - |
| 105 | 0.001 | 0.01 | 0.1 | 30 | 30 | $2.2 \times 10^{-1}$ | $2.3 \times 10^{-1}$ | $2.2 \times 10^{-1}$ | $3.9 \times 10^{-1}$ | -  | OK | - |
| 106 | 0.001 | 0.01 | 0.1 | 30 | 50 | $2.2 \times 10^{-1}$ | $2.5 \times 10^{-1}$ | $2.3 \times 10^{-1}$ | $4.2 \times 10^{-1}$ | -  | OK | - |
| 107 | 0.001 | 0.01 | 0.1 | 50 | 50 | $3.5 \times 10^{-1}$ | $3.7 \times 10^{-1}$ | $3.6 \times 10^{-1}$ | $6.5 \times 10^{-1}$ | -  | -  | - |
| 108 | 0.001 | 0.01 | 1   | 1  | 10 | $9.4 \times 10^{-2}$ | $7.7 \times 10^{-2}$ | $9.5 \times 10^{-2}$ | $1.3 \times 10^{-1}$ | -  | OK | - |
| 109 | 0.001 | 0.01 | 1   | 1  | 30 | $9.4 \times 10^{-2}$ | $8.9 \times 10^{-2}$ | $1.2 \times 10^{-1}$ | $1.4 \times 10^{-1}$ | -  | OK | - |
| 110 | 0.001 | 0.01 | 1   | 1  | 50 | $9.4 \times 10^{-2}$ | $9.3 \times 10^{-2}$ | $1.5 \times 10^{-1}$ | $1.4 \times 10^{-1}$ | -  | OK | - |
| 111 | 0.001 | 0.01 | 1   | 5  | 10 | $1.3 \times 10^{-1}$ | $1.1 \times 10^{-1}$ | $1.2 \times 10^{-1}$ | $1.8 \times 10^{-1}$ | -  | OK | - |
| 112 | 0.001 | 0.01 | 1   | 5  | 30 | $1.3 \times 10^{-1}$ | $1.1 \times 10^{-1}$ | $1.4 \times 10^{-1}$ | $1.8 \times 10^{-1}$ | -  | OK | - |
| 113 | 0.001 | 0.01 | 1   | 5  | 50 | $1.3 \times 10^{-1}$ | $1.2 \times 10^{-1}$ | $1.7 \times 10^{-1}$ | $1.8 \times 10^{-1}$ | -  | OK | - |
| 114 | 0.001 | 0.01 | 1   | 10 | 10 | $1.7 \times 10^{-1}$ | $1.5 \times 10^{-1}$ | $1.4 \times 10^{-1}$ | $2.5 \times 10^{-1}$ | -  | OK | - |
| 115 | 0.001 | 0.01 | 1   | 10 | 30 | $1.7 \times 10^{-1}$ | $1.5 \times 10^{-1}$ | $1.7 \times 10^{-1}$ | $2.7 \times 10^{-1}$ | -  | OK | - |
| 116 | 0.001 | 0.01 | 1   | 10 | 50 | $1.7 \times 10^{-1}$ | $1.6 \times 10^{-1}$ | $2.0 \times 10^{-1}$ | $2.7 \times 10^{-1}$ | -  | OK | - |
| 117 | 0.001 | 0.01 | 1   | 30 | 30 | $2.9 \times 10^{-1}$ | $2.8 \times 10^{-1}$ | $2.8 \times 10^{-1}$ | $5.0 \times 10^{-1}$ | -  | -  | - |
| 118 | 0.001 | 0.01 | 1   | 30 | 50 | $2.9 \times 10^{-1}$ | $3.0 \times 10^{-1}$ | $2.9 \times 10^{-1}$ | $5.2 \times 10^{-1}$ | -  | -  | - |
| 119 | 0.001 | 0.01 | 1   | 50 | 50 | $4.3 \times 10^{-1}$ | $4.3 \times 10^{-1}$ | $4.2 \times 10^{-1}$ | $7.6 \times 10^{-1}$ | -  | -  | - |
| 120 | 0.001 | 0.01 | 5   | 5  | 10 | $4.5 \times 10^{-1}$ | $3.6 \times 10^{-1}$ | $4.0 \times 10^{-1}$ | $5.9 \times 10^{-1}$ | -  | -  | - |
| 121 | 0.001 | 0.01 | 5   | 5  | 30 | $4.5 \times 10^{-1}$ | $3.7 \times 10^{-1}$ | $4.2 \times 10^{-1}$ | $6.3 \times 10^{-1}$ | -  | -  | - |
| 122 | 0.001 | 0.01 | 5   | 5  | 50 | $4.5 \times 10^{-1}$ | $3.7 \times 10^{-1}$ | $4.6 \times 10^{-1}$ | $6.4 \times 10^{-1}$ | -  | -  | - |
| 123 | 0.001 | 0.01 | 5   | 10 | 10 | $4.8 \times 10^{-1}$ | $3.9 \times 10^{-1}$ | $4.2 \times 10^{-1}$ | $6.5 \times 10^{-1}$ | -  | -  | - |
| 124 | 0.001 | 0.01 | 5 | 10 | 30 | $4.8 \times 10^{-1}$ | $4.0 \times 10^{-1}$ | $4.5 \times 10^{-1}$ | $6.9 \times 10^{-1}$ | - | - | - |
| 125 | 0.001 | 0.01 | 5 | 10 | 50 | $4.8 \times 10^{-1}$ | $4.0 \times 10^{-1}$ | $4.9 \times 10^{-1}$ | $7.0 \times 10^{-1}$ | - | - | - |
| 126 | 0.001 | 0.01 | 5 | 30 | 30 | $6.8 \times 10^{-1}$ | $5.7 \times 10^{-1}$ | $5.6 \times 10^{-1}$ | $9.2 \times 10^{-1}$ | - | - | OK |
| 127 | 0.001 | 0.01 | 5 | 30 | 50 | $6.8 \times 10^{-1}$ | $5.7 \times 10^{-1}$ | $5.9 \times 10^{-1}$ | $9.5 \times 10^{-1}$ | - | - | OK |
| 128 | 0.001 | 0.01 | 5 | 50 | 50 | $8.3 \times 10^{-1}$ | $7.2 \times 10^{-1}$ | $7.0 \times 10^{-1}$ | 1.2 | - | - | OK |
| 129 | 0.001 | 0.01 | 10 | 10 | 10 | $9.0 \times 10^{-1}$ | $7.2 \times 10^{-1}$ | $7.8 \times 10^{-1}$ | 1.2 | - | - | OK |
| 130 | 0.001 | 0.01 | 10 | 10 | 30 | $9.0 \times 10^{-1}$ | $7.2 \times 10^{-1}$ | $8.0 \times 10^{-1}$ | 1.2 | - | - | OK |
| 131 | 0.001 | 0.01 | 10 | 10 | 50 | $9.0 \times 10^{-1}$ | $7.3 \times 10^{-1}$ | $8.3 \times 10^{-1}$ | 1.2 | - | - | OK |
| 132 | 0.001 | 0.01 | 10 | 30 | 30 | 1.1 | $8.6 \times 10^{-1}$ | $8.9 \times 10^{-1}$ | 1.4 | - | - | OK |
| 133 | 0.001 | 0.01 | 10 | 30 | 50 | 1.1 | $8.6 \times 10^{-1}$ | $9.3 \times 10^{-1}$ | 1.4 | - | - | OK |
| 134 | 0.001 | 0.01 | 10 | 50 | 50 | 1.3 | 1.0 | 1.0 | 1.7 | - | - | - |
| 135 | 0.001 | 0.01 | 30 | 30 | 30 | 2.7 | 2.1 | 2.3 | 3.5 | - | - | - |
| 136 | 0.001 | 0.01 | 30 | 30 | 50 | 2.7 | 2.1 | 2.4 | 3.5 | - | - | - |
| 137 | 0.001 | 0.01 | 30 | 50 | 50 | 2.8 | 2.3 | 2.4 | 3.7 | - | - | - |
| 138 | 0.001 | 0.01 | 50 | 50 | 50 | 4.5 | 3.6 | 3.9 | 5.8 | - | - | - |
| 139 | 0.001 | 0.1 | 0.1 | 0.1 | 10 | $3.2 \times 10^{-2}$ | $3.5 \times 10^{-2}$ | $5.3 \times 10^{-2}$ | $5.0 \times 10^{-2}$ | OK | - | - |
| 140 | 0.001 | 0.1 | 0.1 | 0.1 | 30 | $3.2 \times 10^{-2}$ | $4.1 \times 10^{-2}$ | $8.2 \times 10^{-2}$ | $5.0 \times 10^{-2}$ | OK | - | - |
| 141 | 0.001 | 0.1 | 0.1 | 0.1 | 50 | $3.2 \times 10^{-2}$ | $5.0 \times 10^{-2}$ | $1.1 \times 10^{-1}$ | $5.0 \times 10^{-2}$ | OK | - | - |
| 142 | 0.001 | 0.1 | 0.1 | 1 | 10 | $4.1 \times 10^{-2}$ | $4.1 \times 10^{-2}$ | $5.7 \times 10^{-2}$ | $6.2 \times 10^{-2}$ | OK | - | - |
| 143 | 0.001 | 0.1 | 0.1 | 1 | 30 | $4.1 \times 10^{-2}$ | $4.5 \times 10^{-2}$ | $8.7 \times 10^{-2}$ | $6.2 \times 10^{-2}$ | OK | - | - |
| 144 | 0.001 | 0.1 | 0.1 | 1 | 50 | $4.1 \times 10^{-2}$ | $5.4 \times 10^{-2}$ | $1.1 \times 10^{-1}$ | $6.2 \times 10^{-2}$ | - | OK | - |
| 145 | 0.001 | 0.1 | 0.1 | 5 | 10 | $6.9 \times 10^{-2}$ | $6.9 \times 10^{-2}$ | $7.5 \times 10^{-2}$ | $1.2 \times 10^{-1}$ | - | OK | - |
| 146 | 0.001 | 0.1 | 0.1 | 5 | 30 | $6.9 \times 10^{-2}$ | $7.7 \times 10^{-2}$ | $1.0 \times 10^{-1}$ | $1.2 \times 10^{-1}$ | - | OK | - |
| 147 | 0.001 | 0.1 | 0.1 | 5 | 50 | $6.9 \times 10^{-2}$ | $8.5 \times 10^{-2}$ | $1.3 \times 10^{-1}$ | $1.2 \times 10^{-1}$ | - | OK | - |
| 148 | 0.001 | 0.1 | 0.1 | 10 | 10 | $1.0 \times 10^{-1}$ | $1.0 \times 10^{-1}$ | $1.0 \times 10^{-1}$ | $1.7 \times 10^{-1}$ | - | OK | - |
| 149 | 0.001 | 0.1 | 0.1 | 10 | 30 | $1.0 \times 10^{-1}$ | $1.1 \times 10^{-1}$ | $1.3 \times 10^{-1}$ | $2.0 \times 10^{-1}$ | - | OK | - |
| 150 | 0.001 | 0.1 | 0.1 | 10 | 50 | $1.0 \times 10^{-1}$ | $1.2 \times 10^{-1}$ | $1.6 \times 10^{-1}$ | $2.0 \times 10^{-1}$ | - | OK | - |
| 151 | 0.001 | 0.1 | 0.1 | 30 | 30 | $2.5 \times 10^{-1}$ | $2.5 \times 10^{-1}$ | $2.4 \times 10^{-1}$ | $4.3 \times 10^{-1}$ | - | OK | - |
| 152 | 0.001 | 0.1 | 0.1 | 30 | 50 | $2.5 \times 10^{-1}$ | $2.7 \times 10^{-1}$ | $2.5 \times 10^{-1}$ | $4.4 \times 10^{-1}$ | - | OK | - |
| 153 | 0.001 | 0.1 | 0.1 | 50 | 50 | $3.9 \times 10^{-1}$ | $4.0 \times 10^{-1}$ | $3.8 \times 10^{-1}$ | $6.9 \times 10^{-1}$ | - | - | - |
| 154 | 0.001 | 0.1 | 1 | 1 | 10 | $1.1 \times 10^{-1}$ | $9.7 \times 10^{-2}$ | $1.2 \times 10^{-1}$ | $1.7 \times 10^{-1}$ | - | OK | - |
| 155 | 0.001 | 0.1 | 1 | 1 | 30 | $1.1 \times 10^{-1}$ | $1.1 \times 10^{-1}$ | $1.5 \times 10^{-1}$ | $1.7 \times 10^{-1}$ | - | OK | - |
| 156 | 0.001 | 0.1 | 1 | 1 | 50 | $1.1 \times 10^{-1}$ | $1.1 \times 10^{-1}$ | $1.8 \times 10^{-1}$ | $1.7 \times 10^{-1}$ | - | OK | - |
| 157 | 0.001 | 0.1 | 1 | 5 | 10 | $1.5 \times 10^{-1}$ | $1.3 \times 10^{-1}$ | $1.4 \times 10^{-1}$ | $2.0 \times 10^{-1}$ | - | OK | - |
| 158 | 0.001 | 0.1 | 1 | 5 | 30 | $1.5 \times 10^{-1}$ | $1.3 \times 10^{-1}$ | $1.7 \times 10^{-1}$ | $2.1 \times 10^{-1}$ | - | OK | - |
| 159 | 0.001 | 0.1 | 1 | 5 | 50 | $1.5 \times 10^{-1}$ | $1.5 \times 10^{-1}$ | $2.0 \times 10^{-1}$ | $2.1 \times 10^{-1}$ | - | OK | - |
| 160 | 0.001 | 0.1 | 1 | 10 | 10 | $2.0 \times 10^{-1}$ | $1.7 \times 10^{-1}$ | $1.7 \times 10^{-1}$ | $2.7 \times 10^{-1}$ | - | OK | - |
| 161 | 0.001 | 0.1 | 1 | 10 | 30 | $2.0 \times 10^{-1}$ | $1.7 \times 10^{-1}$ | $2.0 \times 10^{-1}$ | $2.9 \times 10^{-1}$ | - | OK | - |
| 162 | 0.001 | 0.1 | 1 | 10 | 50 | $2.0 \times 10^{-1}$ | $1.8 \times 10^{-1}$ | $2.2 \times 10^{-1}$ | $2.9 \times 10^{-1}$ | - | OK | - |
| 163 | 0.001 | 0.1 | 1 | 30 | 30 | $3.1 \times 10^{-1}$ | $3.0 \times 10^{-1}$ | $3.1 \times 10^{-1}$ | $5.3 \times 10^{-1}$ | - | - | - |
| 164 | 0.001 | 0.1 | 1 | 30 | 50 | $3.1 \times 10^{-1}$ | $3.2 \times 10^{-1}$ | $3.2 \times 10^{-1}$ | $5.4 \times 10^{-1}$ | - | - | - |
| 165 | 0.001 | 0.1 | 1 | 50 | 50 | $4.5 \times 10^{-1}$ | $4.5 \times 10^{-1}$ | $4.5 \times 10^{-1}$ | $8.0 \times 10^{-1}$ | - | - | - |
| 166 | 0.001 | 0.1 | 5 | 5 | 10 | $4.8 \times 10^{-1}$ | $3.8 \times 10^{-1}$ | $4.3 \times 10^{-1}$ | $6.3 \times 10^{-1}$ | - | - | - |
| 167 | 0.001 | 0.1 | 5 | 5 | 30 | $4.8 \times 10^{-1}$ | $3.8 \times 10^{-1}$ | $4.5 \times 10^{-1}$ | $6.5 \times 10^{-1}$ | - | - | - |
| 168 | 0.001 | 0.1 | 5 | 5 | 50 | $4.8 \times 10^{-1}$ | $3.9 \times 10^{-1}$ | $4.9 \times 10^{-1}$ | $6.7 \times 10^{-1}$ | - | - | - |
| 169 | 0.001 | 0.1 | 5 | 10 | 10 | $5.0 \times 10^{-1}$ | $4.1 \times 10^{-1}$ | $4.4 \times 10^{-1}$ | $6.8 \times 10^{-1}$ | - | - | - |
| 170 | 0.001 | 0.1 | 5 | 10 | 30 | $5.0 \times 10^{-1}$ | $4.2 \times 10^{-1}$ | $4.7 \times 10^{-1}$ | $7.1 \times 10^{-1}$ | - | - | - |
| 171 | 0.001 | 0.1 | 5 | 10 | 50 | $5.0 \times 10^{-1}$ | $4.2 \times 10^{-1}$ | $5.1 \times 10^{-1}$ | $7.3 \times 10^{-1}$ | - | - | - |
| 172 | 0.001 | 0.1 | 5 | 30 | 30 | $7.0 \times 10^{-1}$ | $5.9 \times 10^{-1}$ | $5.9 \times 10^{-1}$ | $9.6 \times 10^{-1}$ | - | - | OK |
| 173 | 0.001 | 0.1 | 5 | 30 | 50 | $7.0 \times 10^{-1}$ | $5.9 \times 10^{-1}$ | $6.2 \times 10^{-1}$ | $9.8 \times 10^{-1}$ | - | - | OK |
| 174 | 0.001 | 0.1 | 5 | 50 | 50 | $8.7 \times 10^{-1}$ | $7.4 \times 10^{-1}$ | $7.3 \times 10^{-1}$ | 1.3 | - | - | OK |
| 175 | 0.001 | 0.1 | 10 | 10 | 10 | $9.3 \times 10^{-1}$ | $7.4 \times 10^{-1}$ | $8.1 \times 10^{-1}$ | 1.2 | - | - | OK |
| 176 | 0.001 | 0.1 | 10 | 10 | 30 | $9.3 \times 10^{-1}$ | $7.4 \times 10^{-1}$ | $8.3 \times 10^{-1}$ | 1.2 | - | - | OK |
| 177 | 0.001 | 0.1 | 10 | 10 | 50 | $9.3 \times 10^{-1}$ | $7.5 \times 10^{-1}$ | $8.5 \times 10^{-1}$ | 1.3 | - | - | OK |
| 178 | 0.001 | 0.1 | 10 | 30 | 30 | 1.1 | $8.9 \times 10^{-1}$ | $9.1 \times 10^{-1}$ | 1.4 | - | - | OK |
| 179 | 0.001 | 0.1 | 10 | 30 | 50 | 1.1 | $8.9 \times 10^{-1}$ | $9.5 \times 10^{-1}$ | 1.5 | - | - | OK |
| 180 | 0.001 | 0.1 | 10 | 50 | 50 | 1.3 | 1.1 | 1.1 | 1.7 | - | - | - |
| 181 | 0.001 | 0.1 | 30 | 30 | 30 | 2.7 | 2.2 | 2.4 | 3.5 | - | - | - |
| 182 | 0.001 | 0.1 | 30 | 30 | 50 | 2.7 | 2.2 | 2.4 | 3.6 | - | - | - |
| 183 | 0.001 | 0.1 | 30 | 50 | 50 | 2.8 | 2.3 | 2.5 | 3.7 | - | - | - |
| 184 | 0.001 | 0.1 | 50 | 50 | 50 | 4.5 | 3.6 | 3.9 | 5.8 | - | - | - |
| 185 | 0.001 | 1 | 1 | 1 | 10 | $3.0 \times 10^{-1}$ | $3.0 \times 10^{-1}$ | $3.9 \times 10^{-1}$ | $4.7 \times 10^{-1}$ | - | OK | - |
| 186 | 0.001 | 1 | 1 | 1 | 30 | $3.1 \times 10^{-1}$ | $3.1 \times 10^{-1}$ | $4.2 \times 10^{-1}$ | $4.8 \times 10^{-1}$ | - | OK | - |
| 187 | 0.001 | 1 | 1 | 1 | 50 | $3.1 \times 10^{-1}$ | $3.2 \times 10^{-1}$ | $4.4 \times 10^{-1}$ | $4.9 \times 10^{-1}$ | - | OK | - |
| 188 | 0.001 | 1 | 1 | 5 | 10 | $3.5 \times 10^{-1}$ | $3.2 \times 10^{-1}$ | $4.1 \times 10^{-1}$ | $5.0 \times 10^{-1}$ | - | - | - |
| 189 | 0.001 | 1 | 1 | 5 | 30 | $3.6 \times 10^{-1}$ | $3.4 \times 10^{-1}$ | $4.4 \times 10^{-1}$ | $5.3 \times 10^{-1}$ | - | - | - |
| 190 | 0.001 | 1 | 1 | 5 | 50 | $3.6 \times 10^{-1}$ | $3.5 \times 10^{-1}$ | $4.7 \times 10^{-1}$ | $5.5 \times 10^{-1}$ | - | - | - |
| 191 | 0.001 | 1 | 1 | 10 | 10 | $4.1 \times 10^{-1}$ | $3.6 \times 10^{-1}$ | $4.3 \times 10^{-1}$ | $5.4 \times 10^{-1}$ | - | - | - |
| 192 | 0.001 | 1 | 1 | 10 | 30 | $4.1 \times 10^{-1}$ | $3.7 \times 10^{-1}$ | $4.5 \times 10^{-1}$ | $5.7 \times 10^{-1}$ | - | - | - |
| 193 | 0.001 | 1 | 1 | 10 | 50 | $4.1 \times 10^{-1}$ | $3.8 \times 10^{-1}$ | $4.9 \times 10^{-1}$ | $6.0 \times 10^{-1}$ | - | - | - |
| 194 | 0.001 | 1 | 1 | 30 | 30 | $5.9 \times 10^{-1}$ | $5.2 \times 10^{-1}$ | $5.5 \times 10^{-1}$ | $8.0 \times 10^{-1}$ | - | - | OK |
| 195 | 0.001 | 1 | 1 | 30 | 50 | $5.9 \times 10^{-1}$ | $5.2 \times 10^{-1}$ | $5.7 \times 10^{-1}$ | $8.6 \times 10^{-1}$ | - | - | OK |
| 196 | 0.001 | 1 | 1 | 50 | 50 | $6.8 \times 10^{-1}$ | $6.7 \times 10^{-1}$ | $6.9 \times 10^{-1}$ | 1.1 | - | - | OK |
| 197 | 0.001 | 1 | 5 | 5 | 10 | $6.8 \times 10^{-1}$ | $5.8 \times 10^{-1}$ | $7.1 \times 10^{-1}$ | $9.1 \times 10^{-1}$ | - | - | OK |
| 198 | 0.001 | 1 | 5 | 5 | 30 | $6.8 \times 10^{-1}$ | $5.9 \times 10^{-1}$ | $7.3 \times 10^{-1}$ | $9.7 \times 10^{-1}$ | - | - | OK |
| 199 | 0.001 | 1 | 5 | 5 | 50 | $6.8 \times 10^{-1}$ | $6.1 \times 10^{-1}$ | $7.7 \times 10^{-1}$ | $9.9 \times 10^{-1}$ | - | - | OK |
| 200 | 0.001 | 1 | 5 | 10 | 10 | $7.1 \times 10^{-1}$ | $6.2 \times 10^{-1}$ | $7.2 \times 10^{-1}$ | $9.3 \times 10^{-1}$ | - | - | OK |
| 201 | 0.001 | 1 | 5 | 10 | 30 | $7.1 \times 10^{-1}$ | $6.2 \times 10^{-1}$ | $7.5 \times 10^{-1}$ | 1.0 | - | - | OK |
| 202 | 0.001 | 1 | 5 | 10 | 50 | $7.1 \times 10^{-1}$ | $6.3 \times 10^{-1}$ | $7.8 \times 10^{-1}$ | 1.1 | - | - | OK |
| 203 | 0.001 | 1 | 5 | 30 | 30 | $9.5 \times 10^{-1}$ | $7.8 \times 10^{-1}$ | $8.4 \times 10^{-1}$ | 1.2 | - | - | OK |
| 204 | 0.001 | 1 | 5 | 30 | 50 | $9.5 \times 10^{-1}$ | $7.8 \times 10^{-1}$ | $8.7 \times 10^{-1}$ | 1.2 | - | - | OK |
| 205 | 0.001 | 1 | 5 | 50 | 50 | 1.1 | $9.4 \times 10^{-1}$ | $9.9 \times 10^{-1}$ | 1.5 | - | - | OK |
| 206 | 0.001 | 1 | 10 | 10 | 10 | 1.1 | $9.5 \times 10^{-1}$ | 1.1 | 1.5 | - | - | OK |
| 207 | 0.001 | 1 | 10 | 10 | 30 | 1.1 | $9.5 \times 10^{-1}$ | 1.1 | 1.5 | - | - | OK |
| 208 | 0.001 | 1 | 10 | 10 | 50 | 1.1 | $9.5 \times 10^{-1}$ | 1.1 | 1.6 | - | - | - |
| 209 | 0.001 | 1 | 10 | 30 | 30 | 1.3 | 1.1 | 1.2 | 1.7 | - | - | - |
| 210 | 0.001 | 1 | 10 | 30 | 50 | 1.3 | 1.1 | 1.2 | 1.8 | - | - | - |
| 211 | 0.001 | 1 | 10 | 50 | 50 | 1.5 | 1.3 | 1.3 | 2.0 | - | - | - |
| 212 | 0.001 | 1 | 30 | 30 | 30 | 2.9 | 2.4 | 2.6 | 3.8 | - | - | - |
| 213 | 0.001 | 1 | 30 | 30 | 50 | 2.9 | 2.4 | 2.7 | 3.9 | - | - | - |
| 214 | 0.001 | 1 | 30 | 50 | 50 | 3.0 | 2.5 | 2.7 | 4.0 | - | - | - |
| 215 | 0.001 | 1 | 50 | 50 | 50 | 4.8 | 3.8 | 4.2 | 6.1 | - | - | - |
| 216 | 0.001 | 5 | 5 | 5 | 10 | 1.5 | 1.5 | 1.9 | 2.2 | - | - | - |
| 217 | 0.001 | 5 | 5 | 5 | 30 | 1.5 | 1.5 | 1.9 | 2.3 | - | - | - |
| 218 | 0.001 | 5 | 5 | 5 | 50 | 1.5 | 1.5 | 2.0 | 2.3 | - | - | - |
| 219 | 0.001 | 5 | 5 | 10 | 10 | 1.6 | 1.5 | 1.9 | 2.3 | - | - | - |
| 220 | 0.001 | 5 | 5 | 10 | 30 | 1.6 | 1.5 | 1.9 | 2.3 | - | - | - |
| 221 | 0.001 | 5 | 5 | 10 | 50 | 1.6 | 1.5 | 2.0 | 2.4 | - | - | - |
| 222 | 0.001 | 5 | 5 | 30 | 30 | 1.8 | 1.7 | 2.0 | 2.5 | - | - | - |
| 223 | 0.001 | 5 | 5 | 30 | 50 | 1.8 | 1.7 | 2.0 | 2.6 | - | - | - |
| 224 | 0.001 | 5 | 5 | 50 | 50 | 2.0 | 1.8 | 2.1 | 2.7 | - | - | - |
| 225 | 0.001 | 5 | 10 | 10 | 10 | 2.0 | 1.9 | 2.3 | 2.9 | - | - | - |
| 226 | 0.001 | 5 | 10 | 10 | 30 | 2.0 | 1.9 | 2.3 | 2.9 | - | - | - |
| 227 | 0.001 | 5 | 10 | 10 | 50 | 2.0 | 1.9 | 2.3 | 2.9 | - | - | - |
| 228 | 0.001 | 5 | 10 | 30 | 30 | 2.2 | 2.0 | 2.4 | 3.1 | - | - | - |
| 229 | 0.001 | 5 | 10 | 30 | 50 | 2.2 | 2.0 | 2.5 | 3.1 | - | - | - |
| 230 | 0.001 | 5 | 10 | 50 | 50 | 2.5 | 2.1 | 2.6 | 3.3 | - | - | - |
| 231 | 0.001 | 5 | 30 | 30 | 30 | 3.8 | 3.3 | 3.9 | 5.0 | - | - | - |
| 232 | 0.001 | 5 | 30 | 30 | 50 | 3.8 | 3.3 | 3.9 | 5.1 | - | - | - |
| 233 | 0.001 | 5 | 30 | 50 | 50 | 4.0 | 3.4 | 4.0 | 5.2 | - | - | - |
| 234 | 0.001 | 5 | 50 | 50 | 50 | 5.5 | 4.7 | 5.4 | 7.4 | - | - | - |
| 235 | 0.01 | 0.01 | 0.01 | 0.1 | 10 | $1.1 \times 10^{-2}$ | $1.5 \times 10^{-2}$ | $2.3 \times 10^{-2}$ | $1.1 \times 10^{-2}$ | - | - | - |
| 236 | 0.01 | 0.01 | 0.01 | 0.1 | 30 | $1.1 \times 10^{-2}$ | $2.4 \times 10^{-2}$ | $4.7 \times 10^{-2}$ | $1.1 \times 10^{-2}$ | - | - | - |
| 237 | 0.01 | 0.01 | 0.01 | 0.1 | 50 | $1.1 \times 10^{-2}$ | $3.3 \times 10^{-2}$ | $7.2 \times 10^{-2}$ | $1.1 \times 10^{-2}$ | OK | - | - |
| 238 | 0.01 | 0.01 | 0.01 | 1 | 10 | $1.7 \times 10^{-2}$ | $2.0 \times 10^{-2}$ | $2.9 \times 10^{-2}$ | $2.5 \times 10^{-2}$ | - | - | - |
| 239 | 0.01 | 0.01 | 0.01 | 1 | 30 | $1.7 \times 10^{-2}$ | $2.8 \times 10^{-2}$ | $5.6 \times 10^{-2}$ | $2.5 \times 10^{-2}$ | OK | - | - |
| 240 | 0.01 | 0.01 | 0.01 | 1 | 50 | $1.7 \times 10^{-2}$ | $3.7 \times 10^{-2}$ | $8.1 \times 10^{-2}$ | $2.5 \times 10^{-2}$ | OK | - | - |
| 241 | 0.01 | 0.01 | 0.01 | 5 | 10 | $4.3 \times 10^{-2}$ | $4.9 \times 10^{-2}$ | $5.0 \times 10^{-2}$ | $7.7 \times 10^{-2}$ | - | - | - |
| 242 | 0.01 | 0.01 | 0.01 | 5 | 30 | $4.3 \times 10^{-2}$ | $5.8 \times 10^{-2}$ | $7.7 \times 10^{-2}$ | $8.8 \times 10^{-2}$ | - | OK | - |
| 243 | 0.01 | 0.01 | 0.01 | 5 | 50 | $4.3 \times 10^{-2}$ | $6.3 \times 10^{-2}$ | $1.0 \times 10^{-1}$ | $8.8 \times 10^{-2}$ | - | OK | - |
| 244 | 0.01 | 0.01 | 0.01 | 10 | 10 | $7.7 \times 10^{-2}$ | $8.3 \times 10^{-2}$ | $7.9 \times 10^{-2}$ | $1.4 \times 10^{-1}$ | - | OK | - |
| 245 | 0.01 | 0.01 | 0.01 | 10 | 30 | $7.7 \times 10^{-2}$ | $9.1 \times 10^{-2}$ | $1.0 \times 10^{-1}$ | $1.7 \times 10^{-1}$ | - | OK | - |
| 246 | 0.01 | 0.01 | 0.01 | 10 | 50 | $7.7 \times 10^{-2}$ | $9.7 \times 10^{-2}$ | $1.3 \times 10^{-1}$ | $1.7 \times 10^{-1}$ | - | OK | - |
| 247 | 0.01 | 0.01 | 0.01 | 30 | 30 | $2.1 \times 10^{-1}$ | $2.3 \times 10^{-1}$ | $2.2 \times 10^{-1}$ | $3.9 \times 10^{-1}$ | - | OK | - |
| 248 | 0.01 | 0.01 | 0.01 | 30 | 50 | $2.1 \times 10^{-1}$ | $2.5 \times 10^{-1}$ | $2.3 \times 10^{-1}$ | $4.1 \times 10^{-1}$ | - | OK | - |
| 249 | 0.01 | 0.01 | 0.01 | 50 | 50 | $3.4 \times 10^{-1}$ | $3.7 \times 10^{-1}$ | $3.6 \times 10^{-1}$ | $6.4 \times 10^{-1}$ | - | - | - |
| 250 | 0.01 | 0.01 | 0.1 | 0.1 | 10 | $1.8 \times 10^{-2}$ | $1.9 \times 10^{-2}$ | $3.2 \times 10^{-2}$ | $2.2 \times 10^{-2}$ | - | - | - |
| 251 | 0.01 | 0.01 | 0.1 | 0.1 | 30 | $1.8 \times 10^{-2}$ | $2.8 \times 10^{-2}$ | $5.8 \times 10^{-2}$ | $2.2 \times 10^{-2}$ | OK | - | - |
| 252 | 0.01 | 0.01 | 0.1 | 0.1 | 50 | $1.8 \times 10^{-2}$ | $3.7 \times 10^{-2}$ | $8.3 \times 10^{-2}$ | $2.2 \times 10^{-2}$ | OK | - | - |
| 253 | 0.01 | 0.01 | 0.1 | 1 | 10 | $2.6 \times 10^{-2}$ | $2.6 \times 10^{-2}$ | $3.5 \times 10^{-2}$ | $3.5 \times 10^{-2}$ | - | - | - |
| 254 | 0.01 | 0.01 | 0.1 | 1 | 30 | $2.6 \times 10^{-2}$ | $3.1 \times 10^{-2}$ | $6.5 \times 10^{-2}$ | $3.5 \times 10^{-2}$ | OK | - | - |
| 255 | 0.01 | 0.01 | 0.1 | 1 | 50 | $2.6 \times 10^{-2}$ | $4.1 \times 10^{-2}$ | $9.0 \times 10^{-2}$ | $3.5 \times 10^{-2}$ | OK | - | - |
| 256 | 0.01 | 0.01 | 0.1 | 5 | 10 | $5.3 \times 10^{-2}$ | $5.5 \times 10^{-2}$ | $5.5 \times 10^{-2}$ | $8.7 \times 10^{-2}$ | - | OK | - |
| 257 | 0.01 | 0.01 | 0.1 | 5 | 30 | $5.3 \times 10^{-2}$ | $6.3 \times 10^{-2}$ | $8.4 \times 10^{-2}$ | $9.5 \times 10^{-2}$ | - | OK | - |
| 258 | 0.01 | 0.01 | 0.1 | 5 | 50 | $5.3 \times 10^{-2}$ | $6.9 \times 10^{-2}$ | $1.1 \times 10^{-1}$ | $9.5 \times 10^{-2}$ | - | OK | - |
| 259 | 0.01 | 0.01 | 0.1 | 10 | 10 | $8.7 \times 10^{-2}$ | $8.8 \times 10^{-2}$ | $8.4 \times 10^{-2}$ | $1.5 \times 10^{-1}$ | - | OK | - |
| 260 | 0.01 | 0.01 | 0.1 | 10 | 30 | $8.7 \times 10^{-2}$ | $9.5 \times 10^{-2}$ | $1.1 \times 10^{-1}$ | $1.7 \times 10^{-1}$ | - | OK | - |
| 261 | 0.01 | 0.01 | 0.1 | 10 | 50 | $8.7 \times 10^{-2}$ | $1.0 \times 10^{-1}$ | $1.4 \times 10^{-1}$ | $1.7 \times 10^{-1}$ | - | OK | - |
| 262 | 0.01 | 0.01 | 0.1 | 30 | 30 | $2.2 \times 10^{-1}$ | $2.3 \times 10^{-1}$ | $2.2 \times 10^{-1}$ | $4.0 \times 10^{-1}$ | - | OK | - |
| 263 | 0.01 | 0.01 | 0.1 | 30 | 50 | $2.2 \times 10^{-1}$ | $2.5 \times 10^{-1}$ | $2.3 \times 10^{-1}$ | $4.2 \times 10^{-1}$ | - | OK | - |
| 264 | 0.01 | 0.01 | 0.1 | 50 | 50 | $3.5 \times 10^{-1}$ | $3.8 \times 10^{-1}$ | $3.6 \times 10^{-1}$ | $6.5 \times 10^{-1}$ | - | - | - |
| 265 | 0.01 | 0.01 | 1 | 1 | 10 | $9.9 \times 10^{-2}$ | $8.3 \times 10^{-2}$ | $1.0 \times 10^{-1}$ | $1.4 \times 10^{-1}$ | - | OK | - |
| 266 | 0.01 | 0.01 | 1 | 1 | 30 | $9.9 \times 10^{-2}$ | $9.5 \times 10^{-2}$ | $1.3 \times 10^{-1}$ | $1.4 \times 10^{-1}$ | - | OK | - |
| 267 | 0.01 | 0.01 | 1 | 1 | 50 | $9.9 \times 10^{-2}$ | $9.9 \times 10^{-2}$ | $1.6 \times 10^{-1}$ | $1.4 \times 10^{-1}$ | - | OK | - |
| 268 | 0.01 | 0.01 | 1 | 5 | 10 | $1.4 \times 10^{-1}$ | $1.1 \times 10^{-1}$ | $1.2 \times 10^{-1}$ | $1.8 \times 10^{-1}$ | - | OK | - |
| 269 | 0.01 | 0.01 | 1 | 5 | 30 | $1.4 \times 10^{-1}$ | $1.2 \times 10^{-1}$ | $1.5 \times 10^{-1}$ | $1.9 \times 10^{-1}$ | - | OK | - |
| 270 | 0.01 | 0.01 | 1 | 5 | 50 | $1.4 \times 10^{-1}$ | $1.3 \times 10^{-1}$ | $1.8 \times 10^{-1}$ | $1.9 \times 10^{-1}$ | - | OK | - |
| 271 | 0.01 | 0.01 | 1 | 10 | 10 | $1.8 \times 10^{-1}$ | $1.5 \times 10^{-1}$ | $1.5 \times 10^{-1}$ | $2.5 \times 10^{-1}$ | - | OK | - |
| 272 | 0.01 | 0.01 | 1 | 10 | 30 | $1.8 \times 10^{-1}$ | $1.6 \times 10^{-1}$ | $1.8 \times 10^{-1}$ | $2.7 \times 10^{-1}$ | - | OK | - |
| 273 | 0.01 | 0.01 | 1 | 10 | 50 | $1.8 \times 10^{-1}$ | $1.6 \times 10^{-1}$ | $2.0 \times 10^{-1}$ | $2.7 \times 10^{-1}$ | - | OK | - |
| 274 | 0.01 | 0.01 | 1 | 30 | 30 | $3.0 \times 10^{-1}$ | $2.9 \times 10^{-1}$ | $2.9 \times 10^{-1}$ | $5.1 \times 10^{-1}$ | - | - | - |
| 275 | 0.01 | 0.01 | 1 | 30 | 50 | $3.0 \times 10^{-1}$ | $3.1 \times 10^{-1}$ | $3.0 \times 10^{-1}$ | $5.2 \times 10^{-1}$ | - | - | - |
| 276 | 0.01 | 0.01 | 1 | 50 | 50 | $4.4 \times 10^{-1}$ | $4.4 \times 10^{-1}$ | $4.3 \times 10^{-1}$ | $7.7 \times 10^{-1}$ | - | - | - |
| 277 | 0.01 | 0.01 | 5 | 5 | 10 | $4.6 \times 10^{-1}$ | $3.7 \times 10^{-1}$ | $4.0 \times 10^{-1}$ | $6.0 \times 10^{-1}$ | - | - | - |
| 278 | 0.01 | 0.01 | 5 | 5 | 30 | $4.6 \times 10^{-1}$ | $3.8 \times 10^{-1}$ | $4.3 \times 10^{-1}$ | $6.4 \times 10^{-1}$ | - | - | - |
| 279 | 0.01 | 0.01 | 5 | 5 | 50 | $4.6 \times 10^{-1}$ | $3.8 \times 10^{-1}$ | $4.7 \times 10^{-1}$ | $6.4 \times 10^{-1}$ | - | - | - |
| 280 | 0.01 | 0.01 | 5 | 10 | 10 | $4.9 \times 10^{-1}$ | $4.0 \times 10^{-1}$ | $4.2 \times 10^{-1}$ | $6.5 \times 10^{-1}$ | - | - | - |
| 281 | 0.01 | 0.01 | 5 | 10 | 30 | $4.9 \times 10^{-1}$ | $4.1 \times 10^{-1}$ | $4.5 \times 10^{-1}$ | $7.0 \times 10^{-1}$ | - | - | - |
| 282 | 0.01 | 0.01 | 5 | 10 | 50 | $4.9 \times 10^{-1}$ | $4.1 \times 10^{-1}$ | $4.9 \times 10^{-1}$ | $7.0 \times 10^{-1}$ | - | - | - |
| 283 | 0.01 | 0.01 | 5 | 30 | 30 | $6.8 \times 10^{-1}$ | $5.8 \times 10^{-1}$ | $5.7 \times 10^{-1}$ | $9.3 \times 10^{-1}$ | - | - | OK |
| 284 | 0.01 | 0.01 | 5 | 30 | 50 | $6.8 \times 10^{-1}$ | $5.8 \times 10^{-1}$ | $6.0 \times 10^{-1}$ | $9.6 \times 10^{-1}$ | - | - | OK |
| 285 | 0.01 | 0.01 | 5 | 50 | 50 | $8.4 \times 10^{-1}$ | $7.2 \times 10^{-1}$ | $7.1 \times 10^{-1}$ | 1.2 | - | - | OK |
| 286 | 0.01 | 0.01 | 10 | 10 | 10 | $9.1 \times 10^{-1}$ | $7.2 \times 10^{-1}$ | $7.9 \times 10^{-1}$ | 1.2 | - | - | OK |
| 287 | 0.01 | 0.01 | 10 | 10 | 30 | $9.1 \times 10^{-1}$ | $7.2 \times 10^{-1}$ | $8.1 \times 10^{-1}$ | 1.2 | - | - | OK |
| 288 | 0.01 | 0.01 | 10 | 10 | 50 | $9.1 \times 10^{-1}$ | $7.4 \times 10^{-1}$ | $8.3 \times 10^{-1}$ | 1.3 | - | - | OK |
| 289 | 0.01 | 0.01 | 10 | 30 | 30 | 1.1 | $8.7 \times 10^{-1}$ | $8.9 \times 10^{-1}$ | 1.4 | - | - | OK |
| 290 | 0.01 | 0.01 | 10 | 30 | 50 | 1.1 | $8.7 \times 10^{-1}$ | $9.3 \times 10^{-1}$ | 1.5 | - | - | OK |
| 291 | 0.01 | 0.01 | 10 | 50 | 50 | 1.3 | 1.1 | 1.0 | 1.7 | - | - | - |
| 292 | 0.01 | 0.01 | 30 | 30 | 30 | 2.7 | 2.2 | 2.3 | 3.5 | - | - | - |
| 293 | 0.01 | 0.01 | 30 | 30 | 50 | 2.7 | 2.2 | 2.4 | 3.5 | - | - | - |
| 294 | 0.01 | 0.01 | 30 | 50 | 50 | 2.8 | 2.3 | 2.4 | 3.7 | - | - | - |
| 295 | 0.01 | 0.01 | 50 | 50 | 50 | 4.5 | 3.6 | 3.9 | 5.8 | - | - | - |
| 296 | 0.01 | 0.1 | 0.1 | 0.1 | 10 | $3.8 \times 10^{-2}$ | $4.1 \times 10^{-2}$ | $5.9 \times 10^{-2}$ | $5.4 \times 10^{-2}$ | OK | - | - |
| 297 | 0.01 | 0.1 | 0.1 | 0.1 | 30 | $3.8 \times 10^{-2}$ | $4.7 \times 10^{-2}$ | $8.7 \times 10^{-2}$ | $5.4 \times 10^{-2}$ | OK | - | - |
| 298 | 0.01 | 0.1 | 0.1 | 0.1 | 50 | $3.8 \times 10^{-2}$ | $5.7 \times 10^{-2}$ | $1.1 \times 10^{-1}$ | $5.4 \times 10^{-2}$ | - | OK | - |
| 299 | 0.01 | 0.1 | 0.1 | 1 | 10 | $4.8 \times 10^{-2}$ | $4.8 \times 10^{-2}$ | $6.3 \times 10^{-2}$ | $6.7 \times 10^{-2}$ | OK | - | - |
| 300 | 0.01 | 0.1 | 0.1 | 1 | 30 | $4.8 \times 10^{-2}$ | $5.1 \times 10^{-2}$ | $9.2 \times 10^{-2}$ | $6.7 \times 10^{-2}$ | - | OK | - |
| 301 | 0.01 | 0.1 | 0.1 | 1 | 50 | $4.8 \times 10^{-2}$ | $6.0 \times 10^{-2}$ | $1.2 \times 10^{-1}$ | $6.7 \times 10^{-2}$ | - | OK | - |
| 302 | 0.01 | 0.1 | 0.1 | 5 | 10 | $7.5 \times 10^{-2}$ | $7.5 \times 10^{-2}$ | $8.1 \times 10^{-2}$ | $1.2 \times 10^{-1}$ | - | OK | - |
| 303 | 0.01 | 0.1 | 0.1 | 5 | 30 | $7.5 \times 10^{-2}$ | $8.3 \times 10^{-2}$ | $1.1 \times 10^{-1}$ | $1.2 \times 10^{-1}$ | - | OK | - |
| 304 | 0.01 | 0.1 | 0.1 | 5 | 50 | $7.5 \times 10^{-2}$ | $9.1 \times 10^{-2}$ | $1.4 \times 10^{-1}$ | $1.2 \times 10^{-1}$ | - | OK | - |
| 305 | 0.01 | 0.1 | 0.1 | 10 | 10 | $1.1 \times 10^{-1}$ | $1.1 \times 10^{-1}$ | $1.1 \times 10^{-1}$ | $1.8 \times 10^{-1}$ | - | OK | - |
| 306 | 0.01 | 0.1 | 0.1 | 10 | 30 | $1.1 \times 10^{-1}$ | $1.2 \times 10^{-1}$ | $1.3 \times 10^{-1}$ | $2.0 \times 10^{-1}$ | - | OK | - |
| 307 | 0.01 | 0.1 | 0.1 | 10 | 50 | $1.1 \times 10^{-1}$ | $1.3 \times 10^{-1}$ | $1.6 \times 10^{-1}$ | $2.0 \times 10^{-1}$ | - | OK | - |
| 308 | 0.01 | 0.1 | 0.1 | 30 | 30 | $2.6 \times 10^{-1}$ | $2.6 \times 10^{-1}$ | $2.5 \times 10^{-1}$ | $4.4 \times 10^{-1}$ | - | OK | - |
| 309 | 0.01 | 0.1 | 0.1 | 30 | 50 | $2.6 \times 10^{-1}$ | $2.8 \times 10^{-1}$ | $2.6 \times 10^{-1}$ | $4.5 \times 10^{-1}$ | - | OK | - |
| 310 | 0.01 | 0.1 | 0.1 | 50 | 50 | $3.9 \times 10^{-1}$ | $4.0 \times 10^{-1}$ | $3.9 \times 10^{-1}$ | $6.9 \times 10^{-1}$ | - | - | - |
| 311 | 0.01 | 0.1 | 1 | 1 | 10 | $1.2 \times 10^{-1}$ | $1.0 \times 10^{-1}$ | $1.3 \times 10^{-1}$ | $1.7 \times 10^{-1}$ | - | OK | - |
| 312 | 0.01 | 0.1 | 1 | 1 | 30 | $1.2 \times 10^{-1}$ | $1.2 \times 10^{-1}$ | $1.5 \times 10^{-1}$ | $1.8 \times 10^{-1}$ | - | OK | - |
| 313 | 0.01 | 0.1 | 1 | 1 | 50 | $1.2 \times 10^{-1}$ | $1.2 \times 10^{-1}$ | $1.9 \times 10^{-1}$ | $1.8 \times 10^{-1}$ | - | OK | - |
| 314 | 0.01 | 0.1 | 1 | 5 | 10 | $1.6 \times 10^{-1}$ | $1.3 \times 10^{-1}$ | $1.5 \times 10^{-1}$ | $2.1 \times 10^{-1}$ | - | OK | - |
| 315 | 0.01 | 0.1 | 1 | 5 | 30 | $1.6 \times 10^{-1}$ | $1.4 \times 10^{-1}$ | $1.8 \times 10^{-1}$ | $2.2 \times 10^{-1}$ | - | OK | - |
| 316 | 0.01 | 0.1 | 1 | 5 | 50 | $1.6 \times 10^{-1}$ | $1.5 \times 10^{-1}$ | $2.1 \times 10^{-1}$ | $2.2 \times 10^{-1}$ | - | OK | - |
| 317 | 0.01 | 0.1 | 1 | 10 | 10 | $2.0 \times 10^{-1}$ | $1.7 \times 10^{-1}$ | $1.8 \times 10^{-1}$ | $2.8 \times 10^{-1}$ | - | OK | - |
| 318 | 0.01 | 0.1 | 1 | 10 | 30 | $2.0 \times 10^{-1}$ | $1.8 \times 10^{-1}$ | $2.0 \times 10^{-1}$ | $3.0 \times 10^{-1}$ | - | OK | - |
| 319 | 0.01 | 0.1 | 1 | 10 | 50 | $2.0 \times 10^{-1}$ | $1.9 \times 10^{-1}$ | $2.3 \times 10^{-1}$ | $3.0 \times 10^{-1}$ | - | OK | - |
| 320 | 0.01 | 0.1 | 1 | 30 | 30 | $3.2 \times 10^{-1}$ | $3.1 \times 10^{-1}$ | $3.1 \times 10^{-1}$ | $5.4 \times 10^{-1}$ | - | - | - |
| 321 | 0.01 | 0.1 | 1 | 30 | 50 | $3.2 \times 10^{-1}$ | $3.3 \times 10^{-1}$ | $3.2 \times 10^{-1}$ | $5.4 \times 10^{-1}$ | - | - | - |
| 322 | 0.01 | 0.1 | 1 | 50 | 50 | $4.6 \times 10^{-1}$ | $4.6 \times 10^{-1}$ | $4.5 \times 10^{-1}$ | $8.0 \times 10^{-1}$ | - | - | - |
| 323 | 0.01 | 0.1 | 5 | 5 | 10 | $4.9 \times 10^{-1}$ | $3.9 \times 10^{-1}$ | $4.3 \times 10^{-1}$ | $6.3 \times 10^{-1}$ | - | - | - |
| 324 | 0.01 | 0.1 | 5 | 5 | 30 | $4.9 \times 10^{-1}$ | $3.9 \times 10^{-1}$ | $4.5 \times 10^{-1}$ | $6.6 \times 10^{-1}$ | - | - | - |
| 325 | 0.01 | 0.1 | 5 | 5 | 50 | $4.9 \times 10^{-1}$ | $4.0 \times 10^{-1}$ | $5.0 \times 10^{-1}$ | $6.8 \times 10^{-1}$ | - | - | - |
| 326 | 0.01 | 0.1 | 5 | 10 | 10 | $5.1 \times 10^{-1}$ | $4.2 \times 10^{-1}$ | $4.5 \times 10^{-1}$ | $6.8 \times 10^{-1}$ | - | - | - |
| 327 | 0.01 | 0.1 | 5 | 10 | 30 | $5.1 \times 10^{-1}$ | $4.3 \times 10^{-1}$ | $4.8 \times 10^{-1}$ | $7.2 \times 10^{-1}$ | - | - | - |
| 328 | 0.01 | 0.1 | 5 | 10 | 50 | $5.1 \times 10^{-1}$ | $4.3 \times 10^{-1}$ | $5.2 \times 10^{-1}$ | $7.4 \times 10^{-1}$ | - | - | - |
| 329 | 0.01 | 0.1 | 5 | 30 | 30 | $7.1 \times 10^{-1}$ | $5.9 \times 10^{-1}$ | $5.9 \times 10^{-1}$ | $9.6 \times 10^{-1}$ | - | - | OK |
| 330 | 0.01 | 0.1 | 5 | 30 | 50 | $7.1 \times 10^{-1}$ | $5.9 \times 10^{-1}$ | $6.2 \times 10^{-1}$ | $9.9 \times 10^{-1}$ | - | - | OK |
| 331 | 0.01 | 0.1 | 5 | 50 | 50 | $8.7 \times 10^{-1}$ | $7.5 \times 10^{-1}$ | $7.4 \times 10^{-1}$ | 1.3 | - | - | OK |
| 332 | 0.01 | 0.1 | 10 | 10 | 10 | $9.4 \times 10^{-1}$ | $7.4 \times 10^{-1}$ | $8.1 \times 10^{-1}$ | 1.2 | - | - | OK |
| 333 | 0.01 | 0.1 | 10 | 10 | 30 | $9.4 \times 10^{-1}$ | $7.4 \times 10^{-1}$ | $8.4 \times 10^{-1}$ | 1.2 | - | - | OK |
| 334 | 0.01 | 0.1 | 10 | 10 | 50 | $9.4 \times 10^{-1}$ | $7.6 \times 10^{-1}$ | $8.6 \times 10^{-1}$ | 1.3 | - | - | OK |
| 335 | 0.01 | 0.1 | 10 | 30 | 30 | 1.1 | $9.0 \times 10^{-1}$ | $9.2 \times 10^{-1}$ | 1.4 | - | - | OK |
| 336 | 0.01 | 0.1 | 10 | 30 | 50 | 1.1 | $9.0 \times 10^{-1}$ | $9.6 \times 10^{-1}$ | 1.5 | - | - | OK |
| 337 | 0.01 | 0.1 | 10 | 50 | 50 | 1.3 | 1.1 | 1.1 | 1.7 | - | - | - |
| 338 | 0.01 | 0.1 | 30 | 30 | 30 | 2.7 | 2.2 | 2.4 | 3.5 | - | - | - |
| 339 | 0.01 | 0.1 | 30 | 30 | 50 | 2.7 | 2.2 | 2.4 | 3.6 | - | - | - |
| 340 | 0.01 | 0.1 | 30 | 50 | 50 | 2.8 | 2.3 | 2.5 | 3.7 | - | - | - |
| 341 | 0.01 | 0.1 | 50 | 50 | 50 | 4.5 | 3.6 | 3.9 | 5.8 | - | - | - |
| 342 | 0.01 | 1 | 1 | 1 | 10 | $3.1 \times 10^{-1}$ | $3.1 \times 10^{-1}$ | $4.0 \times 10^{-1}$ | $4.7 \times 10^{-1}$ | - | OK | - |
| 343 | 0.01 | 1 | 1 | 1 | 30 | $3.2 \times 10^{-1}$ | $3.2 \times 10^{-1}$ | $4.2 \times 10^{-1}$ | $4.9 \times 10^{-1}$ | - | OK | - |
| 344 | 0.01 | 1 | 1 | 1 | 50 | $3.2 \times 10^{-1}$ | $3.3 \times 10^{-1}$ | $4.4 \times 10^{-1}$ | $5.0 \times 10^{-1}$ | - | OK | - |
| 345 | 0.01 | 1 | 1 | 5 | 10 | $3.5 \times 10^{-1}$ | $3.3 \times 10^{-1}$ | $4.1 \times 10^{-1}$ | $5.1 \times 10^{-1}$ | - | - | - |
| 346 | 0.01 | 1 | 1 | 5 | 30 | $3.7 \times 10^{-1}$ | $3.4 \times 10^{-1}$ | $4.4 \times 10^{-1}$ | $5.4 \times 10^{-1}$ | - | - | - |
| 347 | 0.01 | 1 | 1 | 5 | 50 | $3.7 \times 10^{-1}$ | $3.5 \times 10^{-1}$ | $4.7 \times 10^{-1}$ | $5.5 \times 10^{-1}$ | - | - | - |
| 348 | 0.01 | 1 | 1 | 10 | 10 | $4.1 \times 10^{-1}$ | $3.7 \times 10^{-1}$ | $4.3 \times 10^{-1}$ | $5.5 \times 10^{-1}$ | - | - | - |
| 349 | 0.01 | 1 | 1 | 10 | 30 | $4.1 \times 10^{-1}$ | $3.8 \times 10^{-1}$ | $4.6 \times 10^{-1}$ | $5.8 \times 10^{-1}$ | - | - | - |
| 350 | 0.01 | 1 | 1 | 10 | 50 | $4.1 \times 10^{-1}$ | $3.9 \times 10^{-1}$ | $4.9 \times 10^{-1}$ | $6.1 \times 10^{-1}$ | - | - | - |
| 351 | 0.01 | 1 | 1 | 30 | 30 | $5.9 \times 10^{-1}$ | $5.2 \times 10^{-1}$ | $5.6 \times 10^{-1}$ | $8.0 \times 10^{-1}$ | - | - | OK |
| 352 | 0.01 | 1 | 1 | 30 | 50 | $5.9 \times 10^{-1}$ | $5.2 \times 10^{-1}$ | $5.8 \times 10^{-1}$ | $8.7 \times 10^{-1}$ | - | - | OK |
| 353 | 0.01 | 1 | 1 | 50 | 50 | $6.9 \times 10^{-1}$ | $6.7 \times 10^{-1}$ | $6.9 \times 10^{-1}$ | 1.1 | - | - | OK |
| 354 | 0.01 | 1 | 5 | 5 | 10 | $6.8 \times 10^{-1}$ | $5.9 \times 10^{-1}$ | $7.1 \times 10^{-1}$ | $9.2 \times 10^{-1}$ | - | - | OK |
| 355 | 0.01 | 1 | 5 | 5 | 30 | $6.8 \times 10^{-1}$ | $5.9 \times 10^{-1}$ | $7.4 \times 10^{-1}$ | $9.7 \times 10^{-1}$ | - | - | OK |
| 356 | 0.01 | 1 | 5 | 5 | 50 | $6.8 \times 10^{-1}$ | $6.1 \times 10^{-1}$ | $7.8 \times 10^{-1}$ | $9.9 \times 10^{-1}$ | - | - | OK |
| 357 | 0.01 | 1 | 5 | 10 | 10 | $7.2 \times 10^{-1}$ | $6.3 \times 10^{-1}$ | $7.3 \times 10^{-1}$ | $9.4 \times 10^{-1}$ | - | - | OK |
| 358 | 0.01 | 1 | 5 | 10 | 30 | $7.2 \times 10^{-1}$ | $6.3 \times 10^{-1}$ | $7.6 \times 10^{-1}$ | 1.0 | - | - | OK |
| 359 | 0.01 | 1 | 5 | 10 | 50 | $7.2 \times 10^{-1}$ | $6.4 \times 10^{-1}$ | $7.9 \times 10^{-1}$ | 1.1 | - | - | OK |
| 360 | 0.01 | 1 | 5 | 30 | 30 | $9.6 \times 10^{-1}$ | $7.8 \times 10^{-1}$ | $8.4 \times 10^{-1}$ | 1.2 | - | - | OK |
| 361 | 0.01 | 1 | 5 | 30 | 50 | $9.6 \times 10^{-1}$ | $7.8 \times 10^{-1}$ | $8.7 \times 10^{-1}$ | 1.2 | - | - | OK |
| 362 | 0.01 | 1 | 5 | 50 | 50 | 1.1 | $9.4 \times 10^{-1}$ | 1.0 | 1.5 | - | - | OK |
| 363 | 0.01 | 1 | 10 | 10 | 10 | 1.1 | $9.5 \times 10^{-1}$ | 1.1 | 1.5 | - | - | OK |
| 364 | 0.01 | 1 | 10 | 10 | 30 | 1.1 | $9.5 \times 10^{-1}$ | 1.1 | 1.5 | - | - | OK |
| 365 | 0.01 | 1 | 10 | 10 | 50 | 1.1 | $9.6 \times 10^{-1}$ | 1.1 | 1.6 | - | - | - |
| 366 | 0.01 | 1 | 10 | 30 | 30 | 1.3 | 1.1 | 1.2 | 1.7 | - | - | - |
| 367 | 0.01 | 1 | 10 | 30 | 50 | 1.3 | 1.1 | 1.2 | 1.8 | - | - | - |
| 368 | 0.01 | 1 | 10 | 50 | 50 | 1.5 | 1.3 | 1.3 | 2.0 | - | - | - |
| 369 | 0.01 | 1 | 30 | 30 | 30 | 2.9 | 2.4 | 2.6 | 3.8 | - | - | - |
| 370 | 0.01 | 1 | 30 | 30 | 50 | 2.9 | 2.4 | 2.7 | 3.9 | - | - | - |
| 371 | 0.01 | 1 | 30 | 50 | 50 | 3.0 | 2.5 | 2.7 | 4.1 | - | - | - |
| 372 | 0.01 | 1 | 50 | 50 | 50 | 4.8 | 3.8 | 4.2 | 6.1 | - | - | - |
| 373 | 0.01 | 5 | 5 | 5 | 10 | 1.5 | 1.5 | 1.9 | 2.2 | - | - | - |
| 374 | 0.01 | 5 | 5 | 5 | 30 | 1.5 | 1.5 | 1.9 | 2.3 | - | - | - |
| 375 | 0.01 | 5 | 5 | 5 | 50 | 1.5 | 1.5 | 2.0 | 2.3 | - | - | - |
| 376 | 0.01 | 5 | 5 | 10 | 10 | 1.6 | 1.6 | 1.9 | 2.3 | - | - | - |
| 377 | 0.01 | 5 | 5 | 10 | 30 | 1.6 | 1.6 | 2.0 | 2.3 | - | - | - |
| 378 | 0.01 | 5 | 5 | 10 | 50 | 1.6 | 1.6 | 2.0 | 2.4 | - | - | - |
| 379 | 0.01 | 5 | 5 | 30 | 30 | 1.8 | 1.7 | 2.0 | 2.5 | - | - | - |
| 380 | 0.01 | 5 | 5 | 30 | 50 | 1.8 | 1.7 | 2.0 | 2.6 | - | - | - |
| 381 | 0.01 | 5 | 5 | 50 | 50 | 2.0 | 1.8 | 2.1 | 2.7 | - | - | - |
| 382 | 0.01 | 5 | 10 | 10 | 10 | 2.0 | 1.9 | 2.3 | 2.9 | - | - | - |
| 383 | 0.01 | 5 | 10 | 10 | 30 | 2.0 | 1.9 | 2.4 | 2.9 | - | - | - |
| 384 | 0.01 | 5 | 10 | 10 | 50 | 2.0 | 1.9 | 2.4 | 2.9 | - | - | - |
| 385 | 0.01 | 5 | 10 | 30 | 30 | 2.2 | 2.0 | 2.5 | 3.1 | - | - | - |
| 386 | 0.01 | 5 | 10 | 30 | 50 | 2.2 | 2.0 | 2.5 | 3.1 | - | - | - |
| 387 | 0.01 | 5 | 10 | 50 | 50 | 2.5 | 2.1 | 2.6 | 3.3 | - | - | - |
| 388 | 0.01 | 5 | 30 | 30 | 30 | 3.9 | 3.3 | 3.9 | 5.0 | - | - | - |
| 389 | 0.01 | 5 | 30 | 30 | 50 | 3.9 | 3.3 | 3.9 | 5.1 | - | - | - |
| 390 | 0.01 | 5 | 30 | 50 | 50 | 4.0 | 3.4 | 4.0 | 5.2 | - | - | - |
| 391 | 0.01 | 5 | 50 | 50 | 50 | 5.5 | 4.7 | 5.4 | 7.4 | - | - | - |
| 392 | 0.1 | 0.1 | 0.1 | 0.1 | 10 | $1.0 \times 10^{-1}$ | $1.0 \times 10^{-1}$ | $1.1 \times 10^{-1}$ | $1.0 \times 10^{-1}$ | - | OK | - |
| 393 | 0.1 | 0.1 | 0.1 | 0.1 | 30 | $1.0 \times 10^{-1}$ | $1.1 \times 10^{-1}$ | $1.4 \times 10^{-1}$ | $1.0 \times 10^{-1}$ | - | OK | - |
| 394 | 0.1 | 0.1 | 0.1 | 0.1 | 50 | $1.0 \times 10^{-1}$ | $1.2 \times 10^{-1}$ | $1.6 \times 10^{-1}$ | $1.0 \times 10^{-1}$ | - | OK | - |
| 395 | 0.1 | 0.1 | 0.1 | 1 | 10 | $1.1 \times 10^{-1}$ | $1.1 \times 10^{-1}$ | $1.2 \times 10^{-1}$ | $1.1 \times 10^{-1}$ | - | OK | - |
| 396 | 0.1 | 0.1 | 0.1 | 1 | 30 | $1.1 \times 10^{-1}$ | $1.2 \times 10^{-1}$ | $1.5 \times 10^{-1}$ | $1.1 \times 10^{-1}$ | - | OK | - |
| 397 | 0.1 | 0.1 | 0.1 | 1 | 50 | $1.1 \times 10^{-1}$ | $1.3 \times 10^{-1}$ | $1.7 \times 10^{-1}$ | $1.1 \times 10^{-1}$ | - | OK | - |
| 398 | 0.1 | 0.1 | 0.1 | 5 | 10 | $1.3 \times 10^{-1}$ | $1.4 \times 10^{-1}$ | $1.4 \times 10^{-1}$ | $1.7 \times 10^{-1}$ | - | OK | - |
| 399 | 0.1 | 0.1 | 0.1 | 5 | 30 | $1.3 \times 10^{-1}$ | $1.5 \times 10^{-1}$ | $1.7 \times 10^{-1}$ | $1.8 \times 10^{-1}$ | - | OK | - |
| 400 | 0.1 | 0.1 | 0.1 | 5 | 50 | $1.3 \times 10^{-1}$ | $1.5 \times 10^{-1}$ | $1.9 \times 10^{-1}$ | $1.8 \times 10^{-1}$ | - | OK | - |
| 401 | 0.1 | 0.1 | 0.1 | 10 | 10 | $1.7 \times 10^{-1}$ | $1.7 \times 10^{-1}$ | $1.7 \times 10^{-1}$ | $2.2 \times 10^{-1}$ | - | OK | - |
| 402 | 0.1 | 0.1 | 0.1 | 10 | 30 | $1.7 \times 10^{-1}$ | $1.8 \times 10^{-1}$ | $1.9 \times 10^{-1}$ | $2.5 \times 10^{-1}$ | - | OK | - |
| 403 | 0.1 | 0.1 | 0.1 | 10 | 50 | $1.7 \times 10^{-1}$ | $1.9 \times 10^{-1}$ | $2.2 \times 10^{-1}$ | $2.5 \times 10^{-1}$ | - | OK | - |
| 404 | 0.1 | 0.1 | 0.1 | 30 | 30 | $3.0 \times 10^{-1}$ | $3.2 \times 10^{-1}$ | $3.1 \times 10^{-1}$ | $4.7 \times 10^{-1}$ | - | OK | - |
| 405 | 0.1 | 0.1 | 0.1 | 30 | 50 | $3.0 \times 10^{-1}$ | $3.4 \times 10^{-1}$ | $3.2 \times 10^{-1}$ | $5.0 \times 10^{-1}$ | - | - | - |
| 406 | 0.1 | 0.1 | 0.1 | 50 | 50 | $4.3 \times 10^{-1}$ | $4.6 \times 10^{-1}$ | $4.5 \times 10^{-1}$ | $7.2 \times 10^{-1}$ | - | - | - |
| 407 | 0.1 | 0.1 | 1 | 1 | 10 | $1.8 \times 10^{-1}$ | $1.7 \times 10^{-1}$ | $1.8 \times 10^{-1}$ | $2.1 \times 10^{-1}$ | - | OK | - |
| 408 | 0.1 | 0.1 | 1 | 1 | 30 | $1.8 \times 10^{-1}$ | $1.8 \times 10^{-1}$ | $2.1 \times 10^{-1}$ | $2.2 \times 10^{-1}$ | - | OK | - |
| 409 | 0.1 | 0.1 | 1 | 1 | 50 | $1.8 \times 10^{-1}$ | $1.8 \times 10^{-1}$ | $2.5 \times 10^{-1}$ | $2.2 \times 10^{-1}$ | - | OK | - |
| 410 | 0.1 | 0.1 | 1 | 5 | 10 | $2.2 \times 10^{-1}$ | $2.0 \times 10^{-1}$ | $2.0 \times 10^{-1}$ | $2.6 \times 10^{-1}$ | - | OK | - |
| 411 | 0.1 | 0.1 | 1 | 5 | 30 | $2.2 \times 10^{-1}$ | $2.0 \times 10^{-1}$ | $2.3 \times 10^{-1}$ | $2.7 \times 10^{-1}$ | - | OK | - |
| 412 | 0.1 | 0.1 | 1 | 5 | 50 | $2.2 \times 10^{-1}$ | $2.1 \times 10^{-1}$ | $2.6 \times 10^{-1}$ | $2.7 \times 10^{-1}$ | - | OK | - |
| 413 | 0.1 | 0.1 | 1 | 10 | 10 | $2.6 \times 10^{-1}$ | $2.3 \times 10^{-1}$ | $2.3 \times 10^{-1}$ | $3.3 \times 10^{-1}$ | - | OK | - |
| 414 | 0.1 | 0.1 | 1 | 10 | 30 | $2.6 \times 10^{-1}$ | $2.4 \times 10^{-1}$ | $2.6 \times 10^{-1}$ | $3.5 \times 10^{-1}$ | - | OK | - |
| 415 | 0.1 | 0.1 | 1 | 10 | 50 | $2.6 \times 10^{-1}$ | $2.5 \times 10^{-1}$ | $2.9 \times 10^{-1}$ | $3.5 \times 10^{-1}$ | - | OK | - |
| 416 | 0.1 | 0.1 | 1 | 30 | 30 | $3.8 \times 10^{-1}$ | $3.7 \times 10^{-1}$ | $3.7 \times 10^{-1}$ | $5.9 \times 10^{-1}$ | - | - | - |
| 417 | 0.1 | 0.1 | 1 | 30 | 50 | $3.8 \times 10^{-1}$ | $4.0 \times 10^{-1}$ | $3.8 \times 10^{-1}$ | $6.0 \times 10^{-1}$ | - | - | - |
| 418 | 0.1 | 0.1 | 1 | 50 | 50 | $5.3 \times 10^{-1}$ | $5.2 \times 10^{-1}$ | $5.1 \times 10^{-1}$ | $8.5 \times 10^{-1}$ | - | - | OK |
| 419 | 0.1 | 0.1 | 5 | 5 | 10 | $5.4 \times 10^{-1}$ | $4.5 \times 10^{-1}$ | $4.9 \times 10^{-1}$ | $6.8 \times 10^{-1}$ | - | - | - |
| 420 | 0.1 | 0.1 | 5 | 5 | 30 | $5.4 \times 10^{-1}$ | $4.6 \times 10^{-1}$ | $5.1 \times 10^{-1}$ | $7.2 \times 10^{-1}$ | - | - | - |
| 421 | 0.1 | 0.1 | 5 | 5 | 50 | $5.4 \times 10^{-1}$ | $4.6 \times 10^{-1}$ | $5.5 \times 10^{-1}$ | $7.2 \times 10^{-1}$ | - | - | - |
| 422 | 0.1 | 0.1 | 5 | 10 | 10 | $5.7 \times 10^{-1}$ | $4.8 \times 10^{-1}$ | $5.0 \times 10^{-1}$ | $7.3 \times 10^{-1}$ | - | - | - |
| 423 | 0.1 | 0.1 | 5 | 10 | 30 | $5.7 \times 10^{-1}$ | $4.9 \times 10^{-1}$ | $5.3 \times 10^{-1}$ | $7.8 \times 10^{-1}$ | - | - | - |
| 424 | 0.1 | 0.1 | 5 | 10 | 50 | $5.7 \times 10^{-1}$ | $4.9 \times 10^{-1}$ | $5.8 \times 10^{-1}$ | $7.8 \times 10^{-1}$ | - | - | - |
| 425 | 0.1 | 0.1 | 5 | 30 | 30 | $7.6 \times 10^{-1}$ | $6.6 \times 10^{-1}$ | $6.5 \times 10^{-1}$ | 1.0 | - | - | OK |
| 426 | 0.1 | 0.1 | 5 | 30 | 50 | $7.6 \times 10^{-1}$ | $6.6 \times 10^{-1}$ | $6.8 \times 10^{-1}$ | 1.0 | - | - | OK |
| 427 | 0.1 | 0.1 | 5 | 50 | 50 | $9.2 \times 10^{-1}$ | $8.0 \times 10^{-1}$ | $7.9 \times 10^{-1}$ | 1.3 | - | - | OK |
| 428 | 0.1 | 0.1 | 10 | 10 | 10 | $9.9 \times 10^{-1}$ | $8.1 \times 10^{-1}$ | $8.7 \times 10^{-1}$ | 1.2 | - | - | OK |
| 429 | 0.1 | 0.1 | 10 | 10 | 30 | $9.9 \times 10^{-1}$ | $8.1 \times 10^{-1}$ | $8.9 \times 10^{-1}$ | 1.3 | - | - | OK |
| 430 | 0.1 | 0.1 | 10 | 10 | 50 | $9.9 \times 10^{-1}$ | $8.2 \times 10^{-1}$ | $9.2 \times 10^{-1}$ | 1.3 | - | - | OK |
| 431 | 0.1 | 0.1 | 10 | 30 | 30 | 1.2 | $9.5 \times 10^{-1}$ | $9.8 \times 10^{-1}$ | 1.5 | - | - | OK |
| 432 | 0.1 | 0.1 | 10 | 30 | 50 | 1.2 | $9.5 \times 10^{-1}$ | 1.0 | 1.5 | - | - | - |
| 433 | 0.1 | 0.1 | 10 | 50 | 50 | 1.4 | 1.1 | 1.1 | 1.7 | - | - | - |
| 434 | 0.1 | 0.1 | 30 | 30 | 30 | 2.8 | 2.2 | 2.4 | 3.5 | - | - | - |
| 435 | 0.1 | 0.1 | 30 | 30 | 50 | 2.8 | 2.2 | 2.5 | 3.6 | - | - | - |
| 436 | 0.1 | 0.1 | 30 | 50 | 50 | 2.8 | 2.4 | 2.5 | 3.8 | - | - | - |
| 437 | 0.1 | 0.1 | 50 | 50 | 50 | 4.6 | 3.7 | 4.0 | 5.8 | - | - | - |
| 438 | 0.1 | 1 | 1 | 1 | 10 | $3.7 \times 10^{-1}$ | $3.7 \times 10^{-1}$ | $4.5 \times 10^{-1}$ | $5.2 \times 10^{-1}$ | - | - | - |
| 439 | 0.1 | 1 | 1 | 1 | 30 | $3.8 \times 10^{-1}$ | $3.8 \times 10^{-1}$ | $4.8 \times 10^{-1}$ | $5.3 \times 10^{-1}$ | - | - | - |
| 440 | 0.1 | 1 | 1 | 1 | 50 | $3.8 \times 10^{-1}$ | $3.9 \times 10^{-1}$ | $5.0 \times 10^{-1}$ | $5.4 \times 10^{-1}$ | - | - | - |
| 441 | 0.1 | 1 | 1 | 5 | 10 | $4.2 \times 10^{-1}$ | $3.9 \times 10^{-1}$ | $4.7 \times 10^{-1}$ | $5.6 \times 10^{-1}$ | - | - | - |
| 442 | 0.1 | 1 | 1 | 5 | 30 | $4.3 \times 10^{-1}$ | $4.1 \times 10^{-1}$ | $4.9 \times 10^{-1}$ | $5.9 \times 10^{-1}$ | - | - | - |
| 443 | 0.1 | 1 | 1 | 5 | 50 | $4.3 \times 10^{-1}$ | $4.2 \times 10^{-1}$ | $5.3 \times 10^{-1}$ | $6.0 \times 10^{-1}$ | - | - | - |
| 444 | 0.1 | 1 | 1 | 10 | 10 | $4.8 \times 10^{-1}$ | $4.3 \times 10^{-1}$ | $4.9 \times 10^{-1}$ | $6.0 \times 10^{-1}$ | - | - | - |
| 445 | 0.1 | 1 | 1 | 10 | 30 | $4.8 \times 10^{-1}$ | $4.5 \times 10^{-1}$ | $5.1 \times 10^{-1}$ | $6.2 \times 10^{-1}$ | - | - | - |
| 446 | 0.1 | 1 | 1 | 10 | 50 | $4.8 \times 10^{-1}$ | $4.5 \times 10^{-1}$ | $5.5 \times 10^{-1}$ | $6.7 \times 10^{-1}$ | - | - | - |
| 447 | 0.1 | 1 | 1 | 30 | 30 | $6.4 \times 10^{-1}$ | $5.9 \times 10^{-1}$ | $6.2 \times 10^{-1}$ | $8.5 \times 10^{-1}$ | - | - | OK |
| 448 | 0.1 | 1 | 1 | 30 | 50 | $6.4 \times 10^{-1}$ | $5.9 \times 10^{-1}$ | $6.3 \times 10^{-1}$ | $9.3 \times 10^{-1}$ | - | - | OK |
| 449 | 0.1 | 1 | 1 | 50 | 50 | $7.5 \times 10^{-1}$ | $7.4 \times 10^{-1}$ | $7.5 \times 10^{-1}$ | 1.1 | - | - | OK |
| 450 | 0.1 | 1 | 5 | 5 | 10 | $7.4 \times 10^{-1}$ | $6.5 \times 10^{-1}$ | $7.7 \times 10^{-1}$ | $9.7 \times 10^{-1}$ | - | - | OK |
| 451 | 0.1 | 1 | 5 | 5 | 30 | $7.4 \times 10^{-1}$ | $6.6 \times 10^{-1}$ | $7.9 \times 10^{-1}$ | 1.0 | - | - | OK |
| 452 | 0.1 | 1 | 5 | 5 | 50 | $7.4 \times 10^{-1}$ | $6.8 \times 10^{-1}$ | $8.3 \times 10^{-1}$ | 1.0 | - | - | OK |
| 453 | 0.1 | 1 | 5 | 10 | 10 | $7.8 \times 10^{-1}$ | $6.9 \times 10^{-1}$ | $7.8 \times 10^{-1}$ | $9.9 \times 10^{-1}$ | - | - | OK |
| 454 | 0.1 | 1 | 5 | 10 | 30 | $7.8 \times 10^{-1}$ | $6.9 \times 10^{-1}$ | $8.1 \times 10^{-1}$ | 1.1 | - | - | OK |
| 455 | 0.1 | 1 | 5 | 10 | 50 | $7.8 \times 10^{-1}$ | $7.0 \times 10^{-1}$ | $8.4 \times 10^{-1}$ | 1.1 | - | - | OK |
| 456 | 0.1 | 1 | 5 | 30 | 30 | 1.0 | $8.5 \times 10^{-1}$ | $8.9 \times 10^{-1}$ | 1.2 | - | - | OK |
| 457 | 0.1 | 1 | 5 | 30 | 50 | 1.0 | $8.5 \times 10^{-1}$ | $9.3 \times 10^{-1}$ | 1.3 | - | - | OK |
| 458 | 0.1 | 1 | 5 | 50 | 50 | 1.2 | 1.0 | 1.1 | 1.5 | - | - | - |
| 459 | 0.1 | 1 | 10 | 10 | 10 | 1.2 | 1.0 | 1.1 | 1.5 | - | - | - |
| 460 | 0.1 | 1 | 10 | 10 | 30 | 1.2 | 1.0 | 1.2 | 1.5 | - | - | - |
| 461 | 0.1 | 1 | 10 | 10 | 50 | 1.2 | 1.0 | 1.2 | 1.6 | - | - | - |

| 462 | 0.1 | 1 | 10 | 30 | 30 | 1.4 | 1.2 | 1.2 | 1.8 | - | - | - |
|-----|-----|---|----|----|----|-----|-----|-----|-----|---|---|---|
| 463 | 0.1 | 1 | 10 | 30 | 50 | 1.4 | 1.2 | 1.3 | 1.8 | - | - | - |
| 464 | 0.1 | 1 | 10 | 50 | 50 | 1.6 | 1.3 | 1.4 | 2.0 | - | - | - |
| 465 | 0.1 | 1 | 30 | 30 | 30 | 3.0 | 2.4 | 2.7 | 3.8 | - | - | - |
| 466 | 0.1 | 1 | 30 | 30 | 50 | 3.0 | 2.4 | 2.7 | 3.9 | - | - | - |
| 467 | 0.1 | 1 | 30 | 50 | 50 | 3.1 | 2.6 | 2.8 | 4.1 | - | - | - |
| 468 | 0.1 | 1 | 50 | 50 | 50 | 4.9 | 3.9 | 4.2 | 6.1 | - | - | - |
| 469 | 0.1 | 5 | 5  | 5  | 10 | 1.6 | 1.6 | 2.0 | 2.3 | - | - | - |
| 470 | 0.1 | 5 | 5  | 5  | 30 | 1.6 | 1.6 | 2.0 | 2.3 | - | - | - |
| 471 | 0.1 | 5 | 5  | 5  | 50 | 1.6 | 1.6 | 2.0 | 2.4 | - | - | - |
| 472 | 0.1 | 5 | 5  | 10 | 10 | 1.6 | 1.6 | 2.0 | 2.4 | - | - | - |
| 473 | 0.1 | 5 | 5  | 10 | 30 | 1.6 | 1.6 | 2.0 | 2.4 | - | - | - |
| 474 | 0.1 | 5 | 5  | 10 | 50 | 1.6 | 1.6 | 2.0 | 2.5 | - | - | - |
| 475 | 0.1 | 5 | 5  | 30 | 30 | 1.8 | 1.7 | 2.1 | 2.5 | - | - | - |
| 476 | 0.1 | 5 | 5  | 30 | 50 | 1.9 | 1.7 | 2.1 | 2.6 | - | - | - |
| 477 | 0.1 | 5 | 5  | 50 | 50 | 2.1 | 1.9 | 2.2 | 2.8 | - | - | - |
| 478 | 0.1 | 5 | 10 | 10 | 10 | 2.0 | 1.9 | 2.4 | 2.9 | - | - | - |
| 479 | 0.1 | 5 | 10 | 10 | 30 | 2.0 | 1.9 | 2.4 | 2.9 | - | - | - |
| 480 | 0.1 | 5 | 10 | 10 | 50 | 2.0 | 1.9 | 2.4 | 3.0 | - | - | - |
| 481 | 0.1 | 5 | 10 | 30 | 30 | 2.3 | 2.0 | 2.5 | 3.2 | - | - | - |
| 482 | 0.1 | 5 | 10 | 30 | 50 | 2.3 | 2.0 | 2.5 | 3.2 | - | - | - |
| 483 | 0.1 | 5 | 10 | 50 | 50 | 2.5 | 2.2 | 2.6 | 3.3 | - | - | - |
| 484 | 0.1 | 5 | 30 | 30 | 30 | 3.9 | 3.3 | 3.9 | 5.1 | - | - | - |
| 485 | 0.1 | 5 | 30 | 30 | 50 | 3.9 | 3.3 | 4.0 | 5.2 | - | - | - |
| 486 | 0.1 | 5 | 30 | 50 | 50 | 4.1 | 3.5 | 4.0 | 5.3 | - | - | - |
| 487 | 0.1 | 5 | 50 | 50 | 50 | 5.6 | 4.8 | 5.4 | 7.4 | - | - | - |
| 488 | 1 | 1 | 1 | 1 | 10 | 1.0 | 1.0 | 1.0 | 1.0 | - | - | OK |
| 489 | 1 | 1 | 1 | 1 | 30 | 1.0 | 1.0 | 1.0 | 1.0 | - | - | OK |
| 490 | 1 | 1 | 1 | 1 | 50 | 1.0 | 1.0 | 1.1 | 1.0 | - | - | OK |
| 491 | 1 | 1 | 1 | 5 | 10 | 1.0 | 1.0 | 1.0 | 1.1 | - | - | OK |
| 492 | 1 | 1 | 1 | 5 | 30 | 1.0 | 1.0 | 1.1 | 1.1 | - | - | OK |
| 493 | 1 | 1 | 1 | 5 | 50 | 1.0 | 1.0 | 1.1 | 1.1 | - | - | OK |
| 494 | 1 | 1 | 1 | 10 | 10 | 1.1 | 1.1 | 1.1 | 1.1 | - | - | OK |
| 495 | 1 | 1 | 1 | 10 | 30 | 1.1 | 1.1 | 1.1 | 1.1 | - | - | OK |
| 496 | 1 | 1 | 1 | 10 | 50 | 1.1 | 1.1 | 1.1 | 1.1 | - | - | OK |
| 497 | 1 | 1 | 1 | 30 | 30 | 1.2 | 1.2 | 1.2 | 1.4 | - | - | OK |
| 498 | 1 | 1 | 1 | 30 | 50 | 1.2 | 1.2 | 1.2 | 1.4 | - | - | OK |
| 499 | 1 | 1 | 1 | 50 | 50 | 1.3 | 1.4 | 1.3 | 1.6 | - | - | - |
| 500 | 1 | 1 | 5 | 5 | 10 | 1.4 | 1.3 | 1.3 | 1.5 | - | - | OK |
| 501 | 1 | 1 | 5 | 5 | 30 | 1.4 | 1.3 | 1.3 | 1.5 | - | - | - |
| 502 | 1 | 1 | 5 | 5 | 50 | 1.4 | 1.3 | 1.4 | 1.5 | - | - | - |
| 503 | 1 | 1 | 5 | 10 | 10 | 1.4 | 1.3 | 1.3 | 1.5 | - | - | - |
| 504 | 1 | 1 | 5 | 10 | 30 | 1.4 | 1.3 | 1.4 | 1.6 | - | - | - |
| 505 | 1 | 1 | 5 | 10 | 50 | 1.4 | 1.3 | 1.4 | 1.6 | - | - | - |
| 506 | 1 | 1 | 5 | 30 | 30 | 1.6 | 1.5 | 1.5 | 1.8 | - | - | - |
| 507 | 1 | 1 | 5 | 30 | 50 | 1.6 | 1.5 | 1.5 | 1.8 | - | - | - |
| 508 | 1 | 1 | 5 | 50 | 50 | 1.7 | 1.6 | 1.6 | 2.1 | - | - | - |
| 509 | 1 | 1 | 10 | 10 | 10 | 1.8 | 1.6 | 1.7 | 2.0 | - | - | - |
| 510 | 1 | 1 | 10 | 10 | 30 | 1.8 | 1.6 | 1.7 | 2.1 | - | - | - |
| 511 | 1 | 1 | 10 | 10 | 50 | 1.8 | 1.7 | 1.7 | 2.1 | - | - | - |
| 512 | 1 | 1 | 10 | 30 | 30 | 2.0 | 1.8 | 1.8 | 2.3 | - | - | - |
| 513 | 1 | 1 | 10 | 30 | 50 | 2.0 | 1.8 | 1.9 | 2.3 | - | - | - |
| 514 | 1 | 1 | 10 | 50 | 50 | 2.2 | 2.0 | 2.0 | 2.6 | - | - | - |
| 515 | 1 | 1 | 30 | 30 | 30 | 3.6 | 3.1 | 3.3 | 4.3 | - | - | - |
| 516 | 1 | 1 | 30 | 30 | 50 | 3.6 | 3.1 | 3.3 | 4.4 | - | - | - |
| 517 | 1 | 1 | 30 | 50 | 50 | 3.7 | 3.2 | 3.4 | 4.6 | - | - | - |
| 518 | 1 | 1 | 50 | 50 | 50 | 5.4 | 4.5 | 4.8 | 6.6 | - | - | - |
| 519 | 1 | 5 | 5 | 5 | 10 | 2.2 | 2.2 | 2.5 | 2.8 | - | - | - |
| 520 | 1 | 5 | 5 | 5 | 30 | 2.2 | 2.2 | 2.5 | 2.8 | - | - | - |
| 521 | 1 | 5 | 5 | 5 | 50 | 2.2 | 2.2 | 2.6 | 2.9 | - | - | - |
| 522 | 1 | 5 | 5 | 10 | 10 | 2.3 | 2.2 | 2.5 | 2.9 | - | - | - |
| 523 | 1 | 5 | 5 | 10 | 30 | 2.3 | 2.2 | 2.6 | 2.9 | - | - | - |
| 524 | 1 | 5 | 5 | 10 | 50 | 2.3 | 2.2 | 2.6 | 2.9 | - | - | - |
| 525 | 1 | 5 | 5 | 30 | 30 | 2.5 | 2.4 | 2.6 | 3.1 | - | - | - |
| 526 | 1 | 5 | 5 | 30 | 50 | 2.5 | 2.4 | 2.7 | 3.1 | - | - | - |
| 527 | 1 | 5 | 5 | 50 | 50 | 2.7 | 2.5 | 2.8 | 3.3 | - | - | - |
| 528 | 1 | 5 | 10 | 10 | 10 | 2.7 | 2.6 | 2.9 | 3.4 | - | - | - |
| 529 | 1 | 5 | 10 | 10 | 30 | 2.7 | 2.6 | 3.0 | 3.4 | - | - | - |
| 530 | 1 | 5 | 10 | 10 | 50 | 2.7 | 2.6 | 3.0 | 3.4 | - | - | - |
| 531 | 1 | 5 | 10 | 30 | 30 | 2.9 | 2.7 | 3.1 | 3.6 | - | - | - |
| 532 | 1 | 5 | 10 | 30 | 50 | 2.9 | 2.7 | 3.1 | 3.7 | - | - | - |
| 533 | 1 | 5 | 10 | 50 | 50 | 3.1 | 2.8 | 3.2 | 3.7 | - | - | - |
| 534 | 1 | 5 | 30 | 30 | 30 | 4.6 | 4.0 | 4.5 | 5.6 | - | - | - |
| 535 | 1 | 5 | 30 | 30 | 50 | 4.6 | 4.0 | 4.5 | 5.7 | - | - | - |
| 536 | 1 | 5 | 30 | 50 | 50 | 4.7 | 4.1 | 4.6 | 5.8 | - | - | - |
| 537 | 1 | 5 | 50 | 50 | 50 | 6.2 | 5.4 | 6.0 | 7.9 | - | - | - |
| 538 | 5 | 5 | 5 | 5 | 10 | 5.0 | 5.0 | 5.0 | 5.0 | - | - | - |
| 539 | 5 | 5 | 5 | 5 | 30 | 5.0 | 5.0 | 5.0 | 5.0 | - | - | - |
| 540 | 5 | 5 | 5 | 5 | 50 | 5.0 | 5.0 | 5.1 | 5.0 | - | - | - |
| 541 | 5 | 5 | 5 | 10 | 10 | 5.0 | 5.0 | 5.0 | 5.1 | - | - | - |
| 542 | 5 | 5 | 5 | 10 | 30 | 5.0 | 5.0 | 5.1 | 5.1 | - | - | - |
| 543 | 5 | 5 | 5 | 10 | 50 | 5.0 | 5.1 | 5.1 | 5.1 | - | - | - |
| 544 | 5 | 5 | 5 | 30 | 30 | 5.2 | 5.2 | 5.2 | 5.3 | - | - | - |
| 545 | 5 | 5 | 5 | 30 | 50 | 5.2 | 5.2 | 5.2 | 5.3 | - | - | - |
| 546 | 5 | 5 | 5 | 50 | 50 | 5.3 | 5.3 | 5.3 | 5.6 | - | - | - |
| 547 | 5 | 5 | 10 | 10 | 10 | 5.4 | 5.4 | 5.4 | 5.6 | - | - | - |
| 548 | 5 | 5 | 10 | 10 | 30 | 5.4 | 5.4 | 5.4 | 5.6 | - | - | - |
| 549 | 5 | 5 | 10 | 10 | 50 | 5.4 | 5.4 | 5.5 | 5.6 | - | - | - |
| 550 | 5 | 5 | 10 | 30 | 30 | 5.7 | 5.5 | 5.5 | 5.8 | - | - | - |
| 551 | 5 | 5 | 10 | 30 | 50 | 5.7 | 5.5 | 5.5 | 5.9 | - | - | - |
| 552 | 5 | 5 | 10 | 50 | 50 | 5.8 | 5.7 | 5.7 | 6.2 | - | - | - |
| 553 | 5 | 5 | 30 | 30 | 30 | 7.2 | 6.8 | 6.9 | 7.9 | - | - | - |
| 554 | 5 | 5 | 30 | 30 | 50 | 7.2 | 6.8 | 7.0 | 7.9 | - | - | - |
| 555 | 5 | 5 | 30 | 50 | 50 | 7.3 | 6.9 | 7.0 | 8.1 | - | - | - |
| 556 | 5 | 5 | 50 | 50 | 50 | 9.0 | 8.2 | 8.5 | 10.0 | - | - | - |

**Notes.** The final WMF of a terrestrial planet is the median WMF calculated over the 37 planet analogs of the respective planet. If the median WMFs of Venus, Earth, and Mars simultaneously satisfied the respective observational constraints assumed in the fiducial (dry Venus), fiducial (wet Venus), and water worlds hypotheses, the WMF model was considered successful. See Sections 2.2 and 3.3 for more details.